\documentclass{article}

\usepackage[english]{babel}

\usepackage[letterpaper,top=2cm,bottom=2cm,left=3cm,right=3cm,marginparwidth=1.75cm]{geometry}

\usepackage{graphicx}
\usepackage{makecell}
\usepackage[colorlinks=true, allcolors=blue]{hyperref}

\usepackage[utf8]{inputenc}
\usepackage[T1]{fontenc}
\usepackage{lmodern} 
\usepackage{amsmath, amssymb} 
\usepackage{authblk} 

\usepackage{cite}

\hypersetup{
    colorlinks=true,
    linkcolor=blue,
    filecolor=magenta,      
    urlcolor=cyan,
} 

\newcommand{\gr}[1]{\overset{\circ}{#1}{}}

\newcommand{\qm}[1]{``#1''}

\begin{document}

\title{Projective Symmetry and Its Breaking in Quadratic Metric-Affine Gravity}

\author[1]{Carmen Ferrara}
\affil[1]{\small Scuola Superiore Meridionale, Largo San Marcellino 10, I-80138 Napoli, Italy\\
Istituto Nazionale di Fisica Nucleare, Sezione di Napoli,
Complesso Universitario di Monte S. Angelo, Via Cinthia Edificio 6, I-80126 Napoli, Italy }

\author[2]{Mar\'ia Jos\'e Guzm\'an}
\affil[2]{\small Laboratory of Theoretical Physics, Institute of Physics, University of Tartu, W. Ostwaldi 1, 50411 Tartu, Estonia}

\author[2]{Laur J\"{a}rv}

\maketitle

\begin{abstract}
We investigate generalized projective symmetry and its breaking in four-dimensional parity-even metric-affine gravity, considering an action linear in curvature and at most quadratic in torsion and nonmetricity. We derive the action of the generalized projective transformation on the irreducible components of the affine geometry and determine the coupling relations defining the axial, metric-trace, and fully projectively invariant theories. Complete invariance reduces the $12$ gravitational coefficients to a five-parameter family and generates four vectorial Noether identities for the connection equations.
Projective invariance makes the connection operator singular, thus we develop a symmetry-adapted method for solving its vacuum field equations. 
We then analyze couplings to Dirac, electromagnetic, and complex Klein-Gordon fields. Locally exact axial-projective transformations correspond to chiral rotations for massless fermions at the classical level, while a combination of covectors acts as an Abelian connection linking a locally exact vector-projective representative to scalar $U(1)$ symmetry. Standard Maxwell theory is independent of the affine connection, whereas a torsion-dependent Maxwell-like extension can preserve both electromagnetic and projective invariance through compensating Stückelberg fields.  
Exact gravitational projective invariance makes the corresponding connection directions nondynamical and forces matter currents sourcing them to vanish. Controlled explicit breaking lifts these zero modes and converts them into auxiliary fields.  Thus, projective symmetry provides a unified principle for identifying affine gauge modes, constraining matter couplings, and relating symmetry breaking in the gravitational sector to effective matter interactions.
\end{abstract}

\section{Introduction}

General relativity (GR) describes gravity as the dynamics of spacetime curvature, encoded in the Levi-Civita connection associated with the spacetime metric. In this framework, the affine structure is entirely determined by the metric, and the connection is assumed to be both torsion-free and metric-compatible. While this geometrical setting has been extraordinarily successful, both observationally and conceptually, \cite{Will1993,Ni_2016,De_Marchi_2020,Abbott2017-NS,LIGOScientific2021psn,Abbott2022,Stairs_2003,Kramer_2021}, it represents only a restricted realization of the possible geometric structures available on a differentiable manifold. 
Yet, notwithstanding its remarkable empirical success, there exist well-established theoretical and observational motivations to investigate extensions of GR and, more generally, alternative geometric formulations of gravity. From a theoretical perspective, GR still lacks a consistent unification with quantum field theory within a complete quantum theory of gravity. Moreover, its classical evolution generically leads to spacetime singularities, signaling the limitations of the classical framework and pointing toward the necessity of more fundamental descriptions \cite{Hooft1974,Obukhov:2017pxa,Giulini:2003tc,Rovelli:2004tv}. From the observational standpoint, although the standard $\Lambda$CDM paradigm provides an excellent fit to a broad array of cosmological data, \cite{Perivolaropoulos:2021jda,DESI:2024mwx}, it depends on components whose microphysical nature remains largely unknown, including dark matter, \cite{Mayet:2016zxu,Capozziello:2017rvz,Arun:2017uaw}, dark energy, \cite{Carroll:2000fy,Planck:2018vyg}, and the physical mechanism responsible for inflation \cite{Guth:2005zr}. In addition, the model is challenged by persistent high-precision discrepancies, most notably the $H_0$ and $\sigma_8$ tensions \cite{Aloni:2021eaq,DiValentino:2021izs,Pantos:2026koc}. These anomalies have emerged as key observational benchmarks for testing the concordance model and offer a compelling data-driven motivation for exploring extensions of gravitational dynamics.

A more general formulation of gravity is obtained in metric-affine gravity (MAG), where the metric and affine connection are regarded as independent fundamental variables \cite{Ferraris:1982wci,Hehl:1994ue}. In such a framework, the connection is not constrained a priori to coincide with the Levi-Civita one \cite{Einstein1916}, and the geometry of spacetime is characterized not only by curvature, but also by torsion and nonmetricity, which are properties of the linear affine connection. Metric-affine geometry therefore provides a natural arena in which to explore extensions of gravitational dynamics beyond the Riemannian paradigm. The antisymmetric part of the affine connection gives rise to the torsion tensor, which measures the failure of infinitesimal parallelograms to close, while the covariant derivative of the metric defines the nonmetricity tensor, which measures the variation of lengths and angles under parallel transport \cite{Hehl:1994ue}. Together with curvature, these quantities describe the most general local properties of an affine spacetime. This richer geometric structure has attracted renewed interest in recent years, both from the perspective of modified gravity and from the search for alternative geometric formulations of gravitational interactions, including teleparallel, symmetric teleparallel, and more general metric-affine theories \cite{Bahamonde:2022kwg,Aoki:2023sum,Blixt:2020ekl,Blixt:2023kyr,Bahamonde:2024efl}. 

In the metric-affine approach, the gravitational action can be constructed from scalar invariants formed out of the curvature, torsion, and nonmetricity tensors. The simplest curvature contribution is the general curvature scalar built from the full affine connection, which reduces to the Ricci scalar of GR only when torsion and nonmetricity vanish. In addition, one may construct independent quadratic scalars from torsion and nonmetricity \cite{BeltranJimenez:2019odq}. The torsion sector can be described by a torsion scalar obtained from suitable contractions of the torsion tensor and its vectorial trace. Similarly, the nonmetricity sector may be encoded in a nonmetricity scalar formed by combinations of the nonmetricity tensor and the two independent nonmetricity traces.\footnote{The torsion and nonmetricity scalars define the action of teleparallel and symmetric teleparallel equivalents of general relativity (TEGR and STEGR, respectively), and offer alternatives to the familiar curvature structure appearing in GR, while allowing all irreducible components of the affine geometry to contribute simultaneously \cite{Jarv:2018bgs,BeltranJimenez:2019esp, Capozziello:2022zzh}.} A particularly interesting feature of the general metric-affine setting is the possibility of including mixed torsion–nonmetricity invariants. Such terms couple the two non-Riemannian sectors directly and are absent in purely torsional or purely nonmetricity formulations \cite{Iosifidis:2021tvx}. 

In the present work, we consider a geometric Lagrangian composed of the general curvature scalar, a quadratic nonmetricity scalar built from contractions of the nonmetricity tensor and its two vectorial traces, a quadratic torsion scalar constructed from contractions of the torsion tensor and torsion vector, and a mixed scalar involving products between torsion and nonmetricity tensors as well as between the torsion vector and the two nonmetricity vectors. This structure defines a broad class of metric-affine theories in which curvature, torsion, nonmetricity, and their mutual interactions are treated on equal footing. The inclusion of such a general geometric Lagrangian is motivated by several considerations. First, from a purely geometric point of view, once the affine connection is allowed to be independent of the metric, there is no fundamental reason to suppress torsion or nonmetricity, nor their possible couplings. Second, quadratic invariants in torsion and nonmetricity arise naturally in effective gravitational actions and in gauge-theoretic approaches to gravity \cite{Iosifidis:2021tvx,Iosifidis:2019fsh}. Third, the interplay between different non-Riemannian contributions may lead to dynamical cancellations, equivalent reformulations, or new symmetry structures that are not visible in restricted subclasses of metric-affine theories. In this sense, mixed torsion–nonmetricity terms are not merely additional corrections, but may play an essential role in identifying the fundamental degrees of freedom and invariant sectors of the theory. 

A central aspect of this paper is the introduction of the full projective transformation of the affine connection \cite{Sauro:2022proj,Barker:2024dhb}. This kind of transformation of the connection is of particular importance in metric-affine gravity because the connection contains more components than the Levi-Civita connection and may therefore possess gauge-like redundancies. The most familiar example is the one-vector projective transformation, i.e. $\delta_{proj}\Gamma^{\alpha}{}_{\mu\nu}=\delta^\alpha_\mu p_\nu$, for an arbitrary covector $p_\mu$, which also results in a projective invariance. In fact, in MAG theories that use the generalization of the Einstein-Hilbert action using an  independent affine connection, there is a vectorial projective mode that is not determined by the field equations. Such transformations leave certain curvature combinations invariant and play an important role in constraining the coupling between matter and the affine connection \cite{Afonso:2017bxr,Percacci:2020ddy}. On the other hand, the full projective transformations can mix the trace components of torsion and nonmetricity, thereby revealing hidden symmetries among the non-Riemannian sectors of the theory. 
In the theory developed here, we define the full projective transformation of the affine connection and investigate its action on the geometric quantities entering the Lagrangian. In particular, we analyze how the torsion tensor, the nonmetricity tensor, their vectorial traces, and the corresponding quadratic scalars transform. This allows us to identify the combinations of curvature, torsion, nonmetricity, and mixed torsion–nonmetricity terms that remain invariant under the proposed symmetry \cite{Marzo:2021esg}. The resulting invariant structure provides a guiding principle for selecting admissible geometric Lagrangians and for distinguishing physical connection degrees of freedom from pure gauge components. 
The relevance of generalized projective symmetry extends to the coupling between geometry and matter \cite{Sauro:2022proj,Barnich:1993vg,Deser:1963zzc,Rigouzzo:2023sbb}. Since matter can interact directly with the independent affine connection, identifying its gauge freedoms is essential for determining which connection components can support physical matter sources. 

The crown jewel of 20th century fundamental physics is the Standard Model of particle physics, which provides a paradigmatic example of the gauge principle as a constructive guide to fundamental interactions \cite{Einstein1905,Dirac:1927dy,Weyl29}. For a specified field content and its representations, requiring invariance under local $SU(3)_c \times SU(2)_L \times U(1)_Y$ transformations introduces gauge connections through covariant derivatives and determines the form of the minimal couplings between matter and gauge fields \cite{Schwinger:1948iu,Feynman:1949zx,Dyson:1949bp,Weinberg:1967tq}. The non-Abelian structure also dictates the pattern of gauge-boson self-interactions, placing the strong and electroweak interactions within a common theoretical framework \cite{Yang:1954ek,Feynman:1958ty,Glashow:1961tr,Goldstone:1962es,Greenberg:1964pe,ParticleDataGroup:2026mpi}. Together with Lorentz invariance and renormalizability, gauge symmetry therefore guides the construction of the Standard Model Lagrangian, with its coupling constants determined experimentally \cite{Kobayashi:1973fv,Gross1973UltravioletBO,tHooft:1971akt,Fritzsch:1973pi,GargamelleNeutrino:1974khg,ALEPH:2005ab}. The electroweak sector further illustrates how mass generation can be accommodated within this framework: the Higgs mechanism breaks the symmetry spontaneously through the vacuum while preserving gauge invariance of the action \cite{Englert:1964et,Higgs:1964ia,Gell-Mann:1964ewy,Kibble:1967sv,tHooft:1971qjg,Pich:2012sx}. A related perspective underlies gauge approaches to gravity, where curvature and torsion arise as field strengths associated with local linear transformations and translations, respectively, while the Weyl covector plays the role of a connection for local scale transformations \cite{Utiyama:1956sy,Hayashi:1979wj,Ivanenko:1983fts,Hehl:1994ue, Macias:1995hf}. Compared to gauging Lorentz, Poincar\'e, or Weyl symmetry, projective transformations have received much less attention in this respect (but see the papers \cite{Grover:2024whe,Brensinger:2024udu} on Thomas-Whitehead gravity). In the present work, 
we use generalized projective transformations of the affine connection as an organizing principle for constructing metric-affine gravitational actions and their couplings to matter \cite{Sauro:2022proj}. Their interpretation as gauge symmetries requires invariance of the complete gravity-matter action, possibly through simultaneous transformations of matter or compensating fields. Thus, projective symmetry guides the selection of invariant interactions and identifies restrictions on the corresponding matter sources. Its controlled explicit breaking in the gravitational sector allows the affected connection components to become auxiliary fields whose elimination generates effective matter interactions \cite{Weinberg:1978kz}.

Generalized projective transformations provide a systematic criterion for constructing compatible interactions and for identifying restrictions on the matter hypermomentum. Their interpretation as gauge transformations must therefore be assessed for the complete gravitational and matter action. Depending on the coupling, the symmetry may remain intact, survive through simultaneous transformations of the matter fields, or require compensating fields. In other cases, admitting nonvanishing matter currents requires an explicit breaking of the corresponding gravitational symmetry. We investigate these possibilities for Dirac, electromagnetic, and complex Klein–Gordon fields, examining both the construction of invariant interactions and the consistency conditions imposed by the connection equations. In this work, we construct a fully invariant matter coupling under the complete projective transformation of the affine connection. In particular, we analyze the coupling to fermionic fields, the electromagnetic field, and the Klein-Gordon scalar field. For the electromagnetic sector, we propose an extension of the classical Maxwell theory that simultaneously preserves both projective invariance and the $U(1)$ gauge symmetry, achieved through the implementation of the Stückelberg mechanism \cite{Ruegg:2003ps}. This construction guarantees that the full theory, encompassing both the gravitational and matter sectors, shares the same symmetry structure and that the resulting constraints are dynamically consistent. The extension of the symmetry to the matter sector provides a symmetry-based framework for controlling the vectorial components of torsion and nonmetricity and clarifies the role of mixed non-Riemannian contributions in general metric-affine gravity, in this way we can have a symmetry principle rather than by ad hoc restrictions on the connection.

The paper is organized as follows. In Section \ref{sec: Metric_affine}, we introduce the metric-affine framework, fixing our conventions, and review the relevant curvature, torsion, and nonmetricity tensors. In Section \ref{sec:quadratic-MAG} we focus on the theory we consider, the general parity-even quadratic action and its metric and connection field equations. In Section \ref{sec:Proj}, we introduce the full projective transformation of the linear affine connection \cite{Sauro:2022proj}, explaining in detail the geometric interpretation, in Subsection \ref{subsec: gen proj transf} we present its action on the geometrical metric-affine quantities, and in Subsection \ref{subsec:projective-transformations-geometric-sector} we obtain the conditions for invariance of the gravitational action. In Section \ref{sec:Proj Lag} we develop the symmetry-adapted solution of the vacuum connection equations and apply it successively to the axial-projective, metric-trace, and fully generalized projective sectors. Sections \ref{sec: fermions}, \ref{sec: elec field}, and \ref{sec:scalar field} analyze the coupling to Dirac, electromagnetic, and complex Klein-Gordon fields, respectively. In particular, Section \ref{sec: fermions} establishes the relation between axial-projective transformations and classical chiral symmetry; Section \ref{sec: elec field} contrasts ordinary Maxwell theory with a Stückelberg metric-affine field strength; Section \ref{sec:scalar field} identifies the affine vector that acts as an Abelian connection for the scalar field. In Section \ref{sec: full matter sector}, we combine these matter sectors and discuss their simultaneous electromagnetic, chiral, scalar-phase, and projective transformations. Section \ref{sec: full matter invariant lag} studies the coupled gravitational and matter field equations. In Section \ref{sec: Axial Dirac}, we analyze the axial-projective sector, its explicit breaking, and the induced four-fermion interaction, while in Section \ref{sec: vector-u(1)} we perform the analogous analysis for the vector-projective scalar sector and derive the resulting field-dependent current-current interaction. Our conclusions and possible extensions are presented in Section \ref{sec: conclusions}.

\section{Metric-affine geometry}\label{sec: Metric_affine}
Metric-affine geometry provides a natural generalization of Riemannian geometry in which the metric and affine connection are treated as independent geometrical structures. In the Riemannian case, the affine connection is uniquely fixed to be the Levi-Civita connection by imposing vanishing torsion and metric compatibility. Metric-affine geometry relaxes these conditions, allowing the independent connection to possess both torsion and nonmetricity. A gravitational theory is then obtained, once it is specified an action for the metric, for the independent connection, and, when present, for matter fields. A metric-affine geometry is defined by the triplet $\{\mathcal{M},g_{\mu\nu},\Gamma^{\rho}_{\ \mu\nu}\}$, where $\mathcal{M}$ is a four-dimensional spacetime manifold, $g_{\mu\nu}$ is a rank-two symmetric tensor (with 10 independent components), and $\Gamma^{\rho}_{\ \mu\nu}$ is the affine connection (endowed with 64 independent components). In general, the metric and the affine connection do not have to be related; in fact, the former describes distances and angles (which in physics establish the \emph{causal structure} of spacetime), while the latter defines the notion of direction in the sense of \emph{parallel transport} and covariant derivative (which in physics facilitates the coupling of matter to geometry).

We define the covariant derivative $\nabla$ acting on a generic $(1,1)$ tensor as follows:
\begin{equation}
\nabla_\mu A^{\alpha}_{\ \beta}= \partial_\mu A^{\alpha}_{\ \beta} + \Gamma^{\alpha}_{\ \rho\mu}A^{\rho}_{\ \beta} - \Gamma^{\rho}_{\ \beta\mu}A^{\alpha}_{\ \rho}.
\end{equation}
Moreover, in the presence of the metric tensor, the general linear connection can always be decomposed as
\begin{equation}\label{eq: general affine connection}
    \Gamma^{\rho}{}_{\mu\nu} = \gr{\Gamma}{}^{\rho}{}_{\mu\nu} + N^{\rho}{}_{\mu\nu},
\end{equation}
where, in addition to the Levi-Civita connection
\begin{equation}
\gr{\Gamma}{}^{\rho}{}_{\mu\nu} = \dfrac{1}{2} g^{\rho\sigma}( \partial_{\mu} g_{\nu\sigma} + \partial_{\nu} g_{\sigma\mu} - \partial_{\sigma}g_{\mu\nu}),
\end{equation}
we define the tensor $N^\rho{}_{\mu\nu}$ as the difference between the affine connections $\Gamma^{\rho}{}_{\mu\nu}$ and $\gr{\Gamma}{}^{\rho}{}_{\mu\nu}$.
In this framework, $N^\rho{}_{\mu\nu}$ can be uniquely decomposed as 
\begin{equation}
    N^{\rho}{}_{\mu\nu} = K^{\rho}{}_{\mu\nu} + L^{\rho}{}_{\mu\nu},
\end{equation}
where $K^{\rho}_{\ \mu\nu}$ is the contortion tensor and $L^{\rho}_{\ \mu\nu}$ is the disformation tensor, whose explicit expressions are
\begin{subequations}
\begin{align}
    K^{\rho}{}_{\mu\nu} &= \dfrac{1}{2} g^{\rho \alpha}( T_{\mu\alpha\nu} +T_{\nu\alpha\mu} -T_{\alpha\mu\nu} ) = - K_{\mu\ \nu}^{\ \rho} \label{eq: contorsion}, \\
    L^{\rho}{}_{\mu\nu} &= \dfrac{1}{2} g^{\rho\alpha} (Q_{\alpha\mu\nu}- Q_{\mu\alpha\nu} - Q_{\nu\alpha\mu}) = L^{\rho}{}_{\nu\mu} \label{eq: disformation}.
\end{align}
\end{subequations}
The Levi-Civita connection and the disformation tensor are symmetric in their last two indices, while the contortion tensor characterizes the antisymmetric part of the connection through the torsion tensor
\begin{equation}\label{eq: torsion}
    T^{\rho}{}_{\nu\mu} = \Gamma^{\rho}{}_{\mu\nu} - \Gamma^{\rho}{}_{\nu\mu}.
\end{equation}
On the other hand, the disformation tensor is written in terms of the nonmetricity tensor, which is given by 
\begin{equation}\label{eq: nonmetricity}
    Q_{\rho\mu\nu} = \nabla_{\rho} g_{\mu\nu} = \partial_{\rho} g_{\mu\nu} - \Gamma^{\alpha}{}_{\mu\rho}\, g_{\alpha\nu} - \Gamma^{\alpha}{}_{\nu\rho}\, g_{\mu\alpha},
\end{equation}
and represents the deviation of the connection from metric-compatibility. The last property of the general linear connection is the curvature tensor, which reads
\begin{equation}\label{eq: curvaure}
    R^{\sigma}{}_{\rho\mu\nu} =  \partial_{\mu} \Gamma^{\sigma}{}_{\rho\nu} -  \partial_{\nu} \Gamma^{\sigma}{}_{\rho\mu} + \Gamma^{\alpha}{}_{\rho\nu} \Gamma^{\sigma}{}_{\alpha\mu} - \Gamma^{\alpha}{}_{\rho\mu}\Gamma^{\sigma}{}_{\alpha\nu}.
\end{equation}
In the most general case there are three nontrivial rank-2 contractions of the curvature tensor that we denote as $R_{\mu\nu}, \hat{R}_{\mu\nu}$ and $\check{R}_{\mu\nu}$
\begin{equation}
R_{\mu\nu} = R^{\alpha}{}_{\mu\alpha\nu}, \qquad \hat{R}_{\mu\nu} = R^{\alpha}{}_{\alpha\mu\nu}, \qquad \check{R}_{\mu\nu} =g_{\mu\sigma}g^{\alpha\beta}
R^\sigma{}_{\alpha\beta\nu}.
\end{equation}
There is only one nontrivial scalar curvature, given by
\begin{equation}\label{eq: mag ricci scalar}
R = g^{\mu\nu} R_{\mu\nu}.
\end{equation}
Curvature, torsion and nonmetricity characterize distinct geometrical
properties of the affine connection and of the metric-connection pair, and they differently affect the parallel transport of a vector on a manifold:
\begin{itemize}
    \item \emph{curvature} causes a non-null angle when a vector is parallel transported along a closed curve on a non-flat background; 
    \item \emph{torsion} entails a twisted geometry, where the parallel transport of two vectors is antisymmetric by exchanging the transported vectors and the direction of transport. This property results in the non-closure of parallelograms; 
    \item \emph{nonmetricity} alters the length and the angles of the vectors along the parallel transport. 
\end{itemize}

In metric-affine framework, the following geometric relation holds 
\begin{equation}\label{eq: general R}
    R = \gr{R}+T+B_T-Q+B_Q + T_{\rho\mu\nu} Q^{\mu\nu\rho} 
- T_\mu Q^\mu 
+ T_\mu \tilde{Q}^\mu,
\end{equation}
where $R$ is the general Ricci scalar defined with respect to the general linear affine connection \eqref{eq: general affine connection}, and $\gr{R}$ is the Ricci scalar computed from the Levi-Civita part of the connection. Moreover,
\begin{subequations}
\begin{align}    
Q_\alpha&=Q_{\alpha\lambda}{}^\lambda \label{eq:Qa} \,,\\
\tilde{Q}_\alpha&=Q^\lambda_{\ \lambda\alpha},
\label{eq:tildeQa}
\end{align}
\end{subequations}
represent two independent traces of the nonmetricity tensor.
Using Eqs.\ \eqref{eq:Qa} and \eqref{eq:tildeQa}, 
we can define the irreducible decomposition of the nonmetricity tensor in four spacetime dimensions:
\begin{equation}\label{eq:irreducible Q}
Q_{\alpha\mu\nu}
=
\frac{1}{18}
\left(
5Q_\alpha g_{\mu\nu}
-Q_\mu g_{\alpha\nu}
-Q_\nu g_{\alpha\mu}
\right)+
\frac{1}{9}
\left(
-\tilde Q_\alpha g_{\mu\nu}
+2\tilde Q_\mu g_{\alpha\nu}
+2\tilde Q_\nu g_{\alpha\mu}
\right)
+\Omega_{\alpha\mu\nu},
\end{equation}
where
\begin{equation}
\Omega_{\alpha\mu\nu}
=
\Omega_{\alpha\nu\mu},
\qquad
\Omega_{\alpha\mu}{}^\mu=0,
\qquad
\Omega^\alpha{}_{\alpha\mu}=0.
\end{equation}
On the other hand, it is possible to define the trace of the torsion tensor $T^\alpha{}_{\mu\nu}$
\begin{equation}\label{eq:Ta}
   T^{\alpha\mu}{}_{ \alpha}=T^\mu
\end{equation}
as the torsion vector, and 
\begin{equation}
S^\mu=\epsilon^{\mu\nu\rho\sigma}T_{\nu\rho\sigma},
\label{eq:axial-torsion-vector}
\end{equation}
as the axial torsion, a pseudo-vector obtained after the contraction of $T^{\rho}{}_{\mu\nu}$ with the Levi-Civita tensor.
Having introduced these two vectors \eqref{eq:Ta} and \eqref{eq:axial-torsion-vector}, we can proceed to decompose the torsion tensor into its irreducible components. In four spacetime dimensions, the torsion tensor can be decomposed into three irreducible components: a vector, axial-vector, and a purely tensorial part
\begin{equation}\label{eq:irreducible T}
    T^{\rho}_{\ \mu\nu}= \frac{1}{3}(\delta^\rho_\nu T_\mu - \delta^\rho_\mu T_\nu) + \frac{1}{6}\epsilon^\rho_{\ \sigma\mu\nu}S^\sigma + t^\rho_{\ \mu\nu}.
\end{equation}
Finally, $Q$ is the nonmetricity scalar with its related boundary term $B_Q$:
\begin{subequations}
    \begin{align}\label{eq: Q}
        Q &= -\frac{1}{4}Q^{\alpha\rho\nu}Q_{\alpha\rho\nu}+ \frac{1}{2}Q^{\alpha\rho\nu}Q_{\rho\nu\alpha}+\frac{1}{4}Q^{\alpha}Q_{\alpha} -\frac{1}{2} \tilde{Q}^{\alpha}Q_{\alpha}, \\
        B_Q &=\gr{\nabla}_{\alpha}(Q^{\alpha}-\tilde{Q}^{\alpha})\label{eq: boundaryQ},
    \end{align}
\end{subequations}
and $T$ is the torsion scalar, with its related boundary term $B_T$:
\begin{subequations}
\begin{align}\label{eq:general_torsion}
T&=\frac{1}{4} T_{\alpha \mu \nu} T^{\alpha \mu \nu}+\frac{1}{2} T_{\alpha \mu \nu} T^{\mu \alpha \nu}-T_{\alpha} T^{\alpha},\\
 B_T&= 2\gr{\nabla}_\alpha T^{\alpha} \label{eq: boundaryT}.
 \end{align}
\end{subequations}
With all these ingredients, we are ready to build the Lagrangian for metric-affine gravity, which is the subject of the next section.

\section{Parity-even quadratic metric-affine gravity}
\label{sec:quadratic-MAG}

We now introduce the class of metric-affine geometric theories of gravity that will be analyzed throughout this work.
We consider a four-dimensional spacetime endowed with a metric $g_{\mu\nu}$ and an independent affine connection $\Gamma^{\rho}{}_{\mu\nu}$. The latter is allowed to possess both torsion and nonmetricity, whose vector traces are defined by Eqs. \eqref{eq:Qa}, \eqref{eq:tildeQa}, and \eqref{eq:Ta}.
Rather than considering the complete curvature-quadratic metric-affine theory, we restrict attention to a parity-even sector that is linear in the curvature and at most quadratic in torsion and nonmetricity \cite{Iosifidis:2021tvx}. More precisely, we impose the following assumptions: (1) the action is local and diffeomorphism invariant; (2) the curvature enters only linearly through the general Ricci scalar \eqref{eq: mag ricci scalar}; (3) torsion and nonmetricity enter algebraically, at most quadratically, and no additional covariant derivatives of torsion or nonmetricity are introduced; (4) parity-odd contractions involving the Levi-Civita tensor $\epsilon^{\alpha\beta\mu\nu}$ are excluded.
This setup follows from the viewpoint of derivative counting: since
$R\sim L^{-2}$, while $T^{\rho}{}_{\mu\nu}\sim Q^{\rho\mu\nu}\sim L^{-1}$, the invariants $T^{2}$, $Q^{2}$, and $TQ$ have the same dimensionality as the Ricci scalar. By contrast, curvature-squared operators are of order $L^{-4}$ and belong to the next order in a derivative expansion \cite{BeltranJimenez:2019acz}. Moreover, such operators generally supply kinetic terms for components of the independent connection and can activate additional propagating modes. These modes are not automatically healthy: generic non-projectively-invariant higher-curvature sectors may contain ghost-like degrees of freedom \cite{BeltranJimenez:2019acz,BeltranJimenez:2019hrm}. Therefore, our restriction should be understood as defining the algebraic-connection sector of the theory, rather than as a claim that every curvature-quadratic metric-affine theory is pathological \cite{BeltranJimenez:2019hrm, delaCruz-Dombriz:2019tge,Percacci:2020ddy}.

Thus, in the considered case, the parity-even quadratic sector contains eleven independent invariants: three pure-torsion terms, five pure-nonmetricity terms, and three mixed torsion-nonmetricity terms \cite{Iosifidis:2021tvx}. Consequently, we write the gravitational action as
\begin{equation}
\label{RQTlagrangian}
S_{\mathrm g}[g,\Gamma]
=\frac{1}{2\kappa}\int d^4x\sqrt{-g}\mathcal L=
\frac{1}{2\kappa}\int d^4x\sqrt{-g}
\left[
a_0 R+L_{T}+L_{Q}+ L_{TQ}
\right],
\end{equation}
where
\begin{subequations}
\label{eq:quadratic-sectors}
\begin{align}
L_{T}
&=
b_1T_{\rho\mu\nu}T^{\rho\mu\nu}
+b_2T_{\rho\mu\nu}T^{\mu\nu\rho}
+b_3T_\mu T^\mu,
\label{eq:quadratic-sectors-torsion}
\\
L_{Q}
&=
c_1Q_{\rho\mu\nu}Q^{\rho\mu\nu}
+c_2Q_{\nu\mu\rho}Q^{\rho\mu\nu}
+c_3Q_\mu Q^\mu
+c_4\tilde Q_\mu\tilde Q^\mu
+c_5Q_\mu\tilde Q^\mu,
\label{eq:quadratic-sectors-nonmetricity}\\
L_{TQ}
&=
a_1Q_{\rho\mu\nu}T^{\nu\rho\mu}
+a_2Q_\mu T^\mu
+a_3\tilde Q_\mu T^\mu.
\label{eq:quadratic-sectors-mixed}
\end{align}
\end{subequations}
The coefficient $a_0$ fixes the overall gravitational scale, while $b_i$, $c_i$, and $a_i$ control, respectively, the torsional, nonmetric, and mixed sectors. A cosmological constant can be included without affecting the projective-symmetry analysis and will therefore be omitted \cite{Andrei:2024vvy}.
The mixed sector $L_{TQ}$ is particularly important in the present context. Generalized projective transformations mix the vector components of torsion and nonmetricity. Consequently, the space of projectively invariant theories is not, in general, closed if the mixed contractions are set to zero from the outset. Retaining all eleven quadratic invariants is therefore necessary for a complete classification of the projectively invariant parameter subspaces.

Since the metric and affine connection are treated as independent variables, the related Euler-Lagrange tensors arise from variation,
\begin{equation}
\delta S_{\mathrm g}
=
\frac{1}{2\kappa}\int d^4x\sqrt{-g}
\left(
\mathcal M_{\mu\nu}\delta g^{\mu\nu}
+
\mathcal P_{\alpha}{}^{\mu\nu}\delta\Gamma^{\alpha}{}_{\mu\nu}
\right)
+\text{boundary terms}.
\label{eq:variational-definitions}
\end{equation}
The vacuum field equations are therefore
\begin{equation}
\mathcal M_{\mu\nu}=0,
\qquad
\mathcal P_{\alpha}{}^{\mu\nu}=0.
\label{eq:vacuum-field-equations}
\end{equation}
If a matter action $S_{\mathrm m}[g,\Gamma,\Psi]$ is included, we define the energy-momentum and hypermomentum tensors
\begin{equation}
\mathcal{T}_{\mu\nu}
=
-\frac{2}{\sqrt{-g}}
\frac{\delta S_{\mathrm m}}{\delta g^{\mu\nu}},
\qquad
\Delta_{\alpha}{}^{\mu\nu}
=
-\frac{2}{\sqrt{-g}}
\frac{\delta S_{\mathrm m}}
{\delta\Gamma^{\alpha}{}_{\mu\nu}}\label{eq: hyper} .
\end{equation}
With these conventions, the sourced equations become
\begin{align}
\mathcal M_{\mu\nu}
&=\kappa\mathcal{T}_{\mu\nu},\qquad
\mathcal P_{\alpha}{}^{\mu\nu}=
\kappa\Delta_{\alpha}{}^{\mu\nu} \label{eq: pala and hyper}.
\end{align}
We compute the explicit metric and connection Euler–Lagrange tensors of the gravitational action as defined in Eq. \eqref{RQTlagrangian}, obtaining
\begin{align}
\mathcal M_{\mu\nu}\label{eq: metric field equations}
={}&a_0\gr{R}_{\mu\nu}
-\frac12g_{\mu\nu}\Bigl[
a_0\bigl(
\gr{R}+T-Q
+T_{\rho\alpha\beta}Q^{\alpha\beta\rho}
+T_\alpha(\tilde Q^\alpha-Q^\alpha)
\bigr)
+L_T+L_Q+L_{TQ}
\Bigr]\notag\\
&+\gr{\nabla}_\alpha\Bigl[
(a_0+a_1)K^\alpha{}_{(\mu\nu)}
+(a_0-2c_2)L^\alpha{}_{\mu\nu}
+(2c_1+c_2)Q^\alpha{}_{\mu\nu}
\Bigr]\notag\\
&+(a_0+a_3)\gr{\nabla}_{(\mu}T_{\nu)}
+\left(\frac{a_0}{2}+c_5\right)\gr{\nabla}_{(\mu}Q_{\nu)}
+2c_4\gr{\nabla}_{(\mu}\tilde Q_{\nu)}\notag\\
&-\left[
(a_0+a_3)T_\rho
+\left(\frac{a_0}{2}+c_5\right)Q_\rho
+2c_4\tilde Q_\rho
\right]
\bigl(K^\rho{}_{(\mu\nu)}+L^\rho{}_{\mu\nu}\bigr)\notag\\
&-\bigl(K^\rho{}_{\mu\alpha}+L^\rho{}_{\mu\alpha}\bigr)
\left[
\frac{a_0}{2}K^\alpha{}_{\rho\nu}
+a_1K^\alpha{}_{(\rho\nu)}
+(2c_1+c_2)Q^\alpha{}_{\rho\nu}
+\left(\frac{a_0}{2}-2c_2\right)L^\alpha{}_{\rho\nu}
\right]\notag\\
&-\bigl(K^\rho{}_{\nu\alpha}+L^\rho{}_{\nu\alpha}\bigr)
\left[
\frac{a_0}{2}K^\alpha{}_{\rho\mu}
+a_1K^\alpha{}_{(\rho\mu)}
+(2c_1+c_2)Q^\alpha{}_{\rho\mu}
+\left(\frac{a_0}{2}-2c_2\right)L^\alpha{}_{\rho\mu}
\right]\notag\\
&+a_1\left(
Q_{(\mu}{}^\alpha{}_\beta T^\beta{}_{\nu)\alpha}
-Q_{\alpha(\mu|\beta|}T_{\nu)}{}^{\alpha\beta}
\right)\notag\\
&+b_1\left(
2T^{\alpha\beta}{}_{(\mu}T_{\alpha\beta\nu)}
-T_{\mu\alpha\beta}T_\nu{}^{\alpha\beta}
\right)
+b_2T^\alpha{}_{\beta(\mu}T^\beta{}_{\nu)\alpha}\notag\\
&-2c_1Q^{\alpha\beta}{}_{(\mu}Q_{\alpha\beta\nu)}
-c_2Q^{\alpha\beta}{}_{(\mu}Q_{\beta\alpha\nu)}
+c_1Q_{\mu\alpha\beta}Q_\nu{}^{\alpha\beta}\notag\\
&+c_4\left(
2T_{(\mu}\tilde Q_{\nu)}
+Q_{(\mu}\tilde Q_{\nu)}
-\tilde Q_\mu\tilde Q_\nu
\right)
+(a_3+b_3)T_\mu T_\nu\notag\\
&+\left(a_2+\frac{a_3}{2}+c_5\right)Q_{(\mu}T_{\nu)}
+\left(c_3+\frac{c_5}{2}\right)Q_\mu Q_\nu\notag\\
&+g_{\mu\nu}\left[
\left(\frac{a_2-a_0}{2}\right)B_T
+(2c_3+c_5)\gr{\nabla}_\alpha Q^\alpha
-\left(\frac{a_0}{2}+c_5\right)B_Q
\right],
\end{align}
while the related connection field tensor are
\begin{equation}\label{eq: connection field eqs}
    \begin{aligned}
       \mathcal{P}_\alpha{}^{\mu\nu} =& a_0[Q_{\alpha}{}^{\mu\nu} - T_{\alpha}g^{\mu\nu} - \frac{1}{2}Q_\alpha g^{\mu\nu} + T^\mu \delta^\nu_\alpha + \frac{1}{2}Q^{\mu}\delta^{\nu}_{\alpha} - \tilde{Q}^\mu \delta^\nu_\alpha+ g^{\beta\mu}T^{\nu}{}_{\alpha\beta}] + b_1 4 T_{\alpha}{}^{\nu\mu} + 2b_2(T^\mu{}_{\alpha}{}^{\nu} - T^{\nu}{}_{\alpha}{}^\mu) \\ &+ 2b_3 T^{\rho}(\delta^\mu_\alpha\delta^\nu_\rho - \delta^\mu_\rho\delta^\nu_\alpha)
        -[4c_1Q^{\nu\mu}{}_{\alpha} +2c_2(Q^{\mu\nu}{}_{\alpha}+Q_{\alpha}{}^{\mu\nu}) + c_3(4Q^{\nu} \delta^\mu_\alpha)  + 2c_4(\tilde{Q}_\alpha g^{\mu\nu} + \tilde{Q}^\mu\delta^\nu_{\alpha}) \\& + c_5(2\tilde{Q}^\nu \delta^\mu_{\alpha} + Q_\alpha g^{\mu \nu} + Q^\mu \delta^{\nu}_{\alpha})] +
[-a_1(T_{\alpha}{}^{\nu\mu} + T^{\mu\nu}{}_\alpha + Q^{\mu\nu}{}_\alpha - Q^{\nu\mu}{}_{\alpha}) + a_2(-2T^\nu \delta^\mu_\alpha - Q^\mu \delta^\nu_\alpha + Q^\nu \delta^\mu_\alpha) \\ &+ a_3 (-T_\alpha g^{\mu\nu} - T^\mu  \delta^\nu_\alpha + \tilde{Q}^\nu\delta^\mu_\alpha - \tilde{Q}^\mu\delta^\nu_\alpha)] \,.
    \end{aligned}
\end{equation}
Note that in the latter expression torsion and nonmetricity enter only algebraically. This illustrates why the connection in our theory is a nondynamical quantity.

\section{Projective transformations}\label{sec:Proj}
Projective transformations provide a useful way of separating the information contained in an affine connection from the information encoded in its autoparallel trajectories. In metric-affine gravity this distinction is particularly relevant because the metric and the connection are independent variables: a transformation of the connection can preserve a family of curves without preserving the curvature scalar or the gravitational action. Conversely, a symmetry of the action can identify connections whose generic autoparallels are different. These are distinct questions, which we will examine separately.
In Sec. \ref{subsec: gen proj transf}, we first discuss the standard projective
transformation generated by a single covector and then introduce its
four-vector generalization. Autoparallels allow us to identify precisely which
parts of this generalization are projective in the strict geometric sense and which
preserve only null trajectories. In Sec. \ref{subsec:projective-transformations-geometric-sector}, we study the corresponding transformations of the general Ricci scalar and explain how generalized projective invariance can be imposed on the complete quadratic gravitational action. Throughout this section, spacetime is four-dimensional and Lorentzian, the metric is held fixed, and the connection is unrestricted unless stated otherwise.

\subsection{Generalized projective transformations and autoparallel curves}\label{subsec: gen proj transf}

In metric-affine geometry, the metric and affine structures play conceptually distinct roles. The metric $g_{\mu\nu}$ determines lengths, angles, and causal character, whereas the independent connection $\Gamma^{\rho}{}_{\mu\nu}$ determines parallel transport and its associated autoparallel curves.
An autoparallel is a curve whose tangent is parallel transported along itself,
up to a rescaling associated with the choice of parameter.  For a curve
$x^\mu(\tau)$ with tangent $u^\mu=dx^\mu/d\tau$, its equation is
\begin{equation}
u^\mu\nabla_\mu u^\rho
=\frac{d^2x^\rho}{d\tau^2}
+\Gamma^\rho{}_{\mu\nu}u^\mu u^\nu
=f(\tau)u^\rho.
\label{eq:nonaffine-autoparallel}
\end{equation}
The function $f$ vanishes for an affine parameter and can locally be removed
by reparametrizing the curve.
Two connections are said to be \emph{projectively equivalent} when they determine the same unparametrized autoparallel trajectories. Therefore, projective equivalence concerns the images of the curves in spacetime,
rather than a particular parametrization of those images.
This notion is purely kinematical and must be distinguished from projective invariance of an action: a transformation of the connection represents a gauge symmetry of a theory only if the corresponding action is invariant.
The standard metric-affine projective transformation is
\begin{equation}
\Gamma^{\rho}{}_{\mu\nu}
\longrightarrow
\Gamma^{\prime\rho}{}_{\mu\nu}
=\Gamma^{\rho}{}_{\mu\nu}+\delta^{\rho}_{\mu}p_{\nu},
\qquad g'_{\mu\nu}=g_{\mu\nu},
\label{eq:standard-projective-transformation}
\end{equation}
where $p_\nu$ is an arbitrary local covector. Since
$\delta^\rho_\mu \, p_\nu u^\mu u^\nu=p_\nu u^\nu u^\rho$, we obtain
\begin{equation}
u^\mu\nabla'_\mu u^\rho
=(f+p_\mu u^\mu)u^\rho.
\label{eq:standard-projective-autoparallel}
\end{equation}
The additional acceleration is entirely tangent to the curve and therefore
changes only its parametrization. In particular, if $\tau$ is an affine
parameter for $\Gamma$, an affine parameter $\tau'$ for $\Gamma'$ along
the same trajectory satisfies
\begin{equation}
\frac{d\tau'}{d\tau}
=C\exp\!\left[
\int_{\tau_0}^{\tau}
p_\nu\bigl(x(\ell)\bigr)
\frac{dx^\nu}{d\ell}\,d\ell
\right],
\qquad C\neq0.
\label{eq:standard-projective-reparametrization}
\end{equation}
The integral is taken along the chosen curve; no assumption that $p_\mu$ is
an exact covector is required. This establishes the projective character of
Eq.\ \eqref{eq:standard-projective-transformation} directly from the
autoparallel equation.

The basic projective transformation explores only a particular freedom of the independent connection. It changes the torsion vector and the two nonmetricity traces in fixed proportions, while leaving the axial torsion unchanged. A general metric-affine connection, however, contains these components independently, and different matter couplings can probe different combinations of them. Extending the transformation therefore allows us to examine whether additional components of the connection can also be changed without affecting the gravitational action, and whether those freedoms remain compatible with matter.
The generalized transformation is useful precisely because it identifies these restrictions across the vector and axial sectors of the connection. Our analysis will therefore distinguish the geometrical preservation of autoparallel trajectories from invariance of the gravitational action, and subsequently examine the consequences of that invariance for coupling to matter.
To make this extension precise, consider local connection deformations that
contain no derivatives of the transformation parameters, are linear in
covector parameters, and are constructed using only the metric, the
Kronecker tensor, and the Levi-Civita volume tensor. In four dimensions the
independent index structures give
\begin{equation}
\Gamma^{\rho}{}_{\mu\nu}
\longrightarrow
\Gamma^{\prime\rho}{}_{\mu\nu}
=\Gamma^{\rho}{}_{\mu\nu}
+\delta^{\rho}_{\mu}p_{\nu}
+\delta^{\rho}_{\nu}q_{\mu}
+g_{\mu\nu}r^{\rho}
+\epsilon^{\sigma\rho}{}_{\mu\nu}s_{\sigma}.
\label{genprojtr}
\end{equation}
This is the generalized projective transformation considered in
Ref.~\cite{Sauro:2022proj}. The parameters $p_\mu$, $q_\mu$, $r_\mu$,
and $s_\mu$ are independent local covectors, with
$r^\mu=g^{\mu\nu}r_\nu$. The last parameter describes the axial sector.
Its parity characterization depends on the transformation assigned to
$s_\mu$: if it is an ordinary covector, the epsilon contribution is parity
odd; if it is an axial covector, its product with $\epsilon^{\sigma\rho}{}_{\mu\nu}$ transforms as an ordinary tensor.
Thus the presence of an epsilon tensor in the transformation does not, by itself, require parity-odd terms in the gravitational action.
The first additive contribution in Eq.\ \eqref{genprojtr} is the standard projective shift in \eqref{eq:standard-projective-transformation},  while the second has the lower connection indices interchanged \cite{Iosifidis:2018zwo}. The third contribution is a metric-trace deformation: for each fixed $\rho$, its symmetric lower-index part is proportional to $g_{\mu\nu}$.\footnote{The term $g_{\mu\nu} r^{\rho}$ is pure trace with respect to the lower index pair, since for fixed $\rho$ it is proportional to the metric and therefore has vanishing traceless part.} The last term belongs to the axial sector and is antisymmetric in $\mu$ and $\nu$. Its precise parity characterization depends on the transformation assigned to $s_\mu$: if $s_\mu$ is treated as an ordinary covector, the product $\epsilon^{\sigma\rho}{}_{\mu\nu}s_\sigma$ is parity odd; alternatively, $s_\mu$ may be regarded as an axial covector.\footnote{In arbitrary dimension, the axial contribution generalizes to 
\begin{equation*}
\Gamma^{\rho}{}_{\mu\nu}\big|_{s}
=
\epsilon^{\alpha_1\cdots\alpha_{d-3}\rho}{}_{\mu\nu}
s_{\alpha_1\cdots\alpha_{d-3}},
\end{equation*}
where $s_{\alpha_1\cdots\alpha_{d-3}}$ is a $(d-3)$-form. Therefore, $s$ is a covector in $d=4$, while in $d=3$ it reduces to a pseudoscalar, and its contribution changes sign under parity transformation.}

A useful way to understand the geometric content of Eq.\ \eqref{genprojtr} is to decompose the deformation into its symmetric and antisymmetric parts with respect to the lower connection indices
\begin{subequations}
\begin{align}
\Gamma^{\rho}{}_{(\mu\nu)}
&=
\delta^{\rho}_{(\mu}(p+q)_{\nu)}
+g_{\mu\nu}r^\rho,
\\
\Gamma^{\rho}{}_{[\mu\nu]}
&=
\delta^{\rho}_{[\mu}(p-q)_{\nu]}
+\epsilon^{\sigma\rho}{}_{\mu\nu}s_\sigma .
\end{align}
\label{eq:projective-symmetric-antisymmetric}
\end{subequations}
Since the autoparallel equation contains the symmetric product $u^\mu u^\nu$, it probes only $\Gamma^{\rho}{}_{(\mu\nu)}$. Consequently, the combination $p_\mu-q_\mu$ and the axial parameter $s_\mu$ are invisible to the autoparallel trajectories, although they act nontrivially on the torsional components of the connection. For example, the torsion vector transforms in four dimensions as
\begin{equation}\label{eq:torsion-vector-projective}
T'_\mu=T_\mu+3(p_\mu-q_\mu).
\end{equation}

In order to make the projective interpretation explicit, we consider an autoparallel curve $x^\mu(\tau)$, with tangent vector $u^\mu=dx^\mu/d\tau$, satisfying Eq.\ \eqref{eq:nonaffine-autoparallel}.
Under Eq.\ \eqref{genprojtr}, one finds
\begin{equation}
u^\mu\nabla'_\mu u^\rho
=
u^\mu\nabla_\mu u^\rho
+(\delta^{\rho}_{\mu}p_{\nu}
+\delta^{\rho}_{\nu}q_{\mu}
+g_{\mu\nu}r^{\rho}
+\epsilon^{\sigma\rho}{}_{\mu\nu}s_{\sigma})u^\nu u^\mu,
\end{equation}
in which the additional contribution reduces to
\begin{equation}\label{eq:projective-autoparallel-contraction}
\Gamma^\rho_{(\mu\nu)}u^\mu u^\nu=
[(p_\mu+q_\mu)u^\mu]u^\rho
+u^2r^\rho,
\end{equation}
where $u^2=g_{\mu\nu}u^\mu u^\nu$. The Levi-Civita contribution vanishes because ($u^\mu u^\nu$) is symmetric, whereas
$\epsilon^{\sigma\rho}{}_{\mu\nu}$ is antisymmetric in $\mu,\nu$. Hence
\begin{equation}\label{eq:transformed-autoparallel}
u^\mu\nabla'_\mu u^\rho
=
\left[
f+(p_\mu+q_\mu)u^\mu
\right]u^\rho
+u^2r^\rho.
\end{equation}

Equation~\eqref{eq:transformed-autoparallel} separates the four sectors geometrically. The combination $(p_\mu+q_\mu)$ changes only the term tangent to the curve and can therefore be absorbed into a redefinition of its parameter. The complementary combination $(p_\mu-q_\mu)$, as well as $s_\mu$, does not even change the non-affinity function: these sectors leave the autoparallel equation unchanged. By contrast, the $r^\mu$-sector contributes the generally non-tangential term $u^2r^\rho$. Therefore, it changes generic timelike and spacelike autoparallel trajectories. For a null curve, however, $u^2=0$, so this contribution vanishes and the transformed connection determines the same unparametrized null trajectory. Within the family \eqref{genprojtr}, the complete projective structure is therefore preserved when $r_\mu=0$, whereas the null, or light-cone, projective structure is preserved for arbitrary $r_\mu$ \cite{Sauro:2022proj}.

The latter statement does not mean that the conformal structure has been transformed: throughout Eq.\ \eqref{genprojtr}, the metric and hence its null cones remain fixed. What is preserved is the set of unparameterized null autoparallels. The light-cone projective structure and the conformal structure are therefore conceptually distinct, even though both involve the metric null cone \cite{Matveev:2020wif,Sauro:2022proj}.

There is also a minor subtlety associated with the expression \qm{null autoparallel}. In a generic nonmetric geometry, an initially null tangent vector need not remain null under affine parallel transport, since Eq.\ \eqref{eq:nonaffine-autoparallel} gives
\begin{equation}
\frac{d u^2}{d\tau}
=
Q_{\lambda\mu\nu}u^\lambda u^\mu u^\nu
+2fu^2.
\label{eq:nullity-evolution}
\end{equation}
Thus, the null-projective statement applies to autoparallels whose tangent remains null. This property holds, for example, for Weyl-type nonmetricity
$Q_{\lambda\mu\nu}\propto A_\lambda g_{\mu\nu}$, and more generally whenever
$Q_{(\lambda\mu\nu)}u^\lambda u^\mu u^\nu=0$ along the curve.

The terminology \emph{generalized projective transformation} should
therefore be understood with this distinction in mind. Equation~\eqref{genprojtr}
extends the standard projective shift to the full set of vector and axial
deformations under consideration, but arbitrary members of the family are
not projective equivalences for all curves. Even within the $r_\mu=0$
subfamily, path preservation does not establish invariance of an action.
For comparison, if both connections are required to be torsion-free, their
difference must be symmetric: within this family this sets $p_\mu=q_\mu$
and $s_\mu=0$. Requiring full projective equivalence additionally sets
$r_\mu=0$, recovering the familiar symmetric projective shift
$\Gamma^{\prime\rho}{}_{\mu\nu}-\Gamma^\rho{}_{\mu\nu}
=2\delta^\rho_{(\mu}p_{\nu)}$.

Autoparallels serve here as a kinematical diagnostic of these transformations.
They are not assumed to describe the motion of arbitrary physical test
bodies. A point-particle action depending only on the metric gives
Levi-Civita geodesics; matter with a direct coupling to the independent
connection can instead respond to torsion and nonmetricity through its
hypermomentum. The relevant trajectories must be derived from the chosen
matter action and its conservation laws, rather than postulated from
Eq.\ \eqref{eq:nonaffine-autoparallel} alone \cite{Obukhov:2021uor}.
 Whether affine autoparallels admit an independent variational interpretation is a separate question. For example, in Finsler geometry, certain torsion-free metric-affine connections with vectorial nonmetricity can be parametrized by a suitable pseudo-Finsler structure, so that their autoparallels become geodesics of an effective velocity-dependent geometry \cite{Csillag:2026kdc}. In a different setting, torsionful autoparallels were obtained from an anholonomic variational principle, in which the allowed path variations do not commute with differentiation along the worldline \cite{Fiziev:1995te,Kleinert:1996yi,Kleinert:1997hr}. Both constructions require additional geometrical or variational assumptions. We shall not adopt either of them here.
Accordingly, we used autoparallels only to characterize the path-preserving content of the generalized projective transformations.
Whether physical test bodies actually follow these autoparallels depends on the chosen matter action and its coupling to the independent connection. This dynamical question lies beyond the scope of the present work.
Finally, the preceding discussion is kinematical and should not be interpreted as a universal postulate about the motion of physical test bodies. In a generic metric-affine theory, extremals of the metric length functional and autoparallels of the independent connection need not coincide. The relevant trajectories must ultimately be derived from the matter action and its conservation laws. A point-particle action depending only on $g_{\mu\nu}$ yields Levi-Civita geodesics, whereas matter carrying hypermomentum can couple directly to torsion and nonmetricity \cite{Iosifidis:2023mot, Obukhov:2021uor}. 

\subsection{Generalized projective transformations of the geometric sector}
\label{subsec:projective-transformations-geometric-sector}

We now turn from the paths determined by the connection to the scalar
curvature entering the gravitational action \eqref{RQTlagrangian}.
For the basic transformation \eqref{eq:standard-projective-transformation},
the curvature and Ricci tensor transform exactly as
\begin{subequations}
\label{eq:standard-projective-curvature}
\begin{align}
R^{\prime\sigma}{}_{\rho\mu\nu}
&=R^\sigma{}_{\rho\mu\nu}
+\delta^\sigma_\rho
\left(\partial_\mu p_\nu-\partial_\nu p_\mu\right),
\\
R'_{\rho\nu}&=R_{\rho\nu}+2\partial_{[\rho} p_{\nu]}.
\end{align}
\end{subequations}
Only the antisymmetric part of the Ricci tensor changes. Contraction with
the symmetric inverse metric gives the exact, pointwise identity
\begin{equation}
R'=g^{\rho\nu}R'_{\rho\nu}=R.
\label{eq:standard-projective-scalar-invariance}
\end{equation}
This result holds for arbitrary torsion and nonmetricity and for an
arbitrary covector $p_\mu$. It explains the projective freedom of the
metric-affine Einstein-Hilbert term. It also applies to a gravitational
Lagrangian depending on the connection only through an algebraic function
of $R$. Additional connection-dependent terms, including matter couplings,
must be checked separately.

The cancellation has an algebraic origin. For each fixed derivative
index $\nu$, the added term $\delta^\rho_\mu p_\nu$ is proportional
to the identity in the remaining pair of connection indices. It therefore
commutes with the connection in the quadratic curvature terms, leaving
only the curl in Eq.\ \eqref{eq:standard-projective-curvature}.
The same argument applies to the $p_\mu$ contribution in
Eq.\ \eqref{genprojtr}: all terms quadratic in $p_\mu$, and all mixed
terms containing $p_\mu$ and another transformation parameter, cancel.
The other three contributions have different index structures and their
effect on $R$ must be calculated explicitly. 
The transformation is local because $p_\mu$, $q_\mu$,
$r_\mu$, and $s_\mu$ may depend arbitrarily on the spacetime coordinates.
It is finite when every power of these parameters is retained. Although
the transformation of the connection is linear in the four vectors, the
curvature scalar and the quadratic torsion and nonmetricity invariants
also generate terms that are quadratic in them.
Therefore, the finite transformation of the nonmetricity tensor \eqref{eq: nonmetricity} is
\begin{equation}
Q'_{\rho\mu\nu}
=
Q_{\rho\mu\nu}
-2g_{\mu\nu}p_\rho
-g_{\rho\mu}(q_\nu+r_\nu)
-g_{\rho\nu}(q_\mu+r_\mu).
\label{eq:projective-transformation-nonmetricity}
\end{equation}
Consequently, its vector traces \eqref{eq:Qa} and \eqref{eq:tildeQa} transform in four dimensions as
\begin{equation}\label{eq:proj-transf-Q-traces}
Q'_\mu
=
Q_\mu-8p_\mu-2q_\mu-2r_\mu,
\qquad
\tilde Q'_\mu
=
\tilde Q_\mu-2p_\mu-5q_\mu-5r_\mu.
\end{equation}
The signs are opposite to those obtained when nonmetricity is defined with a minus sign \cite{Sauro:2022proj}.

The torsion tensor \eqref{eq: torsion} transforms according to
\begin{equation}
T'^{\rho}{}_{\nu\mu}
=
T^{\rho}{}_{\nu\mu}
+\delta^\rho_\mu(p_\nu-q_\nu)
-\delta^\rho_\nu(p_\mu-q_\mu)
+2\epsilon^{\alpha\rho}{}_{\mu\nu}s_\alpha.
\label{eq:projective-transformation-torsion}
\end{equation}
The vector \eqref{eq:Ta} and axial \eqref{eq:axial-torsion-vector} traces of torsion transform as
\begin{equation}
T'_\mu=T_\mu+3p_\mu-3q_\mu,
\qquad
S'_\mu=S_\mu+12s_\mu.
\label{eq:projective-transformation-torsion-traces}
\end{equation}
These are exact finite transformations. They affect only the vector traces
of nonmetricity and the vector and axial components of torsion. The
remaining irreducible tensor components are unchanged.
These formulas also explain why the four-vector extension is useful.
The two nonmetricity traces respond independently to $p_\mu$ and
$q_\mu+r_\mu$, whereas the torsion trace responds to $p_\mu-q_\mu$.
Consequently, $p_\mu$, $q_\mu$, and $r_\mu$ can be chosen to produce arbitrary independent shifts of $T_\mu$, $Q_\mu$, and $\tilde Q_\mu$.
The parameter $s_\mu$ independently shifts $S_\mu$. By comparison, the standard transformation, recovered for $p_\mu\neq0$ and $q_\mu=r_\mu=s_\mu=0$, changes the three ordinary traces in fixed proportions and leaves the axial trace untouched. If the complete action is invariant under all four transformations, the corresponding four vector sectors can be fixed by gauge choice. This stronger conclusion cannot be drawn from ordinary projective invariance alone. Under the generalized projective transformation \eqref{genprojtr},
the finite transformation of the curvature tensor is
\begin{align}
R'^\sigma{}_{\rho\mu\nu}
=&R^\sigma{}_{\rho\mu\nu}
+\delta^\sigma_\rho
\left(
\partial_\mu p_\nu-\partial_\nu p_\mu
\right)
+\delta^\sigma_\nu\nabla_\mu q_\rho
-\delta^\sigma_\mu\nabla_\nu q_\rho
+g_{\rho\nu}\nabla_\mu r^\sigma
-g_{\rho\mu}\nabla_\nu r^\sigma
+r^\sigma
\left(
Q_{\mu\rho\nu}-Q_{\nu\rho\mu}
\right)
\nonumber\\
&+g_{\alpha\rho}r^\sigma T^\alpha{}_{\mu\nu}
+q_\rho T^\sigma{}_{\mu\nu}
+\nabla_\mu
\left(
\epsilon^{\lambda\sigma}{}_{\rho\nu}s_\lambda
\right)
-\nabla_\nu
\left(
\epsilon^{\lambda\sigma}{}_{\rho\mu}s_\lambda
\right)
+T^\alpha{}_{\mu\nu}
 \epsilon^{\lambda\sigma}{}_{\rho\alpha}s_\lambda
\nonumber\\
&+q_\rho
\left(
\delta^\sigma_\mu q_\nu
-\delta^\sigma_\nu q_\mu
\right)+q_\alpha r^\alpha
\left(
\delta^\sigma_\mu g_{\rho\nu}
-\delta^\sigma_\nu g_{\rho\mu}
\right)
+r^\sigma
\left(
g_{\rho\nu}r_\mu
-g_{\rho\mu}r_\nu
\right)-2q_\rho
 \epsilon^{\lambda\sigma}{}_{\mu\nu}s_\lambda
\nonumber\\
&+s_\lambda q_\alpha
\left(
\delta^\sigma_\mu
 \epsilon^{\lambda\alpha}{}_{\rho\nu}
-\delta^\sigma_\nu
 \epsilon^{\lambda\alpha}{}_{\rho\mu}
\right)
+s_\lambda r^\alpha
\left(
g_{\rho\nu}
 \epsilon^{\lambda\sigma}{}_{\alpha\mu}
-g_{\rho\mu}
 \epsilon^{\lambda\sigma}{}_{\alpha\nu}
\right)
\nonumber\\
&+s_\lambda r^\sigma
\left(
g_{\alpha\mu}
 \epsilon^{\lambda\alpha}{}_{\rho\nu}
-g_{\alpha\nu}
 \epsilon^{\lambda\alpha}{}_{\rho\mu}
\right)+s_\lambda s_\kappa
\left(
\epsilon^{\lambda\alpha}{}_{\rho\nu}
\epsilon^{\kappa\sigma}{}_{\alpha\mu}
-
\epsilon^{\lambda\alpha}{}_{\rho\mu}
\epsilon^{\kappa\sigma}{}_{\alpha\nu}
\right).
\label{eq:finite-generalized projective-curvature-transformation}
\end{align}
The covariant derivatives on the right-hand side refer to the original
connection. The terms containing derivatives of the transformation
parameters, and those containing one transformation parameter multiplied
by torsion or nonmetricity, are linear. The remaining terms are quadratic.
There are no higher-order terms because curvature contains at most two
powers of the connection. As anticipated above, every quadratic or mixed
term containing $p_\mu$ cancels identically.
The contraction of $\sigma$ and $\mu$ in
\eqref{eq:finite-generalized projective-curvature-transformation}
gives the finite transformation of the Ricci tensor
\begin{align}
R'_{\rho\nu}
=& R_{\rho\nu}
+\partial_\rho p_\nu
-\partial_\nu p_\rho
-3\nabla_\nu q_\rho
-q_\rho T_\nu
+g_{\rho\nu}\nabla_\mu r^\mu
-\nabla_\nu r_\rho
+r^\mu Q_{\mu\rho\nu}
+g_{\alpha\rho}r^\mu T^\alpha{}_{\mu\nu}
+\nabla_\mu
\left(
\epsilon^{\lambda\mu}{}_{\rho\nu}s_\lambda
\right)\nonumber\\
&
+T^\alpha{}_{\mu\nu}
 \epsilon^{\lambda\mu}{}_{\rho\alpha}s_\lambda
+3q_\rho q_\nu
+3g_{\rho\nu}q_\alpha r^\alpha
+g_{\rho\nu}r_\alpha r^\alpha
-r_\rho r_\nu\nonumber\\
&+3\epsilon^{\lambda\alpha}{}_{\rho\nu}
 s_\lambda q_\alpha
+3\epsilon^{\lambda\alpha}{}_{\rho\nu}
 s_\lambda r_\alpha+2g_{\rho\nu}s_\alpha s^\alpha
-2s_\rho s_\nu.
\label{eq:finite-generalized projective-ricci-transformation}
\end{align}
The Ricci tensor is generally nonsymmetric. The curl of $p_\mu$, the
displayed derivative of the epsilon term, and the mixed terms proportional to $s_\lambda q_\alpha$ and
$s_\lambda r_\alpha$ are antisymmetric in $\rho,\nu$.
However, the torsion-epsilon contraction need not be antisymmetric,
and the term quadratic in $s_\mu$ is symmetric. The axial sector therefore contributes to the Ricci scalar despite being invisible to autoparallel trajectories.
In Eq.\ \eqref{eq:finite-generalized projective-ricci-transformation}, the
Levi-Civita tensor must remain inside the covariant derivative. Holding
the metric fixed under the transformation does not make it covariantly
constant with respect to a general affine connection. In particular,
one must use
\begin{equation}
\nabla_\mu \left(\epsilon^{\lambda\sigma}{}_{\rho\nu}s_\lambda\right)
=\left(\nabla_\mu\epsilon^{\lambda\sigma}{}_{\rho\nu}\right)s_\lambda
+\epsilon^{\lambda\sigma}{}_{\rho\nu}\nabla_\mu s_\lambda,
\label{eq:projective-epsilon-derivative}
\end{equation}
since nonmetricity generally makes the first term nonzero.
Contracting
\eqref{eq:finite-generalized projective-ricci-transformation}
with $g^{\rho\nu}$ gives
\begin{align}\label{eq:finite-generalized projective-curvature-scalar}
R'
={}&R
+3\gr{\nabla}_\mu(r^\mu-q^\mu)
+2q_\mu T^\mu
-2r_\mu T^\mu
+\frac{3}{2}q^\mu Q_\mu
-\frac{1}{2}r^\mu Q_\mu
-3q^\mu\tilde Q_\mu
-r^\mu\tilde Q_\mu
+s_\mu S^\mu
\nonumber\\
&+3q_\mu q^\mu
+12q_\mu r^\mu
+3r_\mu r^\mu
+6s_\mu s^\mu.
\end{align}

Several consequences follow. The parameter $p_\mu$ drops out exactly,
like in Eq.\ \eqref{eq:standard-projective-scalar-invariance}. The
$q_\mu$ sector, although projective for all autoparallel trajectories,
generically changes $R$. A pure axial shift leaves even the parametrized
autoparallel equation unchanged but gives
\begin{equation}
\left.R'-R\right|_{p=q=r=0}=s_\mu S^\mu+6s_\mu s^\mu.
\label{eq:axial-projective-scalar}
\end{equation}
The $r_\mu$ sector also changes $R$, while preserving only null
autoparallels in general. Consequently, preservation of all paths, or of
their null subset, is insufficient to guarantee invariance of scalar
curvature.

Moreover, the full change in
Eq.\ \eqref{eq:finite-generalized projective-curvature-scalar} is not merely
a total divergence. The trace-dependent and quadratic terms are bulk
contributions. Within this family, demanding invariance of $R$ for
arbitrary connections leaves only the standard $p_\mu$ freedom. Indeed,
the independent $S_\mu$ coefficient requires $s_\mu=0$, the $T_\mu$
coefficient requires $q_\mu=r_\mu$, and the $Q_\mu$ coefficient then
requires $q_\mu=r_\mu=0$. Special choices on a particular background do
not establish an off-shell symmetry of the theory.

The distinction between finite and infinitesimal transformations is
especially useful when testing action invariance. At first order in
$p_\mu$, $q_\mu$, $r_\mu$, and $s_\mu$, the terms quadratic in these parameters are omitted.
The infinitesimal transformation of the curvature
tensor is therefore
\begin{equation}
    \begin{aligned}
        \delta_{proj} R^{\sigma}{}_{\rho\mu\nu}
        =& \delta^{\sigma}_{\rho}(\partial_\mu p_\nu - \partial_\nu p_\mu)+\delta^\sigma_\nu \nabla_\mu q_\rho - \delta^{\sigma}_\mu \nabla_\nu q_\rho + g_{\rho \nu}\nabla_\mu r^\sigma- g_{\rho \mu}\nabla_\nu r^\sigma + r^{\sigma}(Q_{\mu\rho\nu} - Q_{\nu\rho\mu})\\ &+ g_{\alpha\rho}r^\sigma T^{\alpha}{}_{\mu\nu} + q_\rho T^{\sigma}{}_{\ \mu\nu} +\nabla_\mu(\epsilon^{\lambda\sigma}{}_{\rho\nu}s_\lambda)- \nabla_\nu(\epsilon^{\lambda\sigma}{}_{\rho\mu}s_\lambda) + T^{\alpha}{}_{\mu\nu} \epsilon^{\lambda\sigma}{}_{\rho\alpha}s_\lambda;
    \end{aligned}
\end{equation}
contracting the indices $\sigma$ and $\mu$ we get the variation of the general Ricci tensor
\begin{equation}
    \begin{aligned}
        \delta_{proj}R_{\rho\nu} =& \partial_\rho p_\nu - \partial_\nu p_\rho - 3\nabla_\nu q_\rho - q_\rho T_\nu + g_{\rho\nu} \nabla_\mu r^{\mu} - \nabla_\nu r_\rho + r^{\mu}Q_{\mu\rho\nu} + g_{\alpha\rho}r^\mu T^{\alpha}{}_{\mu\nu}\\ &+ \nabla_\mu(\epsilon^{\lambda\mu}{}_{\rho\nu} s_\lambda) + T^{\alpha}{}_{\mu\nu} \epsilon^{\lambda\mu}{}_{\rho\alpha} s_\lambda.
    \end{aligned}
\end{equation}
At first order in the transformation parameters, using Eq.\ \eqref{eq: general affine connection}, the curvature scalar varies as
\begin{equation}
\delta_{\mathrm{proj}}R
=
3\gr{\nabla}_\mu(r^\mu-q^\mu)
+2q_\mu T^\mu
-2r_\mu T^\mu
+\frac{3}{2}q^\mu Q_\mu
-\frac{1}{2}r^\mu Q_\mu
-3q^\mu\tilde Q_\mu
-r^\mu\tilde Q_\mu
+s_\mu S^\mu.
\label{eq:infinitesimal-projective-transformation-curvature-scalar}
\end{equation}
The failure of the Einstein-Hilbert term to possess the full enlarged symmetry does not exclude that symmetry for a more general action. Torsion and nonmetricity transform simultaneously, so their quadratic invariants can cancel the bulk variation of $R$. This is the relevant criterion for the parity-even quadratic metric-affine action considered here~\cite{Iosifidis:2021tvx}. The infinitesimal variation of the torsion-quadratic terms \eqref{eq:quadratic-sectors-torsion} is
\begin{equation}
\begin{aligned}
\delta_{\mathrm{proj}} L_{T}=&\delta_{\mathrm{proj}}
\Big(
b_1T_{\rho\mu\nu}T^{\rho\mu\nu}
+b_2T_{\rho\mu\nu}T^{\mu\nu\rho}
+b_3T_\mu T^\mu
\Big)
\\ =&
(4b_1-2b_2+6b_3)p_\mu T^\mu
-(4b_1-2b_2+6b_3)q_\mu T^\mu
-4(b_1+b_2)s_\mu S^\mu \,,
\end{aligned}
\label{eq:projective-variation-torsion-quadratic-terms}
\end{equation}
the infinitesimal variation of the nonmetricity-quadratic terms \eqref{eq:quadratic-sectors-nonmetricity} is
\begin{equation}
\begin{aligned}
\delta_{\mathrm{proj}} L_Q=&\delta_{\mathrm{proj}}
\Big(c_1Q_{\rho\mu\nu}Q^{\rho\mu\nu}
+c_2Q_{\nu\mu\rho}Q^{\rho\mu\nu}
+c_3Q_\mu Q^\mu+c_4\tilde Q_\mu\tilde Q^\mu +c_5Q_\mu\tilde Q^\mu
\Big)
\\ =&
-(4c_1+16c_3+2c_5)p_\mu Q^\mu
-(4c_2+4c_4+8c_5)p_\mu\tilde Q^\mu
\\
&-(2c_2+4c_3+5c_5)q_\mu Q^\mu-(2c_2+4c_3+5c_5)r_\mu Q^\mu
\\
&-(4c_1+2c_2+10c_4+2c_5)
q_\mu\tilde Q^\mu
-(4c_1+2c_2+10c_4+2c_5)
r_\mu\tilde Q^\mu \,,
\end{aligned}
\label{eq:projective-variation-nonmetricity-quadratic-terms}
\end{equation}
while the infinitesimal variation of the mixed torsion-nonmetricity terms \eqref{eq:quadratic-sectors-mixed} is
\begin{equation}
\begin{aligned}
\delta_{\mathrm{proj}}L_{TQ}=&\delta_{\mathrm{proj}}
\Big(a_1Q_{\rho\mu\nu}T^{\nu\rho\mu}
+a_2Q_\mu T^\mu
+a_3\tilde Q_\mu T^\mu
\Big)
\\=&
-(2a_1+8a_2+2a_3)p_\mu T^\mu+(a_1-2a_2-5a_3)q_\mu T^\mu
\\
&+(a_1-2a_2-5a_3)r_\mu T^\mu+(a_1+3a_2)p_\mu Q^\mu
-(a_1+3a_2)q_\mu Q^\mu
\\
&+(-a_1+3a_3)p_\mu\tilde Q^\mu+(a_1-3a_3)q_\mu\tilde Q^\mu.
\end{aligned}
\label{eq:projective-variation-mixed-quadratic-terms}
\end{equation}
The individual curvature, torsion, nonmetricity, and mixed sectors do not need to be invariant separately. Projective invariance is a property of the complete gravitational action, and cancellations among the different sectors are essential.
In the following variations, $\mathcal L$ denotes the scalar expression inside square brackets in the gravitational action \eqref{RQTlagrangian}; collecting all contributions gives 
\begin{align}\label{eq:fullprojL}
\delta_{\mathrm{proj}}\mathcal L
={}&
p_\mu T^\mu
\left(
-2a_1-8a_2-2a_3
+4b_1-2b_2+6b_3
\right)
+
q_\mu T^\mu
\left(
a_1-2a_2-5a_3
-4b_1+2b_2-6b_3+2a_0
\right)
\nonumber\\
&+
r_\mu T^\mu
\left(
a_1-2a_2-5a_3-2a_0
\right)
\nonumber\\
&+
p_\mu Q^\mu
\left(
a_1+3a_2
-4c_1-16c_3-2c_5
\right)
+
q_\mu Q^\mu
\left(
-a_1-3a_2
-2c_2-4c_3-5c_5
+\tfrac{3}{2}a_0
\right)
\nonumber\\
&+
r_\mu Q^\mu
\left(
-2c_2-4c_3-5c_5
-\tfrac{1}{2}a_0
\right)
\nonumber\\
&+
p_\mu\tilde Q^\mu
\left(
-a_1+3a_3
-4c_2-4c_4-8c_5
\right)
+
q_\mu\tilde Q^\mu
\left(
a_1-3a_3
-4c_1-2c_2-10c_4-2c_5-3a_0
\right)
\nonumber\\
&+
r_\mu\tilde Q^\mu
\left(
-4c_1-2c_2-10c_4-2c_5-a_0
\right)
\nonumber\\
&+
s_\mu S^\mu
\left(
a_0-4b_1-4b_2
\right)
+
3a_0\gr{\nabla}_\mu(r^\mu-q^\mu).
\end{align}
The final contribution is a total divergence
\begin{equation}\label{eq:proj-boundary}
\sqrt{-g}
\gr{\nabla}_\mu(r^\mu-q^\mu)
=
\partial_\mu
\left[
\sqrt{-g}(r^\mu-q^\mu)
\right].
\end{equation}
It does not affect the action when the transformation vectors have compact support or when suitable boundary conditions are imposed. Therefore, the relevant requirement is invariance of the action up to a boundary term.

Because $p_\mu$, $q_\mu$, $r_\mu$, and $s_\mu$ are independent, invariance under the complete generalized projective transformation requires
\begin{subequations}
\label{eq:complete-projective-invariance-conditions}
\begin{align}
-2a_1-8a_2-2a_3+4b_1-2b_2+6b_3&=0,
\\
a_1-2a_2-5a_3-4b_1+2b_2-6b_3+2a_0&=0,
\\
a_1-2a_2-5a_3-2a_0&=0,
\\
a_1+3a_2-4c_1-16c_3-2c_5&=0,
\\
-a_1-3a_2-2c_2-4c_3-5c_5+\frac{3}{2}a_0&=0,
\\
-2c_2-4c_3-5c_5-\frac{1}{2}a_0&=0,
\\
-a_1+3a_3-4c_2-4c_4-8c_5&=0,
\\
a_1-3a_3-4c_1-2c_2-10c_4-2c_5-3a_0&=0,
\\
-4c_1-2c_2-10c_4-2c_5-a_0&=0,
\\
a_0-4b_1-4b_2&=0.
\end{align}
\end{subequations}

Note that for the standard projective transformation alone, $p_\mu$
and $q_\mu=r_\mu=s_\mu=0$, since $R$ is already invariant, the only restrictions are the vanishing of the three coefficients multiplying $p_\mu$ in Eq.\ \eqref{eq:fullprojL}:
\begin{subequations}
\label{eq:standard-projective-invariance-conditions}
\begin{align}
-2a_1-8a_2-2a_3+4b_1-2b_2+6b_3&=0,\\
a_1+3a_2-4c_1-16c_3-2c_5&=0,\\
-a_1+3a_3-4c_2-4c_4-8c_5&=0.
\end{align}
\end{subequations}

Although ten equations are displayed in Eq.\ \eqref{eq:complete-projective-invariance-conditions}, only seven are independent.
Taking $a_0$, $a_3$, $b_3$, $c_4$, and $c_5$ as the independent
parameters, the solution is
\begin{equation}
\begin{aligned}
a_1&=2a_0+3a_3,
&
a_2&=-a_3,
\\
b_1&=\frac{3}{4}a_0-b_3,
&
b_2&=-\frac{1}{2}a_0+b_3,
\\
c_1&=-2c_4+\frac{1}{2}c_5,
&
c_2&=-\frac{1}{2}a_0-c_4-2c_5,
\\
c_3&=
\frac{1}{8}a_0
+\frac{1}{2}c_4
-\frac{1}{4}c_5.
\end{aligned}
\label{eq:complete-projective-invariance-solution}
\end{equation}
The
generalized projective transformations form a continuous additive
family. There are
no additional restrictions on the coupling constants arising from finite transformations. Since generalized projective transformations act by additive shifts of the connection, any finite transformation can be obtained by continuously accumulating infinitesimal shifts. Under the conditions above, the infinitesimal variation of the Lagrangian is a total divergence for any connection configuration, without imposing the field equations. This identity therefore holds at every stage of the transformation, ensuring that the finite variation is also a total divergence. The action is consequently invariant under the boundary conditions given in Eq.\ \eqref{eq:proj-boundary}.
The cancellation of the terms quadratic in $s_\mu$ provides a direct consistency check:
\begin{equation}
6a_0s_\mu s^\mu
-24(b_1+b_2)s_\mu s^\mu=
6(a_0-4b_1-4b_2)s_\mu s^\mu
=0.
\end{equation}
This confirms the positive sign of $6s_\mu s^\mu$ in the finite transformation of the curvature scalar.

For the invariant gravitational theory, generalized projective transformations describe a local gauge freedom: connections related by these transformations are different representatives of the same physical gravitational configuration. This redundancy is reflected in identities among the connection field equations.

We use the complete gravitational Euler-Lagrange tensors
$\mathcal M_{\mu\nu}$ and $\mathcal P_\alpha{}^{\mu\nu}$ already
defined in Eq.\ \eqref{eq:variational-definitions} Sec. \ref{sec:Proj}, with the same normalization.

Since the metric is unchanged by a projective transformation, the term
proportional to $\mathcal{M}_{\mu\nu}$ does not contribute to the action
variation. The connection tensor $\mathcal{P}_\alpha{}^{\mu\nu}$ includes
the curvature, torsion, nonmetricity, and mixed contributions.
The four covectors $p_\mu$, $q_\mu$, $r_\mu$, and $s_\mu$ are independent
local transformation parameters. Their spacetime dependence can be chosen freely, and we take them to have compact support so that boundary terms vanish. If we require the action variation to vanish for every such choice, then the coefficient of each parameter field will vanish pointwise. This gives the off-shell Noether identities \cite{Iosifidis:2021tvx}
\begin{equation}\label{eq:complete-projective-noether-identities}
\mathcal P_\mu{}^{\mu\nu}\equiv0,
\qquad \mathcal P_\nu{}^{\mu\nu}\equiv0,
\qquad g_{\mu\nu}\mathcal P_\alpha{}^{\mu\nu}\equiv0,
\qquad
\epsilon^{\sigma\alpha}{}_{\mu\nu}
\mathcal P_\alpha{}^{\mu\nu}\equiv0.
\end{equation}
These identities hold for arbitrary metric and connection configurations,
without imposing either set of field equations. They express the
redundancy of the connection equations along the projective gauge
directions; they do not imply that the entire tensor $\mathcal{P}_\alpha{}^{\mu\nu}$ vanishes identically.
As a consequence, Eqs.\ \eqref{eq:complete-projective-noether-identities} constrain also the corresponding contractions of the matter hypermomentum. The symmetry consequently determines which sources are compatible with the gravitational action and which require additional fields or symmetry breaking. The matter sectors considered below illustrate both its constructive role in defining invariant interactions and the restrictions that remain when the complete field equations are imposed.
For standard projective invariance alone, only the first identity follows. 
This provides a field-theoretical motivation for the generalization. Whenever the action is invariant under an arbitrary local change of certain connection components, those components describe a redundancy of the theory. Their individual values cannot then be assigned an independent physical meaning. Once matter is included, this interpretation requires the complete action to respect the corresponding symmetry. Matter may couple through invariant combinations, or its fields may transform together with the connection so that the variations of the interaction and kinetic terms cancel. In this way, the transformation properties of torsion and nonmetricity can guide the construction of matter couplings. Nevertheless, the matter fields can also supply a controlled mechanism through which the symmetry is broken, as we will see in  Section \ref{sec: full matter invariant lag}.

The same symmetry also supplies a consistency test. A component absent from the gravitational connection equations cannot provide an independent gravitational response to a matter source. A coupling to that component can consequently impose a restriction on the corresponding matter current.
When the gravitational action has the full generalized projective symmetry,
the four contractions of its connection equation imply the corresponding
conditions on the matter source:
\begin{equation}
\begin{gathered}
\Delta_\mu{}^{\mu\nu}=0,\qquad
\Delta_\nu{}^{\mu\nu}=0,\qquad
g_{\mu\nu}\Delta_\alpha{}^{\mu\nu}=0,\qquad
\epsilon^{\sigma\alpha}{}_{\mu\nu}\Delta_\alpha{}^{\mu\nu}=0.
\end{gathered}
\label{eq:projective-hypermomentum-conditions}
\end{equation}
These are consistency conditions imposed by the sourced field equations.
If the matter action is separately invariant under arbitrary connection
shifts of the form~\eqref{genprojtr}, with its matter fields held fixed,
the same contractions vanish identically. If the matter fields also
transform, their Euler-Lagrange terms enter the matter Noether identities,
and invariance may apply only to a restricted set of transformation
parameters. This distinction is relevant to the matter sectors considered
later in the paper. A coupling that breaks a projective symmetry of the
gravitational action must therefore be checked against the corresponding
contracted connection equation; it does not automatically remove that
equation's constraint on the source.

The usefulness of the generalized transformation is therefore both
geometrical and dynamical. It distinguishes deformations that preserve
all autoparallel trajectories from those that preserve only null ones,
and it tests whether the vector and axial components of the independent
connection represent physical variables or gauge freedom in a specified
theory. The standard covector transformation captures the intrinsic
projective invariance of $R$. The enlarged family allows one to formulate
and analyze stronger symmetry requirements on the complete action,
with explicit consequences for its couplings, its connection equations,
and its admissible matter sources.

\section{Projectively invariant Lagrangians in metric-affine gravity}\label{sec:Proj Lag}
The four vector fields entering the generalized projective
transformation generate four distinct sectors of connection
transformations. One may require invariance under each sector
separately, or under selected combinations of them, thereby
identifying the corresponding invariant subspaces of the parameter space. This classification is an off-shell property of the gravitational action. Here, we start considering the vacuum case, which becomes relevant when the connection field equations are solved.

\subsection{Axial projective sector}\label{sub:axial mode}
We begin with the axial sector, which leads to consider only the transformation with respect to the vector $s_\sigma$, setting $p_\mu = q_\mu =r_\mu=0$. In this case, the connection then transforms according to
\begin{equation}\label{eq:axial-projective-transformation}
\Gamma'{}^\rho{}_{\mu\nu}=\Gamma^\rho{}_{\mu\nu}+
\epsilon^{\sigma\rho}{}_{\mu\nu}s_\sigma ,
\end{equation}
where $s_\mu$ is an arbitrary local covector.
In order to make the geometrical content of this transformation explicit, we use the irreducible decomposition of the torsion tensor \eqref{eq:irreducible T}, obtaining
\begin{equation}
T_{\rho\mu\nu}T^{\rho\mu\nu}
=
\frac{2}{3}T_\mu T^\mu
-\frac{1}{6}S_\mu S^\mu
+t_{\rho\mu\nu}t^{\rho\mu\nu},
\label{eq:first-torsion-quadratic-decomposition}
\end{equation}
\begin{equation}
T_{\rho\mu\nu}T^{\mu\nu\rho}
=
-\frac{1}{3}T_\mu T^\mu
-\frac{1}{6}S_\mu S^\mu
-\frac{1}{2}t_{\rho\mu\nu}t^{\rho\mu\nu},
\label{eq:second-torsion-quadratic-decomposition}
\end{equation}
and
\begin{equation}
Q_{\rho\mu\nu}T^{\nu\rho\mu}
=
\frac{1}{3}Q_\mu T^\mu
-\frac{1}{3}\tilde Q_\mu T^\mu
+
Q_{\rho\mu\nu}t^{\nu\rho\mu}.
\label{eq:mixed-quadratic-decomposition}
\end{equation}
As a consequence, the general system of infinitesimal transformation of the Lagrangian given in Eq.\ \eqref{eq:fullprojL} reduces to
\begin{equation}\label{eq: axial lagrangian transformation}
    \delta_{proj}\mathcal{L} = s^\mu S_\mu(a_0 - 4b_1 - 4b_2),
\end{equation}
and infinitesimal axial projective invariance requires that $a_0 -4(b_1 + b_2)=0$ in Eq.\ \eqref{eq: axial lagrangian transformation}. Now, we can rewrite $a_0=4(b_1 + b_2)$, and inserting it in the general quadratic Lagrangian \eqref{RQTlagrangian}, we obtain
\begin{equation}
    \begin{aligned}
\mathcal{L}=\sqrt{-g}\Bigg[&
4(b_1+b_2)R
+(\frac23 b_1-\frac13 b_2+b_3)
T_\mu T^\mu
-
\frac{1}{6}(b_1+b_2)S_\mu S^\mu
+(b_1-\frac12 b_2)
t_{\rho\mu\nu}t^{\rho\mu\nu}
\\&+c_1Q_{\rho\mu\nu}Q^{\rho\mu\nu}+c_2 Q_{\nu\mu\rho}Q^{\rho\mu\nu}+c_3 Q^\mu Q_\mu+c_4 \tilde Q^\mu \tilde Q_\mu+c_5 Q^\mu \tilde Q_\mu
\\&+(a_2+\frac13 a_1)Q_\mu T^\mu+(a_3-\frac13 a_1)
\tilde Q_\mu T^\mu+a_1 Q_{\rho\mu\nu}t^{\nu\rho\mu}
\Bigg] \,.
\label{eq:axial_Lagrangian}
    \end{aligned}
\end{equation}
Although in the axial transformation \eqref{eq:axial-projective-transformation} the Ricci scalar picks up an extra term, the latter is precisely cancelled by the contributions arising from the transformations of the other terms in the Lagrangian \eqref{eq:axial_Lagrangian}.

In absence of matter, the connection field equations reduce to
\begin{equation}\label{eq:axial-vacuum-connection-equation}
\begin{aligned}
    \mathcal{P}_{\alpha}{}^{\mu\nu}=& [4(b_1+b_2)-2c_2]Q_\alpha{}^{\mu\nu}+
(a_1-4c_1)Q^{\nu\mu}{}_\alpha
-(2c_2+a_1)Q^{\mu\nu}{}_\alpha
\\
&-
[2(b_1+b_2)+c_5]Q_\alpha g^{\mu\nu}
+[2(b_1+b_2)-c_5-a_2]Q^\mu\delta^\nu_\alpha
+(a_2-4c_3)Q^\nu\delta^\mu_\alpha
\\
&-
2c_4\tilde Q_\alpha g^{\mu\nu}
-[4(b_1+b_2)+2c_4+a_3]\tilde Q^\mu\delta^\nu_\alpha
+(a_3-2c_5)\tilde Q^\nu\delta^\mu_\alpha
\\
&+[-\tfrac{8}{3}(b_1+b_2)
+\tfrac13 a_1
-a_3]
T_\alpha g^{\mu\nu}
+[\tfrac43 b_1+\tfrac{10}{3}b_2-2b_3+\tfrac13 a_1-a_3]
T^\mu\delta^\nu_\alpha
\\
&+[\tfrac43 b_1-\tfrac{2}{3} b_2+2b_3
-\tfrac23 a_1
-2a_2]
T^\nu\delta^\mu_\alpha
\\
&+(-4b_1-6b_2)t^{\nu\mu}{}_\alpha
+(4b_1-4b_2-a_1)t_\alpha{}^{\nu\mu}
+
(2b_2-a_1)t^{\mu\nu}{}_\alpha = 0.
\end{aligned}
\end{equation}
The absence of the axial torsion vector $S_\mu$ from Eq.\ \eqref{eq:axial-vacuum-connection-equation} reflects invariance under the axial projective transformation $\delta\Gamma^\rho{}_{\mu\nu}=\epsilon^{\sigma\rho}{}_{\mu\nu}s_\sigma$, which shifts only the axial torsion sector. Consequently, $S_\mu$ remains undetermined by the connection field equations and represents a gauge degree of freedom.
Since $s_\mu$ is arbitrary, axial projective invariance gives the off-shell Noether identity \eqref{eq:complete-projective-noether-identities}
\begin{equation}
\epsilon^{\sigma\alpha}{}_{\mu\nu}
\mathcal P_\alpha{}^{\mu\nu}
\equiv0.
\label{eq:axial-Noether-identity}
\end{equation}
Thus, the four axial projections of the connection equation vanish
identically \eqref{eq:complete-projective-noether-identities}.

The connection field equations can be solved only for the remaining pieces
$Q_{\alpha\mu\nu}$, $T_\mu$, and $t^\rho{}_{\mu\nu}$. In order to do this, we can rewrite $\mathcal{P}_\alpha{}^{\mu\nu}$ in terms of an auxiliary matrix, separating the contributions of $Q_{\alpha\mu\nu}$, $T_\mu$, and $t^\rho{}_{\mu\nu}$: 
\begin{equation}\label{eq:axial-block-operator}
    \mathcal P_\alpha{}^{\mu\nu}
=M_{Q\alpha}{}^{\mu\nu\rho\lambda\kappa}
Q_{\rho\lambda\kappa}
+M_{T\alpha}{}^{\mu\nu\rho}
T_\rho
+ M_{t\alpha}{}^{\mu\nu\rho\lambda\kappa}
t_{\rho\lambda\kappa}
=0.
\end{equation}

The three quantities $M_Q$, $M_T$, and $M_t$ are not independent connection equations. They are the three column blocks of the same linear operator, separated according to the geometric quantity on which each block acts. After choosing bases for the corresponding tensor spaces, Eq.\
\eqref{eq:axial-block-operator} 
can be written schematically as
\begin{equation}
\left(
\begin{array}{ccc}
M_Q & M_T & M_t
\end{array}
\right)
\left(
\begin{array}{c}
Q_{\rho\lambda\kappa}\\
T_\rho\\
t_{\rho\lambda\kappa}
\end{array}
\right)
=0.
\label{eq:axial-combined-block-operator}
\end{equation}
Only the complete sum is required to vanish. In particular, the vector traces of nonmetricity can mix with the torsion vector, while the mixed-symmetry traceless part of nonmetricity can mix with the tensor part of torsion. The full calculation to solve the connection field equations using this new proposed method is given in the Appendix \ref{appendix: axial proj}.
In the end, the unique vacuum
solution of Eq.\ \eqref{eq:axial-block-operator} is
\begin{equation}
Q_{\alpha\mu\nu}=0,
\qquad
T_\mu=0,
\qquad
t^\rho{}_{\mu\nu}=0.
\label{eq:generic-axial-nonaxial-solution}
\end{equation}
If any of the coefficient conditions of Eq.\ \eqref{eq:axial-vector-block} in Appendix \ref{appendix: axial proj} fails, the parameter values lie on an
additional degeneracy subspace. A vanishing determinant in Eq.\
\eqref{eq:axial-vector-block} leaves one or more combinations of
$Q_\mu$, $\tilde Q_\mu$, and $T_\mu$ undetermined. If
$b_1+b_2-c_1-c_2=0$, the totally symmetric traceless part of
nonmetricity is undetermined. If the condition
\eqref{eq:axial-mixed-symmetry-condition} fails, a mixed-symmetry
combination of nonmetricity and tensor torsion remains undetermined.
These additional degeneracies are distinct from the four axial gauge directions that are always present.
The axial Noether identity does not directly solve all equations, but it tells us exactly which part cannot be solved, the axial torsion.
Thus, in absence of matter, after having solved the connection field equations \eqref{eq:axial-vacuum-connection-equation}, obtaining the generic solution
\eqref{eq:generic-axial-nonaxial-solution}, we get that the connection is
metric-compatible and differs from the Levi-Civita connection only
through the axial torsion:
\begin{equation}
\Gamma^\alpha{}_{\mu\nu}
=
\gr{\Gamma}^\alpha{}_{\mu\nu}
-\frac{1}{12}
\epsilon^\alpha{}_{\sigma\mu\nu}S^\sigma.
\label{eq:pure-axial-vacuum-connection}
\end{equation}
In the absence of matter hypermomentum, the generic solution of the connection equations reduces to the Levi-Civita connection up to an axial-projective gauge freedom. Indeed, the axial vector remains undetermined because it transforms as
\begin{equation}
S'_\mu=S_\mu+12s_\mu.
\end{equation}
The gauge choice
\begin{equation}
s_\mu=-\frac{1}{12}S_\mu
\end{equation}
sets $S'_\mu=0$ and brings the connection 
\eqref{eq:pure-axial-vacuum-connection} to the Levi-Civita form.
The remaining axial torsion therefore represents a gauge freedom, and the recovery of GR is the expected outcome on this branch.
This reduction can also be demonstrated directly at the level of the action, before fixing the axial gauge. Evaluating the gravitational Lagrangian on the generic solution of the connection equations gives
\begin{equation}\label{eq:axial-on-shell-Lagrangian}
    \mathcal{L}=\sqrt{-g}[
4(b_1+b_2)R-\frac{1}{6}(b_1+b_2)S_\mu S^\mu].
\end{equation}
For the purely axial connection
\eqref{eq:pure-axial-vacuum-connection}, the curvature scalar is
\begin{equation}
R
=
\gr{R}+\frac{1}{24}S_\mu S^\mu.
\label{eq:pure-axial-curvature-decomposition}
\end{equation}
Now, since the theory is invariant under $\delta \Gamma^\rho{}_{\mu\nu} = \epsilon^{\sigma \rho}{}_{\mu\nu} s_\sigma$, substituting \eqref{eq:pure-axial-curvature-decomposition} into
\eqref{eq:axial-on-shell-Lagrangian}, we obtain that the apparent term $-\frac{1}{6}(b_1+b_2)S_\mu S^\mu$ is canceled by the $S_\mu$ dependence inside the curvature scalar $R$, indeed, for the purely axial connection we have
\begin{equation}
    \mathcal{L}_{\mathrm{eff}} = \sqrt{-g}4(b_1+b_2) \gr{R}.
\end{equation}
So the final on-shell/gauge-reduced Lagrangian is just the Einstein-Hilbert Lagrangian built from the Levi-Civita curvature scalar, with effective coefficient $a_{0 eff}= 4(b_1+b_2)$; $S_\mu$ remains a removable gauge mode.
Provided that $b_1+b_2\neq0$, the remaining vacuum metric equation is
\begin{equation}
4(b_1+b_2)
\left(
\gr{R}_{\mu\nu}
-\frac{1}{2}g_{\mu\nu}\gr{R}
\right)
=0.
\end{equation}
For $a_0=4(b_1+b_2)\neq0$, this is the Einstein-Hilbert Lagrangian. Thus, in the absence of hypermomentum sources, eliminating the connection on the generic vacuum branch yields the metric dynamics of general relativity, independently of the choice of axial gauge.
If $b_1+b_2=0$ and also $a_0=0$, the Einstein-Hilbert interpretation does not apply.

\subsection{Metric-trace projective sector}\label{sub: metric-trace}
In this section, we restrict the generalized projective transformation \eqref{genprojtr} to the sector generated by the covector $r_\mu$,
\begin{equation}
p_\mu=q_\mu=s_\mu=0 \,, \qquad \delta_r\Gamma^\rho{}_{\mu\nu}=g_{\mu\nu}r^\rho. 
\end{equation}
This transformation corresponds to the metric-trace sector of the generalized projective transformation \cite{Sauro:2022proj}.
Since this shift is symmetric in the lower indices, it does not affect torsion,
\begin{equation}
\delta_r T^\rho{}_{\mu\nu}=0 \,,
\end{equation}
and consequently, all three irreducible components of torsion are invariant,
\begin{equation}
T'_\mu=T_\mu,
\qquad
S'_\mu=S_\mu,
\qquad
t'{}^\rho{}_{\mu\nu}=t^\rho{}_{\mu\nu}.
\label{eq:metric-trace-irreducible-torsion-transformations}
\end{equation}
However, it does change the nonmetricity tensor,
\begin{equation}\label{eq: r-nonmetr}
\delta_r Q_{\alpha\mu\nu}
=-g_{\alpha\mu}r_\nu-g_{\alpha\nu}r_\mu \,,
\end{equation}
and consequently,
\begin{equation}\label{eq: transfo traces Q}
\delta_r Q_\alpha=-2r_\alpha,
\qquad
\delta_r\tilde Q_\alpha=-5r_\alpha.
\end{equation}
It follows immediately that
\begin{equation}
\tilde Q'_\alpha-\frac52Q'_\alpha
=
\tilde Q_\alpha-\frac52Q_\alpha.
\label{eq:metric-trace-invariant-nonmetricity-vector}
\end{equation}
Thus, the vector combination
$\tilde Q_\alpha-\frac52Q_\alpha$ is invariant, whereas the remaining vector direction generated by $Q_\alpha$ parametrizes the projective gauge orbit.
For this choice of generalized projective transformation, the
infinitesimal variation of the gravitational Lagrangian \eqref{eq:fullprojL} becomes
\begin{equation}
\begin{aligned}\delta_{proj} \mathcal{L}=& r^\mu T_\mu(a_1 - 2a_2 - 5a_3- 2a_0)
+r^\mu Q_\mu(-2c_2 - 4c_3 - 5c_5- \tfrac{1}{2}a_0)
\\&
+r^\mu\tilde{Q}_\mu(-4c_1 - 2c_2 - 10c_4 - 2c_5- a_0)
+3a_0\gr{\nabla}_\mu r^\mu=0,
\end{aligned}
\end{equation}
thus, the invariance of the action requires that parameter relations are
\begin{subequations}\label{eq:r-invariance-conditions}
\begin{align} 
a_1&=2a_0+2a_2+5a_3\\
c_3&=-\frac{1}{8}(a_0+4c_2+10c_5)\\
c_1&= -\frac{1}{4}(a_0+2c_2+10c_4+2c_5).
\end{align}
\end{subequations}
and the remaining parameters $a_0, a_2, a_3, b_1, b_2, b_3, c_2, c_4, c_5$ are undetermined. 
After imposing the relations
\eqref{eq:r-invariance-conditions}, the invariant gravitational
Lagrangian becomes
\begin{equation}
\begin{aligned}
\mathcal L
=
\sqrt{-g}\Bigg[
&
a_0R
+b_1T_{\rho\mu\nu}T^{\rho\mu\nu}
+b_2T_{\rho\mu\nu}T^{\mu\nu\rho}
+b_3T_\mu T^\mu
\\
&-
\frac{1}{4}
\left(
a_0+2c_2+10c_4+2c_5
\right)
Q_{\rho\mu\nu}Q^{\rho\mu\nu}
+
c_2Q_{\nu\mu\rho}Q^{\rho\mu\nu}
\\
&
-
\frac{1}{8}
\left(
a_0+4c_2+10c_5
\right)
Q_\mu Q^\mu
+
c_4\tilde Q_\mu\tilde Q^\mu
+c_5Q_\mu\tilde Q^\mu
\\
&+
\left(
2a_0+2a_2+5a_3
\right)
Q_{\rho\mu\nu}T^{\nu\rho\mu}
+a_2Q_\mu T^\mu
+a_3\tilde Q_\mu T^\mu
\Bigg].
\end{aligned}
\label{eq:r-invariant-lagrangian}
\end{equation}
In vacuum, the connection field equation obtained from
\eqref{eq:r-invariant-lagrangian} is
\begin{equation}
\begin{aligned}
\mathcal P_\alpha{}^{\mu\nu}
={}&
a_0\Bigg[
Q_\alpha{}^{\mu\nu}
-T_\alpha g^{\mu\nu}
-\frac{1}{2}Q_\alpha g^{\mu\nu}
+T^\mu\delta^\nu_\alpha
+\frac{1}{2}Q^\mu\delta^\nu_\alpha
-\tilde Q^\mu\delta^\nu_\alpha
+g^{\beta\mu}T^\nu{}_{\alpha\beta}
\Bigg]
\\
&+
4b_1T_\alpha{}^{\nu\mu}
+
2b_2
\left(
T^\mu{}_\alpha{}^\nu
-T^\nu{}_\alpha{}^\mu
\right)
+
2b_3T^\rho
\left(
\delta^\mu_\alpha\delta^\nu_\rho
-\delta^\mu_\rho\delta^\nu_\alpha
\right)
+
\left(
a_0+2c_2+10c_4+2c_5
\right)
Q^{\nu\mu}{}_\alpha
\\
&-
2c_2
\left(
Q^{\mu\nu}{}_\alpha
+Q_\alpha{}^{\mu\nu}
\right)
+
\frac{1}{2}
\left(
a_0+4c_2+10c_5
\right)
Q^\nu\delta^\mu_\alpha
-
2c_4
\left(
\tilde Q_\alpha g^{\mu\nu}
+\tilde Q^\mu\delta^\nu_\alpha
\right)
\\
&-
c_5
\left(
2\tilde Q^\nu\delta^\mu_\alpha
+Q_\alpha g^{\mu\nu}
+Q^\mu\delta^\nu_\alpha
\right)
-
\left(
2a_0+2a_2+5a_3
\right)
\Big(
T_\alpha{}^{\nu\mu}
+T^{\mu\nu}{}_\alpha
+Q^{\mu\nu}{}_\alpha
-Q^{\nu\mu}{}_\alpha
\Big)
\\
&+
a_2
\left(
-2T^\nu\delta^\mu_\alpha
-Q^\mu\delta^\nu_\alpha
+Q^\nu\delta^\mu_\alpha
\right)
+
a_3
\left(
-T_\alpha g^{\mu\nu}
-T^\mu\delta^\nu_\alpha
+\tilde Q^\nu\delta^\mu_\alpha
-\tilde Q^\mu\delta^\nu_\alpha
\right)
=0.
\end{aligned}
\label{eq:r-reduced-connection-equation}
\end{equation}
At this stage, however, we do not decompose the nonmetricity tensor.
The complete tensor $Q_{\rho\lambda\kappa}$ is retained as a single forty-component unknown.

Substitution of \eqref{eq:irreducible T} into
\eqref{eq:r-reduced-connection-equation} gives
\begin{equation}
\begin{aligned}
\mathcal P_\alpha{}^{\mu\nu}
={}&
M_{Q\alpha}{}^{\mu\nu\rho\lambda\kappa}
Q_{\rho\lambda\kappa}
+
M_{T\alpha}{}^{\mu\nu\rho}T_\rho
+
M_{S\alpha}{}^{\mu\nu\rho}S_\rho
+
M_{t\alpha}{}^{\mu\nu\rho\lambda\kappa}
t_{\rho\lambda\kappa}
=0.
\end{aligned}
\label{eq:r-combined-block-equation}
\end{equation}
Equivalently,
\begin{equation}
\begin{pmatrix}
M_Q & M_T & M_S & M_t
\end{pmatrix}
\begin{pmatrix}
Q_{\rho\lambda\kappa}
\\[1mm]
T_\rho
\\[1mm]
S_\rho
\\[1mm]
t_{\rho\lambda\kappa}
\end{pmatrix}
=0.
\label{eq:r-combined-operator}
\end{equation}
Equation \eqref{eq:r-combined-operator} is one coupled linear system.
In particular, it does not imply the separate equations
$M_QQ=0$, $M_TT=0$, $M_SS=0$, and $M_tt=0$.
The different contributions may cancel one another, and the solution
must be determined from the rank and kernel of the complete operator.\\
Using the procedure explained in Appendix \ref{appendix: metric-trace} and combining the axial, vector, and tensor equations, we obtain the generic vacuum
solution of Eq \eqref{eq:r-combined-block-equation}:
\begin{equation}
S^\mu=0,
\qquad
T^\mu=0,
\qquad
t_{\alpha\mu\nu}=0,
\qquad
\Omega_{\alpha\mu\nu}=0,
\qquad
\tilde Q^\mu=\frac{5}{2}Q^\mu.
\label{eq:r-generic-sector-solution}
\end{equation}
It follows that the complete torsion tensor vanishes
\begin{equation}
T^\rho{}_{\mu\nu}=0.
\end{equation}
The nonmetricity tensor reduces to
\begin{equation}
Q_{\alpha\mu\nu}
=
\frac{1}{2}
\left(
Q_\mu g_{\alpha\nu}
+
Q_\nu g_{\alpha\mu}
\right).
\label{eq:r-final-nonmetricity}
\end{equation}
Therefore, the corresponding affine connection is 
\begin{equation}
\Gamma^\alpha{}_{\mu\nu}
=
\gr{\Gamma}^\alpha{}_{\mu\nu}
+
g_{\mu\nu}r^\alpha,
\label{eq:r-final-connection}
\end{equation}
and vector $r^\alpha$ remains arbitrary. It is not an additional
physical field determined by the connection equations, but the
undetermined connection component associated with the projective symmetry
\begin{equation}
\delta_r\Gamma^\alpha{}_{\mu\nu}
=
g_{\mu\nu}r^\alpha.
\end{equation}
For generic parameters, the complete connection operator has rank
\begin{equation}
\operatorname{rank}
\begin{pmatrix}
M_Q & M_T & M_S & M_t
\end{pmatrix}
=60,
\end{equation}
and its four-dimensional right kernel is precisely the family of
projective deformations appearing in
\eqref{eq:r-final-connection}.
A convenient gauge choice is
\begin{equation}
r^\alpha=0.
\end{equation}
In this gauge,
\begin{equation}
\Gamma^\alpha{}_{\mu\nu}
=
\gr{\Gamma}^\alpha{}_{\mu\nu},
\qquad
T^\rho{}_{\mu\nu}=0,
\qquad
Q_{\alpha\mu\nu}=0.
\end{equation}
Consequently, up to the boundary term generated by the projective
transformation, the gauge-reduced vacuum Lagrangian is
\begin{equation}
\mathcal L_{\mathrm{eff}}
=
\sqrt{-g}\,a_0\gr{R},
\label{eq:r-effective-lagrangian}
\end{equation}
and the related metric field equations are
\begin{equation}
    a_0(\gr{R}_{\mu\nu} - \frac{1}{2}g_{\mu\nu}\gr{R})=0,
\end{equation}
and for $a_0\neq0$ these equations coincide with the GR ones.
The generic conclusion is therefore that the independent connection
does not carry a vacuum degree of freedom in this sector; as expected as in Sec.\ \ref{sub:axial mode}. It is fixed
to the Levi-Civita connection up to the vector mode $r^\alpha$. Additional undetermined connection components can arise
only when the axial coefficient, the vector determinant, or the
restricted tensor operator becomes degenerate.

\subsection{Full projective invariant Lagrangian}\label{sub:full vacuum solution}
In this section, we consider the full projective transformation Eq.\ \eqref{genprojtr}. 
At the infinitesimal level, we apply Eqs. \eqref{genprojtr} and the full projective transformations of the torsion and nonmetricity tensors 
\eqref{eq:projective-transformation-torsion} and
\eqref{eq:projective-transformation-nonmetricity}, with their respective traces, Eqs.\ \eqref{eq:proj-transf-Q-traces} and \eqref{eq:projective-transformation-torsion-traces}, into the variation of the gravitational action \eqref{RQTlagrangian} and we collect the independent contractions multiplying
$p_\mu,q_\mu,r_\mu,s_\mu$. The invariance is guaranteed a Levi-Civita total divergence, since the derivative contribution that remains is
\begin{equation}
3a_0\sqrt{-g}\,\gr{\nabla}_\mu(r^\mu-q^\mu)
=
3a_0\partial_\mu\left[\sqrt{-g}(r^\mu-q^\mu)\right].
\label{eq:projective-boundary-term}
\end{equation}
For compactly supported transformation covectors, or for boundary conditions
that remove this surface term, the variation of the gravitational action \eqref{RQTlagrangian}, with respect to the generalized projective transformation \eqref{genprojtr} leads to the following combination of the coefficients:
\begin{subequations}
\label{eq:full-invariance-linear-system}
\begin{align}
-2a_1-8a_2-2a_3+4b_1-2b_2+6b_3&=0,
\\
a_1-2a_2-5a_3-4b_1+2b_2-6b_3+2a_0&=0,
\\
a_1-2a_2-5a_3-2a_0&=0,
\\
a_1+3a_2-4c_1-16c_3-2c_5&=0,
\\
-a_1-3a_2-2c_2-4c_3-5c_5+\tfrac32a_0&=0,
\\
-2c_2-4c_3-5c_5-\tfrac12a_0&=0,
\\
-a_1+3a_3-4c_2-4c_4-8c_5&=0,
\\
a_1-3a_3-4c_1-2c_2-10c_4-2c_5-3a_0&=0,
\\
-4c_1-2c_2-10c_4-2c_5-a_0&=0,
\\
a_0-4b_1-4b_2&=0.
\end{align}
\end{subequations}
Then, only seven equations in
\eqref{eq:full-invariance-linear-system} are independent; thus solving them according to the chosen independent parameters, $a_0$, $a_3$, $b_3$, $c_4$, $c_5$, fixes the remaining coefficients
\begin{align}\label{eq: invariant coeff}
a_1 &= 2a_0+3a_3,\\
a_2 &= -a_3,\\
b_1 &= \tfrac34 a_0-b_3,\\
b_2 &= -\tfrac12 a_0+b_3,\\
c_1 &= -2c_4+\tfrac12 c_5,\\
c_2 &= -\tfrac12 a_0-c_4-2c_5,\\
c_3 &= \tfrac18 a_0+\tfrac12 c_4-\tfrac14 c_5.
\end{align}
Therefore, the invariant action is
\begin{align}\label{eq:fully-invariant-expanded-action}
S_{\mathrm{inv}}
=\frac{1}{2\kappa}
\int d^4x\,\sqrt{-g}\Bigg[
& a_0 R
+\left(\frac34 a_0-b_3\right)
T_{\rho\mu\nu}T^{\rho\mu\nu}
+\left(-\frac12 a_0+b_3\right)
T_{\rho\mu\nu}T^{\mu\nu\rho}
+b_3 T_\rho T^\rho
\nonumber\\
&+\left(-2c_4+\frac12 c_5\right)
Q_{\rho\mu\nu}Q^{\rho\mu\nu}
+\left(-\frac12 a_0-c_4-2c_5\right)
Q_{\nu\mu\rho}Q^{\rho\mu\nu}
\nonumber\\
&+\left(\frac18 a_0+\frac12 c_4-\frac14 c_5\right)
Q^\mu Q_\mu
+c_4 \tilde Q^\mu \tilde Q_\mu
+c_5 Q^\mu \tilde Q_\mu
\nonumber\\
&+\left(2a_0+3a_3\right)
Q_{\rho\mu\nu}T^{\nu\rho\mu}
-a_3 Q_\mu T^\mu
+a_3 \tilde Q_\mu T^\mu
\Bigg] .
\end{align}
The independent parameters may thus be taken to be
$a_0,a_3,b_3,c_4,c_5$.  Because the transformations form a continuous
additive family, infinitesimal invariance for arbitrary fields and arbitrary
transformation covectors integrates to invariance under the corresponding
finite transformation; no additional algebraic coupling condition appears.
Varying the general action with respect to the
connection gives
\begin{align}
\mathcal P_\alpha{}^{\mu\nu}
&=
a_0\Big(
Q_\alpha{}^{\mu\nu}
-T_\alpha g^{\mu\nu}
-\frac12Q_\alpha g^{\mu\nu}
+T^\mu\delta^\nu_\alpha
+\frac12Q^\mu\delta^\nu_\alpha
-\tilde Q^\mu\delta^\nu_\alpha
+g^{\beta\mu}T^\nu{}_{\alpha\beta}
\Big)
\nonumber\\
&+4b_1T_\alpha{}^{\nu\mu}
+2b_2\left(
T^\mu{}_{\alpha}{}^\nu
-T^\nu{}_{\alpha}{}^\mu
\right)
+2b_3T^\rho
\left(
\delta^\mu_\alpha\delta^\nu_\rho
-\delta^\mu_\rho\delta^\nu_\alpha
\right)
\nonumber\\
&-4c_1Q^{\nu\mu}{}_\alpha
-2c_2\left(
Q^{\mu\nu}{}_\alpha+Q_\alpha{}^{\mu\nu}
\right)
-4c_3Q^\nu\delta^\mu_\alpha
\nonumber\\
&-2c_4\left(
\tilde Q_\alpha g^{\mu\nu}
+\tilde Q^\mu\delta^\nu_\alpha
\right)
-c_5\left(
2\tilde Q^\nu\delta^\mu_\alpha
+Q_\alpha g^{\mu\nu}
+Q^\mu\delta^\nu_\alpha
\right)
\nonumber\\
&-a_1\left(
T_\alpha{}^{\nu\mu}
+T^{\mu\nu}{}_\alpha
+Q^{\mu\nu}{}_\alpha
-Q^{\nu\mu}{}_\alpha
\right)
\nonumber\\
&+a_2\left(
-2T^\nu\delta^\mu_\alpha
-Q^\mu\delta^\nu_\alpha
+Q^\nu\delta^\mu_\alpha
\right)
+a_3\left(
-T_\alpha g^{\mu\nu}
-T^\mu\delta^\nu_\alpha
+\tilde Q^\nu\delta^\mu_\alpha
-\tilde Q^\mu\delta^\nu_\alpha
\right).
\label{eq:general-explicit-connection-equation}
\end{align}
This expression is algebraic in torsion and nonmetricity.
Now, we substitute the torsion decomposition
\eqref{eq:irreducible T} into
\eqref{eq:general-explicit-connection-equation}, 
and we write the connection field equations as a single coupled block equation, analogous to the previous sections \eqref{eq:r-combined-block-equation}
\begin{align}
\mathcal P_\alpha{}^{\mu\nu}
={}&
M_{Q\alpha}{}^{\mu\nu\rho\lambda\kappa}
Q_{\rho\lambda\kappa}
+M_{T\alpha}{}^{\mu\nu\rho}T_\rho
+
M_{S\alpha}{}^{\mu\nu\rho}S_\rho
+M_{t\alpha}{}^{\mu\nu\rho\lambda\kappa}
t_{\rho\lambda\kappa}
=0.
\label{eq:complete-combined-block-equation}
\end{align}
Equivalently,
\begin{equation}
\begin{pmatrix}
M_Q&M_T&M_S&M_t
\end{pmatrix}
\begin{pmatrix}
Q_{\rho\lambda\kappa}\\[1mm]
T_\rho\\[1mm]
S_\rho\\[1mm]
t_{\rho\lambda\kappa}
\end{pmatrix}
=0.
\label{eq:combined-row-operator}
\end{equation}
The four columns act respectively on $40$, $4$, $4$, and $16$ components,
so the block row accounts for all $64$ independent components of the affine
connection.
Equation \eqref{eq:combined-row-operator} does not imply that the four columns vanish separately. The vector
traces contained in the nonmetricity column can mix with the torsion-vector
column, while the mixed-symmetry nonmetricity can mix with tensor torsion.
The symmetry relations of the fully projective theory imply
\begin{equation}
M_{T\alpha}{}^{\mu\nu\rho}T_\rho\equiv0,
\qquad
M_{S\alpha}{}^{\mu\nu\rho}S_\rho\equiv0,
\label{eq:vanishing-vector-axial-columns}
\end{equation}
and the remaining equation becomes
\begin{equation}
\mathcal P_\alpha{}^{\mu\nu}
=
M_{Q\alpha}{}^{\mu\nu\rho\lambda\kappa}
Q_{\rho\lambda\kappa}
+M_{t\alpha}{}^{\mu\nu\rho\lambda\kappa}
t_{\rho\lambda\kappa}
=0.
\label{eq:full-reduced-combined-equation}
\end{equation}
The solution of Eq.\ \eqref{eq:full-reduced-combined-equation} is contained in Appendix \ref{appendix: full proj}. We obtain that only the vectors $T_\mu,Q_\mu,\tilde Q_\mu,S_\mu$ remain arbitrary, since they are precisely the projective gauge components. As a consequence, the general affine connection can be written as:
\begin{equation}
    \Gamma^\rho{}_{\mu\nu}=
\mathring\Gamma^\rho{}_{\mu\nu}
+\delta^\rho_\mu p_\nu
+\delta^\rho_\nu q_\mu+g_{\mu\nu}r^\rho
+\epsilon^{\sigma\rho}{}_{\mu\nu}s_\sigma.
\end{equation}
and after the gauge choice $p_\mu=q_\mu=0$, $r^\alpha=0$, and $s_\sigma=0$, 
we get that 
\begin{equation}
\Gamma^\rho{}_{\mu\nu}=\mathring\Gamma^\rho{}_{\mu\nu}
\label{eq:Levi-Civita-gauge}
\end{equation}
the linear affine connection lies on the generalized projective gauge orbit of the Levi-Civita connection. 
Thus, the effective Lagrangian becomes
\begin{equation}
\mathcal{L}_{\rm eff}
=\sqrt{-g}\,a_0 \gr{R} .
\end{equation}
up to a boundary term, being both torsion and nonmetricity are nondynamical, as said in the previous cases.  Equivalently, the solved connection is gauge-related
to the Levi-Civita connection, and projective invariance guarantees that the
action has the same value on every representative of this gauge orbit.
The remaining vacuum metric equation is
\begin{equation}
a_0
\left(
\mathring R_{\mu\nu}
-\frac12g_{\mu\nu}\mathring R
\right)=0.
\label{eq:vacuum-Einstein-equation}
\end{equation}
For $a_0\neq0$, this is the vacuum Einstein equation.  If $a_0=0$, no
Einstein-Hilbert kinetic term remains and the equivalence with GR does not follow.

\section{Coupling to fermions in metric-affine gravity}\label{sec: fermions}

Having identified the projectively invariant actions of the gravitational sector, in the following sections, we investigate the role of projective transformations on the independent connection in the coupling with matter. Our motivation is to prove that projectively invariant gravitational sectors can consistently couple to matter, preserving such invariance. A natural starting point is Dirac fermions, since their covariant derivative contains the spin connection.

\subsection{Dirac action in metric-affine spacetime}
Let us consider the tetrad field $e^\mu_a$, such that $V^\mu= e^\mu_a V^a$, with the spin connection $\omega^a_{\ b\mu}$ \cite{Karananas:2021zkl,Rigouzzo:2023sbb,Brensinger:2024udu,Vignolo:2021frk}:
\begin{equation}\label{eq:omega}
    \omega^a_{\ b\mu} = e^a_\lambda e^\nu_b \Gamma^{\lambda}_{\ \nu\mu} + e^a_{\lambda}\partial_\mu e^\lambda_b = e^a_\lambda e^\nu_b (\gr{\Gamma}^{\lambda}_{\ \nu\mu} + K^{\lambda}_{\ \nu\mu} + L^{\lambda}_{\ \nu\mu}) + e^a_{\lambda}\partial_\mu e^\lambda_b  =  \gr{\omega}^a_{\ b \mu} + e^a_\lambda e^\nu_b (K^{\lambda}_{\ \nu\mu} + L^{\lambda}_{\ \nu\mu}) \,.
\end{equation}
Now, we can construct the covariant derivative of the Dirac spinor $\psi$:
\begin{equation}
    D_\mu \psi = \partial_\mu \psi - \frac{1}{8}\omega_{ab\mu}[\gamma^a, \gamma^b] \psi= \partial_\mu \psi - \frac{1}{8}(\gr{\omega}_{\lambda\nu\mu} + K_{\lambda\nu\mu} + L_{\lambda\nu\mu})[\gamma^\lambda,\gamma^\nu] \psi,
\end{equation}
having converted the flat Lorentz indices to curved ones in Eq.\ \eqref{eq:omega}. Thus, the minimally coupled Dirac Lagrangian, considering also the Hermitian conjugate form is
\begin{equation}
    \mathcal{L}_D = \frac{i}{2}\sqrt{-g} \left[\bar{\psi} \gamma^\mu D_\mu \psi - (\overline{{D}_\mu {\psi}})\gamma^\mu \psi \right] - \sqrt{-g}m \bar{\psi}\psi.
\end{equation}
Let's examine the $\gamma^\mu D_\mu \psi$ term,
\begin{equation}
\begin{aligned}
    \gamma^\mu D_\mu \psi &=\gamma^\mu\partial_\mu \psi - \frac{1}{8}(\gr{\omega}_{\lambda\nu\mu} + K_{\lambda\nu\mu} + L_{\lambda\nu\mu}) \gamma^\mu[\gamma^\lambda,\gamma^\nu]\psi\\ & =\gamma^\mu \partial_\mu \psi - \frac{1}{8}\gr{\omega}_{\lambda\nu\mu}\gamma^\mu[\gamma^\lambda,\gamma^\nu]\psi + (\frac{1}{2}\gamma^\nu T_\nu - i\frac{1}{8}\gamma^\nu  \gamma^5  S_\nu + \frac{1}{4}Q_\nu \gamma^\nu - \frac{1}{4}\tilde{Q}_\nu \gamma^\nu)\psi,
    \end{aligned}
\end{equation}
where the properties of $\gamma$ matrices have been used:
\begin{subequations}
    \begin{align}
        \{\gamma^\mu, \gamma^\nu\} &= 2g^{\mu\nu}\\
        \gamma^\mu\gamma^\nu\gamma^\rho&= g^{\mu\nu}\gamma^\rho + g^{\nu\rho}\gamma^\mu - g^{\mu\rho}\gamma^\nu -i\epsilon^{\sigma\mu\nu\rho}\gamma_\sigma\gamma^5.
    \end{align}
\end{subequations}
On the other hand, the conjugate $\overline{D_\mu{\psi}}\gamma^\mu$ is
\begin{equation}
\begin{aligned}    \overline{D_\mu{\psi}}\gamma^\mu&= \partial_\mu \bar{\psi}\gamma^\mu  + \frac{1}{8}\bar{\psi} \gr{\omega}_{\lambda\nu\mu}[\gamma^\lambda,\gamma^\nu]\gamma^\mu + (\frac{1}{2} T_\nu + \frac{1}{4}Q_\nu- \frac{1}{4}\tilde{Q}_\nu )\bar{\psi}\gamma^\nu + i \frac{1}{8}S_\nu \bar{\psi}\gamma^\nu\gamma^5 \\ &= \gr{D}_\mu\bar{\psi}\gamma^\mu + (\frac{1}{2} T_\nu + \frac{1}{4}Q_\nu- \frac{1}{4}\tilde{Q}_\nu )\bar{\psi}\gamma^\nu + i \frac{1}{8}S_\nu \bar{\psi}\gamma^\nu\gamma^5.
    \end{aligned}
\end{equation}
Furthermore, focusing on the Dirac adjoint of the axial torsion, we can write
\begin{equation}
\Big[\overline{-\frac{i}{8}S_\nu \gamma^\nu \gamma^5 \psi}\Big]
\equiv
\left(
-\frac{i}{8}S_\nu \gamma^\nu \gamma^5 \psi
\right)^\dagger \gamma^0 = \frac{i}{8}S_\nu
\psi^\dagger
\left(\gamma^\nu \gamma^5\right)^\dagger
\gamma^0 = \frac{i}{8}S_\nu
\bar\psi\gamma^0
\left(\gamma^\nu \gamma^5\right)^\dagger
\gamma^0 = \frac{i}{8}S_\nu
\bar\psi\gamma^\nu\gamma^5, 
\end{equation}
where we have used the properties of the Dirac matrices:
    \begin{align}
        \{\gamma^\nu, \gamma^5\} &= \gamma^\nu \gamma^5 + \gamma^5\gamma^\nu =0; \\
        (\gamma^\nu\gamma^5)^\dagger&=(\gamma^5)^\dagger(\gamma^\nu)^\dagger, \quad (\gamma^5)^\dagger =\gamma^5, \quad (\gamma^\nu)^\dagger=\gamma^0\gamma^\nu\gamma^0.
  \end{align}
Thus, the final form of  our Dirac Lagrangian is \cite{Rigouzzo:2023sbb}
\begin{equation}\label{eq: Dirac Lagrangian}
     \mathcal{L}_D =\frac{i}{2}\sqrt{-g}
\Big[
\bar\psi\gamma^\mu\gr{D}_\mu\psi
-(\gr{D}_\mu\bar\psi)\gamma^\mu\psi
\Big] 
 +\sqrt{-g} \frac{1}{8}S_\nu \bar{\psi} \gamma^\nu \gamma^5 \psi  -m\sqrt{-g}\,\bar\psi\psi  .
\end{equation}

We can also identify the current $J^\mu = \bar{\psi}\gamma^\mu \psi$ and \textit{axial current} $J^{5\mu}=\bar{\psi} \gamma^\mu \gamma^5 \psi$. The current $J^\mu$ is always conserved if the $\psi(x)$ satisfies the Dirac equations, while the axial current $J^{5\mu}$ is conserved, according to the Noether theorem, if and only if the fermions are massless. The two currents $J^\mu$ and $J^{5\mu}$ are the Noether currents corresponding to the transformations:
\begin{equation}
    \psi(x)\rightarrow e^{i\alpha(x)}\psi;  \quad \psi(x)\rightarrow e^{i\alpha(x)\gamma^5}\psi.
\end{equation}
The first of these transformations is a symmetry of the Dirac Lagrangian, while the second one, called \textit{chiral transformation} is a symmetry of the derivative term in the Dirac Lagrangian but not of the mass term \cite{Peskin:1995ev}. Thus, from this point onward we set $m=0$.

\subsection{Projective axial invariance and chiral symmetry}

Now we consider a chiral transformation on $\psi$
\begin{equation}\label{eq: chiral trans}
    \psi \longrightarrow \psi' = e^{i\alpha(x)\gamma^5}\psi
\end{equation}
such that the infinitesimal variation of the Dirac spinor is
\begin{equation}
    \delta \psi = i\alpha(x)\gamma^5 \psi.
\end{equation}
At the same time, let us consider the full projective transformation of the linear affine connection given in Eq.\ \eqref{genprojtr}
and let us apply both transformations, the chiral of the spinor and the projective of the connection, to the Dirac Lagrangian \eqref{eq: Dirac Lagrangian}, as said, setting $m=0$.
The only part that transforms is the axial torsion
\begin{equation}
    \delta_{proj}(S^{\mu}) = 12 s^{\mu},
\end{equation}
and the projective transformation of the axial torsion and the chiral transformation of the spinor leads to
\begin{equation}
S^\mu\to S^\mu+12s^\mu,\qquad
\psi\to \psi+i\alpha\gamma^5\psi.
\end{equation}
At the first order
\begin{align}
    \delta_{tot} J^{5\nu} &= \delta \bar{\psi} \gamma^\nu \gamma^5 \psi + \bar{\psi} \gamma^\nu \gamma^5 \delta\psi = (i \alpha \bar{\psi}\gamma^5)\gamma^\nu \gamma^5 \psi + \bar{\psi} \gamma^\nu \gamma^5(i\alpha\gamma^5\psi) = 0 \,, \\
    \delta_{tot}(S_\nu J^{5\nu}) &= \delta_{proj}(S_\nu) J^{5\nu} + S_\nu \delta(J^{5\nu}) = \delta_{proj}(S_\nu) J^{5\nu} = 12s_\nu J^{5\nu}
\end{align}
thus
\begin{equation}
\label{eq:Dirac connection variation}
    \delta_{tot}(\sqrt{-g}\frac{1}{8}S_\nu  J^{5\nu}) = \sqrt{-g} \frac{3}{2} s_\nu J^{5\nu}.
\end{equation}
On the other hand, the Levi-Civita term, that is a kinetic term, transforms only according to the chiral transformation
\begin{equation}
\begin{aligned}
    \delta_{tot}\Big(\frac{i}{2}\sqrt{-g}
\Big[
\bar\psi\gamma^\mu\gr{D}_\mu\psi
-(\gr{D}_\mu\bar\psi)\gamma^\mu\psi
\Big]\Big) &= \frac{i}{2}\sqrt{-g}\Big(\delta\bar{\psi}\gamma^\mu\gr{D}_\mu\psi + \bar{\psi} \gamma^\mu \gr{D}_\mu\delta\psi-(\gr{D}_\mu\delta\bar\psi)\gamma^\mu\psi - (\gr{D}_\mu\bar\psi)\gamma^\mu\delta\psi
\Big) \\ &=- \sqrt{-g}
(\partial_\mu\alpha)
\bar\psi\gamma^\mu\gamma^5\psi = -\sqrt{-g}
(\partial_\mu\alpha) J^{5\mu},
\end{aligned}
\end{equation}
since the Levi-Civita covariant derivative satisfies
\begin{equation}
\gr{D}_\mu\gamma^5=0,
\qquad
\gr{D}_\mu\gamma^\nu=0,
\end{equation}
where the second identity includes the Levi-Civita connection acting on the spacetime index of $\gamma^\nu$. These compatibility relations are used in calculating the variation of the kinetic term.
Combining the contributions from the chiral rotation and the axial-projective transformation, we obtain
\begin{equation}
\begin{aligned}
\delta_{tot}\mathcal{L}_D
&=
-\sqrt{-g}\,(\partial_\mu\alpha)J^{5\mu}
+\sqrt{-g}\,\frac{3}{2}s_\mu J^{5\mu}
=
-\sqrt{-g}\left(
\partial_\mu\alpha-\frac{3}{2}s_\mu
\right)J^{5\mu}.
\end{aligned}
\end{equation}
The two contributions cancel for arbitrary spinor configurations when the transformation parameters satisfy
\begin{equation}\label{eq: axial as chiral transf}
\partial_\mu\alpha=\frac{3}{2}s_\mu.
\end{equation}
Under this condition, the combined transformation leaves the massless Dirac Lagrangian invariant. Thus, a local chiral rotation with parameter $\alpha$ compensates an axial-projective shift with $s_\mu=\frac{2}{3}\partial_\mu\alpha$. This correspondence applies to the particular, local representative of axial-projective transformations: the chiral current sources
the projective axial component of the connection \cite{Sauro:2022proj}.

\section{Coupling with the electromagnetic field}\label{sec: elec field} 

The coupling of the connection to an electromagnetic field provides a complementary arena to test the role of projective transformations on the independent connection.\footnote{Throughout this subsection, bold symbols denote differential forms, while their indexed components are written without boldface.} The standard Maxwell field strength $\mathbf{F}=d\mathbf{A}$, written in terms of the potential $\mathbf{A}=A_\mu dx^\mu$, is independent of the affine connection and preserves generalized projective invariance automatically. Since in components, the field strength is $F_{\mu\nu}=(d\textbf{A})_{\mu\nu}=\partial_\mu A_\nu-\partial_\nu A_\mu$, our action for the electromagnetic field acquires the form 
\begin{equation}
     S_{EM} = -\frac{1}{2} \int d\mathbf{A}\wedge \star d\mathbf{A} = - \frac{1}{4}\int d^4x\sqrt{-g} F_{\mu\nu}F^{\mu\nu}.
\end{equation}
Now, we introduce the covariant derivative of the vector field as $\nabla_\mu A_\nu = \partial_\mu A_\nu - \Gamma^{\rho}{}_{\ \nu\mu}A_\rho$, so that 
\begin{equation}
    (d\textbf{A})_{\mu\nu}= \nabla_\mu A_\nu-\nabla_\nu A_\mu+T^\rho{}_{\mu\nu}A_\rho = \mathcal{F}_{\mu\nu}
\end{equation}
defines a field strength $\mathcal{F}_{\mu\nu}$, where the torsion tensor \eqref{eq: torsion} cancels the antisymmetric contributions from the covariant derivatives of $\nabla_\mu A_\nu$. An explicit proof of its invariance under Eq.\ \eqref{genprojtr} is seen through the following computation
\begin{equation}
    \delta_{proj}(\mathcal{F}_{\mu\nu})= \delta_{proj}(\nabla_\mu A_\nu-\nabla_\nu A_\mu+T^\rho{}_{\mu\nu}A_\rho)=(-\delta \Gamma^{\rho}_{\ \nu\mu} +
\delta\Gamma^\rho{}_{\mu\nu})A_\rho + \delta T^{\rho}_{\ \mu\nu}A_\rho = 0.
\label{Fpro}
\end{equation}
Thus, the electromagnetic action minimally coupled to MAG is invariant under the full infinitesimal projective transformation and the $U(1)$ gauge symmetry of the four-potential $A_\mu$ is preserved \cite{Delhom:2020hkb}. Henceforth, the action for the covariant corrected field strength $\mathcal{F}_{\mu\nu}$ is  \cite{Delhom:2020hkb}
\begin{equation}
\begin{aligned}
   S_{EM}  &=-\frac{1}{4}\int d^4x\sqrt{-g}
\mathcal{F}_{\mu\nu}\mathcal{F}^{\mu\nu} \\ &=-\frac{1}{4}\int d^4x\sqrt{-g} g^{\mu\alpha}g^{\nu\beta}(\nabla_\mu A_\nu-\nabla_\nu A_\mu+T^\rho{}_{\mu\nu}A_\rho)(\nabla_\alpha A_\beta -\nabla_\beta A_\alpha + T_{\sigma\alpha\beta}A^\sigma).
\end{aligned}
\end{equation}

Now, we propose to keep a direct coupling of the electromagnetic field strength to metric-affine connection and not subtract torsion,
\begin{equation}\label{eq: tilde fmn}
\tilde{F}_{\mu\nu}
=\nabla_\mu A_\nu-\nabla_\nu A_\mu \,.
\end{equation}
The electromagnetic potential one-form and its ordinary field-strength two-form
are
\begin{equation}
\mathbf{A}=A_\mu\,dx^\mu,
\qquad
\mathbf{f}=d\mathbf{A}
=\frac{1}{2}f_{\mu\nu}\,dx^\mu\wedge dx^\nu,
\qquad
f_{\mu\nu}=\partial_\mu A_\nu-\partial_\nu A_\mu.
\end{equation}
The notation $f_{\mu\nu}$ denotes the ordinary electromagnetic field
strength used below to distinguish it from the torsion-dependent extension.
The corresponding two-form of the field \eqref{eq: tilde fmn} is
\begin{equation}
\tilde{\mathbf{F}}
=\frac{1}{2}\tilde{F}_{\mu\nu}\,dx^\mu\wedge dx^\nu.
\end{equation}
Its Hodge dual is the two-form $\star\tilde{\mathbf{F}}$.
We denote
the components of this dual by $(\star\tilde F)_{\mu\nu}$, so that
\begin{equation}
\star\widetilde{\mathbf{F}}
=\frac{1}{2}(\star\tilde F)_{\mu\nu}\,dx^\mu\wedge dx^\nu,
\qquad
(\star\tilde F)^{\mu\nu}
=\frac{1}{2}\epsilon^{\mu\nu\rho\sigma}\tilde F_{\rho\sigma},
\end{equation}
where $\epsilon^{\mu\nu\rho\sigma}$ is the Levi-Civita tensor. Thus the
star denotes Hodge duality, whereas boldface distinguishes a differential
form from its components.
Using the torsion-dependent field strength, the dual components become
\begin{equation}
(\star\tilde F)^{\mu\nu}
=\tilde f^{\mu\nu}
-\frac{1}{2}\epsilon^{\mu\nu\alpha\beta}
T^\rho{}_{\alpha\beta}A_\rho,
\end{equation}
where, retaining the notation for the dual of the ordinary field strength,
\begin{equation}
\tilde f^{\mu\nu}
=\frac{1}{2}\epsilon^{\mu\nu\alpha\beta}f_{\alpha\beta}.
\end{equation}
This will allow a nontrivial study of its dependence on the projective transformation laws. At this point, our action becomes 
\begin{align}\label{eq: ele hodge}
    S_{EM}&=-\frac{1}{2}\int \tilde{\mathbf{F}}\wedge \star \tilde{\mathbf{F}}= \frac{1}{4}\int d^4x\,\sqrt{-g}\,
(\star\tilde F)_{\mu\nu}(\star\tilde F)^{\mu\nu}\notag
\\&=\frac{1}{4} \int d^4x\sqrt{-g}(
\tilde{f}_{\mu\nu}-\frac{1}{2}\epsilon_{\mu\nu\alpha\beta}T^\rho{}^{\alpha\beta}A_\rho
)(\tilde{f}^{\mu\nu}-\frac{1}{2}\epsilon^{\mu\nu\gamma\delta}T^\sigma{}_{\gamma\delta}A_\sigma).
\end{align}
Now, keeping in mind the full projective transformation of the connection of Eq.\ \eqref{genprojtr}, we decompose the torsion tensor into its irreducible components, in four dimensions it is given in Eq.\ \eqref{eq:irreducible T}. 
Thus, the dual field-strength components then read
\begin{equation}
(\star\tilde F)_{\mu\nu}
=\tilde f_{\mu\nu}
-\frac{1}{3}\epsilon_{\mu\nu\alpha\beta}T^\alpha A^\beta
+\frac{1}{6}(A_\mu S_\nu-A_\nu S_\mu)
-\tilde t^\rho{}_{\mu\nu}A_\rho,
\end{equation}
where the dual of the tensor torsion component is
\begin{equation}
\tilde t^\rho{}_{\mu\nu}
=\frac{1}{2}\epsilon_{\mu\nu}{}^{\alpha\beta}
t^\rho{}_{\alpha\beta}.
\end{equation}
Here $T_\mu$ and $S_\mu$ are the components of the torsion trace and
axial covector,
\begin{equation}
\mathbf{T}=T_\mu\,dx^\mu,
\qquad
\mathbf{S}=S_\mu\,dx^\mu.
\end{equation}
In the Lorentzian signature, for a two-form in four dimensions,
\begin{equation}
\star(\star\tilde{\mathbf{F}})
=-\tilde{\mathbf{F}},
\end{equation}
and the corresponding component contraction satisfies
\begin{equation}
(\star\tilde F)_{\mu\nu}(\star\tilde F)^{\mu\nu}
=-\tilde F_{\mu\nu}\tilde F^{\mu\nu}.
\end{equation}
Consequently, Eq.\ \eqref{eq: ele hodge} has the usual Maxwell sign when
written in terms of $\tilde{\mathbf{F}}$ while $\tilde{F}_{\mu\nu}$ plays the role of a Maxwell-like field strength, 
\begin{equation}
\begin{aligned}
    S_{EM} =&-\frac{1}{2}\int
\tilde{\mathbf{F}}\wedge\star\tilde{\mathbf{F}}=-\frac{1}{4} \int d^4x \sqrt{-g}\tilde{F}^{\mu\nu}\tilde{F}_{\mu\nu} =-\frac{1}{4} \int d^4x \sqrt{-g}(f_{\mu\nu} - T^{\rho}_{\ \mu\nu}A_\rho)(f^{\mu\nu} - T_\sigma{}^{\mu\nu}A^\sigma) \\ =&-\frac{1}{4} \int d^4 x \sqrt{-g}[f_{\mu\nu} - \frac{1}{3}(A_\nu T_\mu-A_\mu T_\nu)-\frac{1}{6}\epsilon^\rho{}_{\sigma\mu\nu}A_\rho S^\sigma-t^\rho{}_{\mu\nu}A_\rho]\\ &\hspace{2.5cm} \times [f^{\mu\nu} - \frac{1}{3}(A^\nu T^\mu-A^\mu T^\nu)-\frac{1}{6}\epsilon^{\rho\sigma\mu\nu}A_\rho S_\sigma-t_\rho{}^{\mu\nu}A^\rho].
\label{Maxwell-like Lagrangian}
\end{aligned}
\end{equation}

Now, we can apply the projective transformation given in Eq.\ \eqref{genprojtr} only to the torsion and the axial vectors
\begin{equation}
    \delta_{proj}\tilde{F}_{\mu\nu} =-\frac{1}{3}(A_\nu \delta_{proj} T_\mu - A_\mu \delta_{proj} T_\nu) - \frac{1}{6} \epsilon^\rho{}_{\sigma\mu\nu}A_\rho \delta_{proj} S^\sigma = A_\mu (p_\nu-q_\nu)-A_\nu (p_\mu-q_\mu)-2\epsilon^\rho{}_{\sigma\mu\nu}A_\rho s^\sigma,
\end{equation}
since $\delta t^\rho{}_{\mu\nu}=0$. This Lagrangian is thus not invariant under the full projective transformation, and the torsion-dependent coupling breaks the ordinary $U(1)$ gauge invariance,
\begin{equation}
    \delta_\lambda \tilde{F}_{\mu\nu}=- T^{\rho}_{\ \mu\nu}\delta_{\lambda} A_\rho = -T^{\rho}_{\ \mu\nu}\partial_\rho \lambda.
\end{equation}
Nevertheless, we will reconstruct the full invariance under both symmetries, using the Stückelberg method, i.e. introducing auxiliary fields.
Since
\begin{equation}
    \delta T_\mu=3(p_\mu-q_\mu), \qquad
\delta S_\mu=12s_\mu,
\end{equation}
we can recast the fields introducing the auxiliary fields $P_\mu$ and $X_\mu$, such that $\delta P_\mu = (p_\mu-q_\mu)$ and $\delta X_\mu = s_\mu$. We define the auxiliary torsion and axial vectors as
\begin{equation}
    \tilde{T}^\mu = T^\mu - 3P^\mu \qquad \tilde{S}^\mu = S^\mu - 12 X^\mu, 
\end{equation}
so that
\begin{subequations}
\begin{align}
    \delta \tilde{T}^\mu &= \delta T^\mu - 3\delta P^\mu =0,\\
    \delta \tilde{S}^\mu &=\delta S^\mu - 12\delta X^\mu =0.
\end{align}
\end{subequations}
Then, the connection-invariant field strength is
\begin{equation}
    \tilde{F}_{\mu\nu}= f_{\mu\nu} - \frac{1}{3}(\tilde{T}_\mu A_\nu - \tilde{T}_\nu A_\mu)  -\frac{1}{6}\epsilon^\rho{}_{\sigma\mu\nu}A_\rho\tilde{S}^\sigma-t^\rho{}_{\mu\nu}A_\rho.
\end{equation}
Finally, applying the same procedure to restore the $U(1)$, we introduce a St\"{u}ckelberg field \cite{Ruegg:2003ps}
\begin{equation}
    \delta_\lambda\chi=\lambda
\end{equation}
and we define the vector $B_\mu = A_\mu-\partial_\mu \chi$ such that
\begin{equation}
    \delta_\lambda B_\mu=\partial_\mu\lambda-\partial_\mu\lambda=0 \,.
\end{equation}
Since $f_{\mu\nu}$ is already $U(1)$ invariant, we can define the fully compensated Maxwell-like field strength
\begin{equation}\label{eq: maxwell-like}
\begin{aligned}
    \tilde{F}_{\mu\nu} =& f_{\mu\nu} - \frac{1}{3}(B_\nu \tilde{T}_\mu - B_\mu \tilde{T}_\nu) -\frac{1}{6}\epsilon^\rho{}_{\sigma\mu\nu}B_\rho\tilde{S}^\sigma-t^\rho{}_{\mu\nu}B_\rho \\ =& f_{\mu\nu} - \frac{1}{3}[(A_\nu -\partial_\nu \chi)(T_{\mu} - 3P_\mu)-(A_\mu -\partial_\mu \chi)(T_{\nu} - 3P_\nu)] \\ &- \frac{1}{6} \epsilon^\rho{}_{\sigma\mu\nu}(A_\rho-\partial_\rho\chi)(S^\sigma - 12X^\sigma)- t^\rho{}_{\mu\nu}(A_\rho-\partial_\rho\chi)
\end{aligned}
\end{equation}
By construction,
\begin{equation}
\delta_{proj} \tilde{F}_{\mu\nu}=0,\qquad \delta_\lambda \tilde{F}_{\mu\nu}=0.
\end{equation}
The Maxwell-like Lagrangian \eqref{Maxwell-like Lagrangian} with the field strength \eqref{eq: maxwell-like} is invariant under both the connection-shift transformation and the ordinary electromagnetic $U(1)$ gauge transformation. The gauge-invariant information is carried by the auxiliary combinations $\tilde{T}_\mu$, $\tilde{S}_\mu$ and by the invariant tensor torsion $t^\rho{}_{\mu\nu}$ since they make the symmetries manifest. In particular, under the full projective transformation, we always have the freedom to set
\begin{equation}
P_\mu=0,\qquad X_\mu=0,
\end{equation}
recovering the original variables, or alternatively apply the gauge-fixing to the vector and axial torsion pieces themselves. At the same time, for the $B_\mu$ field, we can always choose the gauge
\begin{equation}
\chi=0,
\end{equation}
which gives $B_\mu=A_\mu$, since $\chi\rightarrow \chi + \lambda$. 
The standard Maxwell field strength $\mathbf{F}=d\mathbf{A}$is independent of the affine connection and preserves both generalized projective invariance and electromagnetic $U(1)$ symmetry. Introducing a direct torsion coupling through $\tilde F_{\mu\nu}=2\nabla_{[\mu}A_{\nu]}$ breaks both symmetries. To restore them, we introduce compensating fields $P_\mu$, $X_\mu$, and $\chi$, forming the invariant combinations $\tilde T_\mu=T_\mu-3P_\mu$, $\tilde S_\mu=S_\mu-12X_\mu$, and $B_\mu=A_\mu-\partial_\mu\chi$. Together with the projectively invariant tensor torsion, these combinations define a compensated field strength whose action preserves both generalized projective invariance and electromagnetic $U(1)$ symmetry, while retaining direct coupling to torsion. The gauge choices $P_\mu=X_\mu=\chi=0$ recover the original not subtracted torsion field-strength expression.

\section{Coupling with complex Klein-Gordon fields}\label{sec:scalar field}

Now we consider the coupling to the connection of a complex scalar field. Since the covariant derivatives coincide with the partial derivatives, a direct coupling to the connection requires an additional prescription. We consider a real coupling to the torsion and nonmetricity vectors, defining the covariant derivative acting on the scalar field $\phi$ as
\begin{equation}\label{eq:new der scal}
\mathcal{D}_\mu \phi = \gr{\mathcal{D}}_\mu \phi + q(T_\mu + \frac{1}{2}Q_\mu) \phi=\partial_\mu \phi + q(T_\mu + \frac{1}{2}Q_\mu) \phi = \partial_\mu \phi  + q\mathcal{G}_\mu \phi.
\end{equation}
$\mathcal{D}$ represents the gauge covariant derivative applied on the scalar field alone. The previous definition of the covariant derivative works as an analogy to a gauge covariant derivative: $q \in \mathbb{R}-\{0\}$ is the affine $U(1)$ charge of the complex scalar field. It specifies its coupling to the geometric vector $\mathcal{G}_\mu$.
The complex scalar fields transform under local $U(1)$ \cite{Pauli:1934xm,Pich:2012sx} as
\begin{equation}\label{eq: U(1)}
    \phi \rightarrow \phi'=e^{i\beta(x)}\phi,
\end{equation}
therefore, under this symmetry we have in the Minkowski spacetime
\begin{equation}
    \partial_\mu \phi' = e^{i\beta}(\partial_\mu \phi + i\partial_\mu \beta\phi).
\end{equation}
In our case where the covariant derivative is defined as in Eq.\ \eqref{eq:new der scal}, then its action on the transformed scalar field is
\begin{equation}
\mathcal{D}'_\mu \phi'= \partial_\mu \phi' + q\mathcal{G}_\mu' \phi'.
\end{equation}
Assembling all the previous expressions together, we find that the complete transformation is
\begin{equation}
    \mathcal{D}'_\mu \phi'=e^{i\beta(x)} [\mathcal{D}_\mu + (i\partial_\mu \beta + q\delta \mathcal{G}_\mu)]\phi=e^{i\beta(x)}\mathcal{D}_\mu \phi.
\end{equation}
To consistently define covariance in this model, we need to match $\mathcal{G}_{\mu}$ and $\beta$ as
\begin{equation}
    \delta \mathcal{G}_\mu=\mathcal{G}_\mu' - \mathcal{G}_\mu = -\frac{i}{q}\partial_\mu \beta.
\end{equation}

In order to match the infinitesimal generalized projective transformation of the connection in Eq.\ \eqref{genprojtr} to the previous variation, then we must compute the variation in the vector $\mathcal{G}_{\mu}$ which depends directly on the connection through the vector torsion and nonmetricity, yielding
\begin{equation}\label{eq: proj Gmu}
\delta_{proj}\mathcal{G}_\mu = \delta_{proj}T_\mu + \frac{1}{2}\delta_{proj} Q_\mu = 3(p_\mu - q_\mu) - \frac{1}{2}(8p_\mu + 2q_\mu + 2 r_\mu) = -p_\mu - 4q_\mu - r_\mu.
\end{equation}
In this way, we obtain a relation among the free functions of the projective transformation and the phase $\beta$, which is
\begin{equation}
\delta_{proj}\mathcal{G}_\mu =  -p_\mu - 4q_\mu - r_\mu =-\frac{i}{q}\partial_\mu \beta.
\end{equation}
However, for our connection being consistent with the $U(1)$ phase symmetry, the term $iq\mathcal{G}_{\mu}$ must be anti-Hermitian. Therefore, we propose a modified definition of the covariant derivative in the following way
\begin{equation}\label{eq: derivative KG field}
    \mathcal{D}_\mu \phi = \partial_\mu \phi + i q \mathcal{G}_\mu \phi, \qquad \mathcal{G}_\mu=T_\mu + \frac{1}{2}Q_\mu,\qquad q \in \mathbb{R},
\end{equation}
so that the previous relation for $\delta \mathcal{G}_{\mu}$ is, instead,
\begin{equation}\label{eq: proj as u(1)}
    \delta \mathcal{G}_\mu=-(p_\mu +4q_\mu + r_\mu) = -\frac{1}{q}\partial_\mu \beta.
\end{equation}
The derivative of the complex conjugate scalar field is
\begin{equation}
    (\mathcal{D}_\mu \phi)^\dagger = \partial_\mu \phi^\dagger - i q \mathcal{G}_\mu \phi^\dagger,
\end{equation}
and consequently, the locally invariant action is
\begin{equation} \label{eq: S_phi}
    \mathcal{S}_\phi = - \int d^4x \sqrt{-g}[g^{\mu\nu}(\mathcal{D}_\mu \phi)^\dagger (\mathcal{D}_\nu \phi) + m^2 \phi\phi^\dagger]
\end{equation}
Our $U(1)$ invariant Lagrangian $\mathcal{L}_\phi$ does not lose its invariance under projective transformation, given that we introduce a modified covariant derivative \eqref{eq: derivative KG field}. In particular, the matter action is invariant under the first three vectors of generalized projective transformations for which $p_\mu +4q_\mu + r_\mu = \frac{1}{q}\partial_\mu \beta$ and $\phi\rightarrow e^{i\beta}\phi.$
In principle, we can also insert the four-potential of the electromagnetic field in our definition of the covariant derivative Eq.\ \eqref{eq: derivative KG field}, which will be analyzed in the next section.

\section{Full projectively invariant interacting matter sector}\label{sec: full matter sector}

The preceding sections identify how fermionic, electromagnetic, and complex scalar sectors respond to transformations in the connection by transforming their Lagrangians in Sections \ref{sec: fermions}, \ref{sec: elec field} and \ref{sec:scalar field}, respectively. This establishes the symmetry content of the combined matter sector before coupling to the gravitational connection equations.

\subsection{Introduction of electric charge to the scalar field}
The Lagrangian of the full matter sector is written then as a sum of the massless Dirac \eqref{eq: Dirac Lagrangian}, electromagnetic \eqref{eq: maxwell-like}, and scalar \eqref{eq: S_phi} terms, 
\begin{equation}
\begin{aligned}
    \mathcal{L}_M =& \mathcal{L}_D + \mathcal{L}_{EM} + \mathcal{L}_\phi \\ =& \frac{i}{2}\sqrt{-g}
\Big[
\bar\psi\gamma^\mu\gr{D}_\mu\psi
-(\gr{D}_\mu\bar\psi)\gamma^\mu\psi
\Big] +\sqrt{-g}\frac{1}{8}S_\nu  J^{5\nu} -\frac{1}{4}\sqrt{-g}
\mathcal{F}_{\mu\nu}\mathcal{F}^{\mu\nu} \\ &-\sqrt{-g} [g^{\mu\nu}(\mathcal{D}_\mu \phi)^\dagger (\mathcal{D}_\nu \phi) + m^2 \phi\phi^\dagger].
\end{aligned}  
\end{equation}
This Lagrangian is invariant under the full projective transformation \eqref{genprojtr} of the linear affine connection, and to preserve this invariance the matter fields have to transform as 
\begin{subequations}
    \begin{align}
     \psi' &\rightarrow e^{i\alpha(x)\gamma^5}\psi, & \partial_\mu \alpha &= \frac{3}{2} s_\mu \\ 
     \phi &\rightarrow e^{i\beta(x)}\phi, &  \frac{1}{q}\partial_\mu \beta &=p_\mu +4q_\mu + r_\mu \,.
   \end{align}  
\end{subequations}
Thus, at the matter-action level, the transformations act like two Abelian gauge symmetries, in which the symmetry is controlled by the parameters of the gauge transformations $\alpha$ and $\beta$. 

To recover a full equivalence with the classical electrodynamics, we can also couple the four-potential vector $A_\mu$ to the complex scalar field $\phi$. In order to do it, we have to \qm{extend} our definition of the derivative given in Eq.\ \eqref{eq:new der scal}, including also $A_\mu$ and the corresponding charge $e$, 
\begin{equation}
    \mathbb{D}_\mu \phi = \partial_\mu \phi + i q \mathcal{G}_\mu \phi + ie A_\mu \phi.
\end{equation}
We have that the four-potential $A_\mu$ is such that
\begin{equation}
    A_\mu\rightarrow A_\mu -\frac{1}{e}\partial_\mu \lambda(x).
\end{equation}
The potential $A_\mu$ does not modify the geometrical sector, as it is introduced as an independent internal gauge field rather than as part of the spacetime connection. The scalar field is therefore coupled to two distinct Abelian gauge structures: $U(1)_{grav}$, with the geometrical gauge connection $\mathcal{G}_\mu$ and, $U(1)_{EM}$ with the electromagnetic four-potential $A_\mu$ as its gauge field.
The full combined gauge transformation is
\begin{equation}
    \phi
\rightarrow \phi'=
e^{i[\beta(x)+\lambda(x)]}\phi 
\end{equation}
which implies
\begin{equation}
\mathbb{D}_{\mu}\phi \rightarrow \mathbb{D}_{\mu}\phi'= 
e^{i[\beta(x)+\lambda(x)]}
\mathbb{D}_{\mu}\phi \,.
\end{equation}
Thus, the full symmetry is $U(1)_{grav} \times U(1)_{EM}$, whose charges are respectively $q$ and $e$. Finally, the full covariant derivative for the scalar field is
\begin{equation}
    \mathbb{D}_\mu \phi = [\partial_\mu + i q (T_\mu + \frac{1}{2}Q_\mu) + ie A_\mu] \phi.
\end{equation}

\subsection{Full matter sector with compensated electromagnetic variation}

In this section, we utilize the covariant derivative definitions proposed in Secs. \ref{sec: fermions}, \ref{sec: elec field}, and \ref{sec:scalar field} for the Dirac, electromagnetic and complex scalar field, respectively.
In particular, the electromagnetic field strength will be considered as defined in Eq.\ \eqref{Fpro}. Altogether, the full Lagrangian is
\begin{equation}
\begin{aligned}
    I_{M} =& \int d^4x[\mathcal{L}_D + \tilde{\mathcal{L}}_{EM} + \mathcal{L}_\phi] \\ =& \int d^4 x\sqrt{-g}\Big\{\frac{i}{2}
\Big[
\bar\psi\gamma^\mu\gr{D}_\mu\psi
-(\gr{D}_\mu\bar\psi)\gamma^\mu\psi
\Big] +\frac{1}{8}S_\nu  J^{5\nu} -\frac{1}{4} \tilde{F}^{\mu\nu}\tilde{F}_{\mu\nu} \\ &\hspace{2cm} -[g^{\mu\nu}(\mathcal{D}_\mu \phi)^\dagger (\mathcal{D}_\nu \phi) + m^2 \phi\phi^\dagger] \Big\},
\end{aligned}
\end{equation}
it is worth noticing that the Maxwell-like strength field in this case has been written as in Eq.\ \eqref{eq: maxwell-like}.
This Lagrangian is invariant under the full projective transformation of the linear affine connection, and to preserve this invariance the matter fields have to transform as
\begin{subequations}
    \begin{align}
     \psi &\rightarrow e^{i\alpha(x)\gamma^5}\psi \,,&  \partial_\mu \alpha &= \frac{3}{2} s_\mu \\ 
     \phi &\rightarrow e^{i\beta(x)}\phi \,,& \frac{1}{q}\partial_\mu \beta &=  p_\mu +4q_\mu + r_\mu \\
     A_\mu &\rightarrow A_\mu + \partial_\mu \lambda \,, \quad \chi \rightarrow \chi +\lambda \,,& B_\mu &= A_\mu - \partial_\mu \chi \rightarrow B_\mu.
   \end{align}  
\end{subequations}
In this case, we have a Stückelberg realization of electromagnetic gauge invariance.
With these assumptions, the full matter Lagrangian is invariant under diffeomorphisms, local Lorentz transformations, the $U(1)$ symmetry of the electromagnetic field with the Stückelberg field $\chi$, and the full projective connection symmetry that compensates the local chiral symmetry of the Dirac field $\partial_\mu \alpha = \frac{3}{2} s_\mu$, the local phase symmetry of the $U(1)_{grav}$ sector given by $p_\mu +4q_\mu + r_\mu = \frac{1}{q}\partial_\mu \beta$. Roughly speaking, this theory is invariant under a combined geometric-gauge symmetry in which the affine connection shift is compensated by Stückelberg-like vector fields, a chiral rotation of the spinor, and a local phase rotation of the scalar.\\
Now, as in other case, we will also insert the coupling with the four-potential $A_\mu$ in the scalar sector. 
The transformation rules become 
\begin{subequations}
    \begin{align}
    \psi &\rightarrow e^{i\alpha(x)\gamma^5}\psi,  &  \partial_\mu \alpha &= \frac{3}{2} s_\mu, \\ 
     \phi &\rightarrow e^{i[\beta(x)+\lambda(x)]}\phi, &  \frac{1}{q}\partial_\mu \beta &= p_\mu +4q_\mu + r_\mu , \\
     A_\mu &\rightarrow A_\mu - \frac{1}{e}\partial_\mu \lambda, \quad \chi \rightarrow \chi - \frac{1}{e}\lambda , \quad \tilde{F}_{\mu\nu} \rightarrow \tilde{F}_{\mu\nu}, & B_\mu&=A_\mu - \partial_\mu \chi \rightarrow B_\mu.
    \end{align}
\end{subequations}
Therefore the scalar sector and the Maxwell-like sector are simultaneously invariant under the ordinary electromagnetic gauge symmetry and the projective affine symmetry.
It is important moreover to highlight that in this case the Maxwell-like field $\tilde{F}_{\mu\nu}$ is the projectively invariant auxiliary field entering in the Maxwell kinetic term.

\section{Full projective invariant Lagrangians}\label{sec: full matter invariant lag}
In Sec. \ref{sec:Proj Lag}, we identified classes of purely gravitational Lagrangians that are invariant, up to a boundary term, under the full infinitesimal generalized projective transformation \eqref{genprojtr}. We now extend this analysis to theories coupled to matter. Since matter fields may interact directly with the independent affine connection through their hypermomentum, projective invariance of the gravitational sector does not, by itself, guarantee invariance of the complete action. Indeed, the projective Noether identities impose nontrivial consistency conditions on the corresponding matter currents. We examine these conditions for Dirac and complex Klein-Gordon fields, focusing respectively on the axial- and vector-projective sectors. We also investigate controlled departures from exact invariance, for which the corresponding projective modes cease to be pure gauge and become auxiliary fields whose elimination generates effective matter self-interactions.

\subsection{Axial and chiral symmetry}\label{sec: Axial Dirac}
We consider the full Lagrangian, given by the gravitational and the Dirac matter sectors,
\begin{equation}
\begin{aligned}
    \mathcal{L}=&\mathcal{L}_{grav} + \mathcal{L}_D \\=& \frac{\sqrt{-g}}{2\kappa}(a_0R + b_1 T_{\rho\mu\nu}T^{\rho\mu\nu} + b_2T_{\rho\mu\nu}T^{\mu\nu\rho} + b_3 T_{\rho}T^{\rho} + c_1 Q_{\rho\mu\nu} Q^{\rho\mu\nu} + c_2 Q_{\nu\mu\rho} Q^{\rho\mu\nu} + c_3 Q^{\mu}Q_{\mu} + c_4 \tilde{Q}^{\mu} \tilde{Q}_{\mu} \\ &+ c_5 Q^{\mu} \tilde{Q}_{\mu} + a_1 Q_{\rho \mu\nu}T^{\nu\rho\mu} + a_2 Q_{\mu}T^{\mu} + a_3 \tilde{Q}_\mu T^\mu) + \frac{i}{2}\sqrt{-g}
\Big[\bar\psi\gamma^\mu\gr{D}_\mu\psi
-(\gr{D}_\mu\bar\psi)\gamma^\mu\psi
\Big] +\sqrt{-g}\frac{1}{8}S_\nu  J^{5\nu},
    \end{aligned}
\end{equation}
where $J^{5\mu}=\bar{\psi} \gamma^\mu \gamma^5 \psi$ is the axial current. As demonstrated in Sec. \ref{sec: fermions}, the Dirac Lagrangian is invariant under the full projective transformation of the connection \eqref{genprojtr} and the chiral transformation \eqref{eq: chiral trans}, if the condition \eqref{eq: axial as chiral transf} holds. 
Thus, the Dirac action is automatically invariant under
the transformations generated by $p_\mu$, $q_\mu$, and $r_\mu$, because these
transformations do not change $S_\mu$. In the axial sector, however, the Dirac
action is invariant only when $s_\mu$ is locally the gradient of the chiral
parameter, as in \eqref{eq: axial as chiral transf}. A generic covector
$s_\mu$ cannot be written as the gradient of a scalar. Consequently, minimally
coupled massless Dirac matter preserves the locally exact axial-chiral
symmetry, rather than the full arbitrary covector axial projective symmetry.
The complete connection field equation is therefore
\begin{equation}
\mathcal P_\alpha{}^{\mu\nu}
=
\kappa\Delta_\alpha{}^{\mu\nu}.
\label{eq:complete-sourced-connection-equation}
\end{equation}
From the definition \eqref{eq:axial-torsion-vector}, denoting with $\delta_\Gamma$ the variation with respect to the general affine connection \eqref{eq: general affine connection} and $\delta_s$ the axial projective transformation \eqref{eq:axial-projective-transformation} of the covector $s_\alpha$, we have
\begin{equation}
\delta_\Gamma S_\sigma
=
-2\epsilon_{\sigma\alpha}{}^{\mu\nu}
\delta\Gamma^\alpha{}_{\mu\nu},
\label{eq:axial-vector-connection-variation-full}
\end{equation}
and
\begin{equation}
\delta_s\Gamma^\alpha{}_{\mu\nu}
=
\epsilon^{\rho\alpha}{}_{\mu\nu}s_\rho
\end{equation}
into \eqref{eq:axial-vector-connection-variation-full} gives
\begin{equation}
\begin{aligned}
\delta_sS_\sigma
&=
-2\epsilon_{\sigma\alpha}{}^{\mu\nu}
\epsilon^{\rho\alpha}{}_{\mu\nu}s_\rho
=12s_\sigma.
\end{aligned}
\end{equation}
The connection variation of the Dirac action is
\begin{equation}
\begin{aligned}
\delta_\Gamma S_D
&=
\int d^4x\,\sqrt{-g}\,
\frac{1}{8}J^{5\sigma}\delta_\Gamma S_\sigma
=
-\frac{1}{4}
\int d^4x\,\sqrt{-g}\,
\epsilon_{\sigma\alpha}{}^{\mu\nu}
J^{5\sigma}
\delta\Gamma^\alpha{}_{\mu\nu}.
\end{aligned}
\label{eq:explicit-dirac-connection-variation}
\end{equation}
Thus, we have that the Dirac hypermomentum is given by
\begin{equation}
\Delta_\alpha{}^{\mu\nu}
=
\frac{1}{2}
\epsilon_{\sigma\alpha}{}^{\mu\nu}
J^{5\sigma}.
\label{eq:explicit-dirac-hypermomentum}
\end{equation}
As a consequence the connection field equation becomes
\begin{equation}
    \mathcal{P}_{\alpha}{}^{\mu\nu}=\kappa\Delta_\alpha{}^{\mu\nu}=\frac{\kappa}{2}\epsilon_{\sigma\alpha}{}^{\mu\nu}
J^{5\sigma}
= \frac{\kappa}{2} \epsilon_{\sigma\alpha}{}^{\mu\nu}\bar{\psi} \gamma^\sigma \gamma^5 \psi \,,
\end{equation}
while its axial projection is
\begin{equation}
\begin{aligned}
\epsilon^{\lambda\alpha}{}_{\mu\nu}
\Delta_\alpha{}^{\mu\nu}
&=
\frac{1}{2}
\epsilon^{\lambda\alpha}{}_{\mu\nu}
\epsilon_{\sigma\alpha}{}^{\mu\nu}
J^{5\sigma}
=-3J^{5\lambda}.
\end{aligned}
\label{eq:axial-projection-dirac-hypermomentum}
\end{equation}
Hence the axial projection of the connection equation is
\begin{equation}
    \lambda S^\sigma = - 3\kappa J^{5\sigma},
\end{equation}
since 
\begin{equation}
\epsilon^{\lambda\alpha}{}_{\mu\nu}
\mathcal P_\alpha{}^{\mu\nu}
=
\left(a_0-4b_1-4b_2\right)S^\lambda.
\label{eq:complete-gravitational-axial-projection}
\end{equation}
Here $a_0-4b_1-4b_2=\lambda$.
Contracting the sourced connection equation
\eqref{eq:complete-sourced-connection-equation} with
$\epsilon^{\lambda\alpha}{}_{\mu\nu}$ gives
\begin{equation}
\left(a_0-4b_1-4b_2\right)S^\lambda
=
-3\kappa J^{5\lambda}.
\end{equation}
The complete axial connection equation is therefore
\begin{equation}
\left(a_0-4b_1-4b_2\right)S^\lambda
+
3\kappa J^{5\lambda}
=0.
\label{eq:complete-axial-connection-equation-expanded}
\end{equation}
At the axial projectively invariant point $\lambda=0$, it becomes
\begin{equation}
    J^{5\lambda}=0
\end{equation}
So a fully axial-projectively invariant gravitational action cannot be coupled consistently to a Dirac field  with nonzero axial current through this connection equation. The gravitational axial connection mode has no equation-producing quadratic term and acts as a Lagrange multiplier enforcing $J^{5\lambda}=0$. 
Moreover, since 
\begin{equation}
    \delta \Gamma^{\alpha}{}_{\mu\nu} = \frac{2}{3} \epsilon^{\sigma\alpha}{}_{\mu\nu}\partial_\sigma \alpha.
\end{equation}
the general projective transformation becomes
\begin{equation}
    \delta \Gamma^\alpha{}_{\mu\nu}=\delta^\alpha_\mu p_\nu + \delta^\alpha_\nu q_\mu + g_{\mu\nu}r^\alpha + \frac{2}{3} \epsilon^{\sigma\alpha}{}_{\mu\nu}\partial_\sigma \alpha.
\end{equation}

To isolate the axial generalized projective sector, we temporarily restrict the transformation to $p_\mu=q_\mu=r_\mu=0$, so that only the $s_\mu$ direction is varied. Requiring invariance under this axial transformation gives that $\lambda$ in Eq.\ \eqref{eq:complete-axial-connection-equation-expanded} must be trivial, so that we can rewrite it as 
\begin{equation}
    b_2=\frac{a_0}{4}-b_1,
\end{equation}
without any restrictions on the other coefficients $b_3,c_1,c_2,c_3,c_4,c_5,a_1,a_2,$ or $a_3$.
After imposing this condition, the action becomes
\begin{align}
\mathcal{S}=&\frac{1}{2\kappa}\int d^4x\,\sqrt{-g}\,
\Bigg[{}
a_0R
+b_1T_{\rho\mu\nu}T^{\rho\mu\nu}
+\left(\frac{a_0}{4}-b_1\right)
T_{\rho\mu\nu}T^{\mu\nu\rho}
+b_3T_\mu T^\mu
\nonumber\\
&\hspace{2.5cm}+c_1Q_{\rho\mu\nu}Q^{\rho\mu\nu}
+c_2Q_{\nu\mu\rho}Q^{\rho\mu\nu}
+c_3Q_\mu Q^\mu
+c_4\tilde Q_\mu\tilde Q^\mu
+c_5Q_\mu\tilde Q^\mu
\nonumber\\
&\hspace{2.5cm}+a_1Q_{\rho\mu\nu}T^{\nu\rho\mu}
+a_2Q_\mu T^\mu
+a_3\tilde Q_\mu T^\mu
\Bigg]\\ &+\int d^4x\,\sqrt{-g}\,
\Bigg\{
\frac{i}{2}
\left[
\bar\psi\gamma^\mu\gr{D}_{\mu}\psi
-\left(\gr{D}_{\mu}\bar\psi\right)
\gamma^\mu\psi
\right]
+\frac18S_\mu J^{5\mu}
\Bigg\}
\end{align}
Since the Dirac hypermomentum is purely axial, the three vector projections of the connection equations and the remaining tensorial projections are unsourced. 
The Noether identity becomes
\begin{align}
&-\frac{1}{3\kappa\sqrt{-g}}
\partial_{\lambda}
\Bigg\{
\sqrt{-g}
\Big[
\left(
a_{0}-4b_{1}-4b_{2}
\right)S^{\lambda}
+
3\kappa
\bar{\psi}\gamma^{\lambda}\gamma^{5}\psi
\Big]
\Bigg\}
\nonumber\\
&\qquad +
i\bar{\psi}\gamma^{5}
\left[
i\gamma^{\mu}
\gr{D}_{\mu}\psi
+
\frac{1}{8}
S_{\mu}\gamma^{\mu}\gamma^{5}\psi
\right]
+i\left[
-i
\left(
\gr{D}_{\mu}\bar{\psi}
\right)\gamma^{\mu}
+
\frac{1}{8}
S_{\mu}\bar{\psi}\gamma^{\mu}\gamma^{5}
\right]
\gamma^{5}\psi =0.
\end{align}
When the Dirac equations are satisfied, the Noether identity gives
\begin{equation}
\frac{1}{\sqrt{-g}}
\partial_{\lambda}
\left( \sqrt{-g}\,
\bar{\psi}\gamma^{\lambda}\gamma^{5}\psi \right) =0.
\end{equation}

We now return to the general action and consider the complementary branch $\lambda\neq0$. In this case, the gravitational action is not invariant under the
axial-projective transformation. In fact, 
under the 
\begin{equation}
\delta_{proj}\Gamma^{\rho}{}_{\mu\nu}
=
\epsilon^{\sigma\rho}{}_{\mu\nu}s_{\sigma},
\qquad
s_{\mu}
=
\frac{2}{3}\partial_{\mu}\alpha,
\end{equation}
the variation of the gravitational action is
\begin{align}
\delta_{proj}S_{\mathrm g}
&=
\frac{1}{2\kappa}
\int d^{4}x\,\sqrt{-g}\,
\epsilon^{\lambda\alpha}{}_{\mu\nu}
P_{\alpha}{}^{\mu\nu}s_{\lambda}
=
\frac{1}{2\kappa}
\int d^{4}x\,\sqrt{-g}\,
\lambda S^{\lambda}s_{\lambda}
=
\frac{\lambda}{3\kappa}
\int d^{4}x\,\sqrt{-g}\,
S^{\lambda}\partial_{\lambda}\alpha.
\end{align}
After integration by parts and neglecting the boundary contribution,
this becomes
\begin{equation}
\delta_{proj}S_{\mathrm g}
=
-\frac{\lambda}{3\kappa}
\int d^{4}x\,\alpha\,
\partial_{\lambda}
\left(
\sqrt{-g}\,S^{\lambda}
\right).
\end{equation}
The Dirac Lagrangian remains invariant under the simultaneous
axial-projective and chiral transformations, thus the variation of the complete action is
\begin{equation}
    \delta_{proj}S_{\mathrm{tot}}
=-\frac{\lambda}{3\kappa}
\int d^{4}x\,\alpha\,
\partial_{\lambda}
\left(
\sqrt{-g}\,S^{\lambda}
\right)
\neq 0.
\end{equation}
Therefore, for $\lambda\neq0$, the total action does not possess an
exact axial-projective-chiral symmetry. Instead, the corresponding
explicitly broken Noether relation is 
\begin{align}
-\frac{1}{3\kappa\sqrt{-g}}
\partial_{\lambda}
\Bigg\{
\sqrt{-g}
\left[
\lambda S^{\lambda}
+
3\kappa
\bar{\psi}\gamma^{\lambda}\gamma^{5}\psi
\right]
\Bigg\} 
\hspace{5cm}&
\nonumber\\
+
i\bar{\psi}\gamma^{5}
\left[
i\gamma^{\mu}
\gr{D}_{\mu}\psi
+
\frac{1}{8}
S_{\mu}\gamma^{\mu}\gamma^{5}\psi
\right]
+i\left[
-i
\left(
\gr{D}_{\mu}\bar{\psi}
\right)\gamma^{\mu}
+
\frac{1}{8}
S_{\mu}\bar{\psi}\gamma^{\mu}\gamma^{5}
\right]
\gamma^{5}\psi
&\equiv
-\frac{\lambda}{3\kappa\sqrt{-g}}
\partial_{\lambda}
\left(
\sqrt{-g}\,S^{\lambda}
\right).
\end{align}
Contracting the complete connection field equation 
\begin{equation}
\mathcal{P}_{\alpha}{}^{\mu\nu}
=
\frac{\kappa}{2}
\epsilon_{\sigma\alpha}{}^{\mu\nu}
\bar{\psi}\gamma^{\sigma}\gamma^{5}\psi
\end{equation}
with
$\epsilon^{\lambda\alpha}{}_{\mu\nu}$ gives
\begin{align}
\epsilon^{\lambda\alpha}{}_{\mu\nu}
P_{\alpha}{}^{\mu\nu}
&=
\frac{\kappa}{2}
\epsilon^{\lambda\alpha}{}_{\mu\nu}
\epsilon_{\sigma\alpha}{}^{\mu\nu}
\bar{\psi}\gamma^{\sigma}\gamma^{5}\psi
=
-3\kappa
\bar{\psi}\gamma^{\lambda}\gamma^{5}\psi.
\end{align}
Since
\begin{equation}
\epsilon^{\lambda\alpha}{}_{\mu\nu}
\mathcal{P}_{\alpha}{}^{\mu\nu}
=
\lambda S^{\lambda},
\end{equation}
the axial connection equation is
\begin{equation}\label{eq:axial connection equation}
\lambda S^{\lambda}
=
-3\kappa
\bar{\psi}\gamma^{\lambda}\gamma^{5}\psi.
\end{equation}
For $\lambda\neq0$, the axial torsion is therefore 
\begin{equation}\label{eq: S come j5}
    S^{\lambda}
=
-\frac{3\kappa}{\lambda}
\bar{\psi}\gamma^{\lambda}\gamma^{5}\psi=-\frac{3\kappa}{\lambda}J^{5\lambda} \,.
\end{equation}
The part of the total Lagrangian density that depends on the axial
torsion is
\begin{equation}
\mathcal L_{\mathrm{axial}}
=
\sqrt{-g}
\left[
\frac{\lambda}{48\kappa}S_{\mu}S^{\mu}
+
\frac{1}{8}S_{\mu}
\bar{\psi}\gamma^{\mu}\gamma^{5}\psi
\right].
\end{equation}
If we complete the square, we get
\begin{align}
\mathcal L_{\mathrm{axial}}
={}&
\frac{\sqrt{-g}\,\lambda}{48\kappa}
\left(
S_{\mu}
+
\frac{3\kappa}{\lambda}
\bar{\psi}\gamma_{\mu}\gamma^{5}\psi
\right)
\left(
S^{\mu}
+
\frac{3\kappa}{\lambda}
\bar{\psi}\gamma^{\mu}\gamma^{5}\psi
\right)
-
\frac{3\kappa\sqrt{-g}}{16\lambda}
\left(
\bar{\psi}\gamma_{\mu}\gamma^{5}\psi
\right)
\left(
\bar{\psi}\gamma^{\mu}\gamma^{5}\psi
\right).
\end{align}
After eliminating $S_{\mu}$ through its algebraic field equation, the
induced four-fermion interaction is
\begin{equation}\label{eq: four fermions interaction}
\mathcal L_{\mathrm{4\psi}}
=
-\frac{3\kappa\sqrt{-g}}{16\lambda}
\left(
\bar{\psi}\gamma_{\mu}\gamma^{5}\psi
\right)
\left(
\bar{\psi}\gamma^{\mu}\gamma^{5}\psi
\right)= -
\frac{3\kappa\sqrt{-g}}{16\lambda}
J_{5\mu}J_5^\mu .
\end{equation}
Thus, the fundamental Lagrangian is
\begin{align}
\mathcal{L}_{\mathrm{fund}}
={}&
\frac{\sqrt{-g}}{2\kappa}
\Big[
a_{0}R(g,\Gamma)
+b_{1}T_{\rho\mu\nu}T^{\rho\mu\nu}
+b_{2}T_{\rho\mu\nu}T^{\mu\nu\rho}
+b_{3}T_{\mu}T^{\mu}
\nonumber\\
&\qquad
+c_{1}Q_{\rho\mu\nu}Q^{\rho\mu\nu}
+c_{2}Q_{\nu\mu\rho}Q^{\rho\mu\nu}
+c_{3}Q_{\mu}Q^{\mu}
+c_{4}\tilde{Q}_{\mu}\tilde{Q}^{\mu}
+c_{5}Q_{\mu}\tilde{Q}^{\mu}
\nonumber\\
&\qquad
+a_{1}Q_{\rho\mu\nu}T^{\nu\rho\mu}
+a_{2}Q_{\mu}T^{\mu}
+a_{3}\tilde{Q}_{\mu}T^{\mu}
\Big]
\nonumber\\
&+
\frac{i}{2}\sqrt{-g}
\Big[
\bar{\psi}\gamma^{\mu}\gr{D}_{\mu}\psi
-
\left(\gr{D}_{\mu}\bar{\psi}\right)
\gamma^{\mu}\psi
\Big]
+
\frac{\sqrt{-g}}{8}S_{\mu}J_{5}^{\mu}.
\end{align}

\subsubsection{Adopting trace gauge}

Let us specialize to the subclass of gravitational theories that retains exact generalized projective invariance under the $p_\mu$, $q_\mu$, and $r_\mu$ transformations, considering the axial $s_\mu$ symmetry explicitly broken by $\lambda\neq0$. Consequently, the three vector trace equations are Noether identities and the corresponding traces may be gauge-fixed, whereas the remaining unsourced tensor sectors must still be eliminated through their homogeneous algebraic field equations.
The components of the connection field equations \eqref{eq:complete-sourced-connection-equation}, are
\begin{equation}
    \mathcal P_\mu{}^{\mu\nu}\equiv0,
\qquad
\mathcal P_\nu{}^{\mu\nu}\equiv0,
\qquad
g_{\mu\nu}\mathcal P_\alpha{}^{\mu\nu}\equiv0,
\end{equation}
which is equivalent to the coefficient conditions
\begin{align}
b_3&=\frac{2a_0-2b_1+b_2}{3},
&c_3&=\frac{11a_0-20c_1+4c_2}{72},
\nonumber\\
c_4&=-\frac{a_0+8c_1+2c_2}{18},
&c_5&=\frac{-2a_0+2c_1-4c_2}{9},
\nonumber\\
a_2&=\frac{2a_0-a_1}{3},
&a_3&=\frac{a_1-2a_0}{3}.
\label{eq:ec-like-invariant-coefficients}
\end{align}
The Noether identities ensure that the quantities $Q_\mu$, $\tilde{Q}_\mu$, and $T_\mu$ are removable variables, thus they are redundant. It is
therefore consistent to obtain the remaining equations by varying the
gauge-fixed action within the trace-free sector.
The gauge
\begin{equation}\label{eq:ec-like-trace-gauge}
    T_\mu=Q_\mu=\tilde Q_\mu=0.
\end{equation}
removes $12$ connection components, corresponding to $3$
four-vectors. It leaves the $4$ axial torsion components, $16$
non-axial torsion components, and $32$ trace-free nonmetricity
components. The latter $48$ components are invariant under the
$3$ vector shifts and must be treated through their own equations.
The curvature scalar, in the gauge \eqref{eq:ec-like-trace-gauge}, obeys
\begin{align}
R={}&\gr{R}
+\frac14T_{\rho\mu\nu}T^{\rho\mu\nu}
-\frac12T_{\rho\mu\nu}T^{\mu\nu\rho}
+\frac14Q_{\rho\mu\nu}Q^{\rho\mu\nu}
-\frac12Q_{\nu\mu\rho}Q^{\rho\mu\nu}
+Q_{\rho\mu\nu}T^{\nu\rho\mu}.
\label{eq:ec-like-curvature-in-gauge}
\end{align}
 Its derivative term is a
Levi-Civita divergence of the vector traces and vanishes in the chosen
gauge. Hence the remaining connection variables appear without derivatives
and are auxiliary. General algebraic connection solutions in quadratic
metric-affine theories are discussed in Ref.\cite{Iosifidis:2021tvx}; here the
invertibility is required only after removing the vector gauge directions.
Using \eqref{eq:ec-like-curvature-in-gauge}, the gauge-fixed density is
\begin{align}
\mathcal L=&\frac{\sqrt{-g}}{2\kappa}
\Big[
a_0\gr{R}
+\left(b_1+\frac{a_0}{4}\right)
T_{\rho\mu\nu}T^{\rho\mu\nu}
+\left(b_2-\frac{a_0}{2}\right)
T_{\rho\mu\nu}T^{\mu\nu\rho}
\nonumber\\
& \hspace{1cm}
+\left(c_1+\frac{a_0}{4}\right)
Q_{\rho\mu\nu}Q^{\rho\mu\nu}
+\left(c_2-\frac{a_0}{2}\right)
Q_{\nu\mu\rho}Q^{\rho\mu\nu}
+(a_0+a_1)Q_{\rho\mu\nu}T^{\nu\rho\mu}
\Big]
\nonumber\\
&
+\frac{i\sqrt{-g}}2
\left[
\bar\psi\gamma^\mu\gr{D}_\mu\psi
-\left(\gr{D}_\mu\bar\psi\right)\gamma^\mu\psi
\right]
+\frac{\sqrt{-g}}8S_\mu\bar\psi\gamma^\mu\gamma^5\psi.
\label{eq:ec-like-gauge-fixed-lagrangian}
\end{align}
Varying the antisymmetric tensor $T_{\rho\mu\nu}=T_{\rho[\mu\nu]}$ and the
symmetric tensor $Q_{\rho\mu\nu}=Q_{\rho(\mu\nu)}$, subject to
\eqref{eq:ec-like-trace-gauge}, gives
\begin{align}
2\left(b_1+\frac{a_0}{4}\right)T_{\rho\mu\nu}
+\left(b_2-\frac{a_0}{2}\right)
\left(T_{\mu\nu\rho}+T_{\nu\rho\mu}\right)+\frac{a_0+a_1}{2}
\left(Q_{\mu\nu\rho}-Q_{\nu\mu\rho}\right)
-\frac\kappa4\epsilon_{\rho\mu\nu\sigma}
\bar\psi\gamma^\sigma\gamma^5\psi &=0,
\label{eq:ec-like-torsion-equation}\\
2\left(c_1+\frac{a_0}{4}\right)Q_{\rho\mu\nu}
+\left(c_2-\frac{a_0}{2}\right)
\left(Q_{\mu\nu\rho}+Q_{\nu\mu\rho}\right)+\frac{a_0+a_1}{2}
\left(T_{\nu\rho\mu}+T_{\mu\rho\nu}\right)&=0.
\label{eq:ec-like-nonmetricity-equation}
\end{align}
All contractions of these equations vanish in the trace gauge, so the
trace-free variations introduce no missing trace equations.
The completely symmetric part of
\eqref{eq:ec-like-nonmetricity-equation} is
\begin{equation}
\big[4(c_1+c_2)-a_0\big]Q_{(\rho\mu\nu)}=0.
\label{eq:ec-like-symmetric-nonmetricity}
\end{equation}
On the branch
\begin{equation}
4(c_1+c_2)-a_0\neq0,
\label{eq:ec-like-first-nondegeneracy}
\end{equation}
it follows that
\begin{equation}
Q_{(\rho\mu\nu)}=0.
\label{eq:ec-like-nonmetricity-cyclic-identity}
\end{equation}
Using this identity and subtracting the axial equation from
\eqref{eq:ec-like-torsion-equation}, the remaining equations become
\begin{align}
(a_0+2b_1-b_2)
\left(T_{\rho\mu\nu}-\frac16\epsilon_{\rho\mu\nu\sigma}S^\sigma\right)+\frac{a_0+a_1}{2}
\left(Q_{\mu\nu\rho}-Q_{\nu\mu\rho}\right)&=0,
\label{eq:ec-like-mixed-torsion-equation}\\
(a_0+2c_1-c_2)Q_{\rho\mu\nu}
+\frac{a_0+a_1}{2}
\left(T_{\nu\rho\mu}+T_{\mu\rho\nu}\right)&=0.
\label{eq:ec-like-mixed-nonmetricity-equation}
\end{align}
The explicitly displayed subtraction in
\eqref{eq:ec-like-mixed-torsion-equation} removes only the completely
antisymmetric part of torsion. It introduces no additional tensor field.

For completeness, the elimination can be performed without dividing by
either diagonal coefficient in these two equations. Permuting the indices
in \eqref{eq:ec-like-mixed-nonmetricity-equation}, taking the difference,
since $T_{[\rho\mu\nu]}
=\frac16\epsilon_{\rho\mu\nu\sigma}S^\sigma$, we have
\begin{align}
&(a_0+2c_1-c_2)
\left(Q_{\mu\nu\rho}-Q_{\nu\mu\rho}\right)
+\frac{3(a_0+a_1)}2
\left(T_{\rho\mu\nu}-\frac16\epsilon_{\rho\mu\nu\sigma}S^\sigma\right)=0.
\label{eq:ec-like-permuted-nonmetricity-equation}
\end{align}
Eliminating the nonmetricity difference between this equation and
Eq.\ \eqref{eq:ec-like-mixed-torsion-equation} yields
\begin{equation}
\Big[
4(a_0+2b_1-b_2)(a_0+2c_1-c_2)-3(a_0+a_1)^2
\Big]\left(
T_{\rho\mu\nu}-\frac16\epsilon_{\rho\mu\nu\sigma}S^\sigma
\right)=0.
\label{eq:ec-like-eliminated-torsion}
\end{equation}
The corresponding sum of permutations of
Eq.\ \eqref{eq:ec-like-mixed-torsion-equation} gives
\begin{equation}
(a_0+2b_1-b_2)
\left(T_{\nu\rho\mu}+T_{\mu\rho\nu}\right)
+\frac{3(a_0+a_1)}2Q_{\rho\mu\nu}=0.
\label{eq:ec-like-permuted-torsion-equation}
\end{equation}
Combining it with \eqref{eq:ec-like-mixed-nonmetricity-equation} gives
\begin{equation}
\Big[
4(a_0+2b_1-b_2)(a_0+2c_1-c_2)-3(a_0+a_1)^2
\Big]Q_{\rho\mu\nu}=0.
\label{eq:ec-like-eliminated-nonmetricity}
\end{equation}
Thus, on the additional nondegenerate branch
\begin{equation}
4(a_0+2b_1-b_2)(a_0+2c_1-c_2)-3(a_0+a_1)^2\neq0,
\label{eq:ec-like-second-nondegeneracy}
\end{equation}
the solution is
\begin{equation}
Q_{\rho\mu\nu}=0,
\qquad
T_{\rho\mu\nu}=\frac16\epsilon_{\rho\mu\nu\sigma}S^\sigma.
\label{eq:ec-like-pure-axial-solution}
\end{equation}
The assumptions \eqref{eq:ec-like-first-nondegeneracy} and
\eqref{eq:ec-like-second-nondegeneracy} guarantee a unique solution for the
unsourced tensor sector after vector gauge fixing.
Now, we can exploit the axial connection equation \eqref{eq:axial connection equation} and eliminate the axial torsion
\begin{equation}
    S^{\mu}
=-\frac{3\kappa}{\lambda}J_{5}^{\mu},
\qquad
\hat{\Gamma}^{\rho}{}_{\mu\nu}
=
\Gamma^{\rho}{}_{\mu\nu}
-\frac{1}{12}
\epsilon^{\sigma\rho}{}_{\mu\nu}S_{\sigma},
\qquad
\hat{S}_{\mu}=0,
\end{equation}
we can rewrite the fundamental metric-affine Lagrangian as:
\begin{align}
\mathcal{L}_{\mathrm{eff}}
={}&
\frac{\sqrt{-g}}{2\kappa}
\Big[
a_{0}R(g,\hat{\Gamma})
+b_{1}\hat{T}_{\rho\mu\nu}
          \hat{T}^{\rho\mu\nu}
+b_{2}\hat{T}_{\rho\mu\nu}
          \hat{T}^{\mu\nu\rho}
+b_{3}\hat{T}_{\mu}\hat{T}^{\mu}
\nonumber\\
&\quad
+c_{1}\hat{Q}_{\rho\mu\nu}
          \hat{Q}^{\rho\mu\nu}
+c_{2}\hat{Q}_{\nu\mu\rho}
          \hat{Q}^{\rho\mu\nu}
+c_{3}\hat{Q}_{\mu}\hat{Q}^{\mu}
+c_{4}\hat{\tilde{Q}}_{\mu}
          \hat{\tilde{Q}}^{\mu}
+c_{5}\hat{Q}_{\mu}
          \hat{\tilde{Q}}^{\mu}
+a_{1}\hat{Q}_{\rho\mu\nu}
          \hat{T}^{\nu\rho\mu}
+a_{2}\hat{Q}_{\mu}\hat{T}^{\mu}
+a_{3}\hat{\tilde{Q}}_{\mu}
          \hat{T}^{\mu}
\Big]
\nonumber\\
&+
\frac{i}{2}\sqrt{-g}
\Big[
\bar{\psi}\gamma^{\mu}\gr{D}_{\mu}\psi
-
\left(\gr{D}_{\mu}\bar{\psi}\right)
\gamma^{\mu}\psi
\Big]
-\frac{3\kappa\sqrt{-g}}{16\lambda}
\left(
\bar{\psi}\gamma_{\mu}\gamma^{5}\psi
\right)
\left(
\bar{\psi}\gamma^{\mu}\gamma^{5}\psi
\right).
\end{align}
Thus, this Lagrangian, after having inserted the solutions given in Eq.\ \eqref{eq:ec-like-pure-axial-solution}, is equivalent to an Einstein-Cartan-like theory
\begin{align}
\mathcal L_{\mathrm{eff}}&=
\frac{a_0\sqrt{-g}}{2\kappa}\gr{R}+\frac{i\sqrt{-g}}2
\Big[
\bar\psi\gamma^\mu\gr{D}_\mu\psi
-\left(\gr{D}_\mu\bar\psi\right)\gamma^\mu\psi
\Big]
-\frac{3\kappa\sqrt{-g}}{16\lambda}
\left(\bar\psi\gamma_\mu\gamma^5\psi\right)
\left(\bar\psi\gamma^\mu\gamma^5\psi\right).
\end{align}
\label{eq:ec-like-final-effective-lagrangian}

In summary, the situation is the following.
While the Dirac action is insensitive to the projective transformations generated by $p_\mu$, $q_\mu$, and $r_\mu$, its invariance under the axial transformation requires the projective parameter to be locally exact and related to the chiral parameter through $s_\mu=\frac{2}{3}\partial_\mu\alpha$.
The global chiral symmetry survives for $\lambda\neq 0$. The broken symmetry is the local combined axial-projective–chiral symmetry. In particular, the fermionic matter preserves only a locally exact axial-chiral symmetry of the arbitrary covector axial-projective transformations.
The axial projection of the Dirac hypermomentum leads to the connection equation $\left(a_0-4b_1-4b_2\right)S^\mu=-3\kappa J^{5\mu}.$
This equation identifies the combination $\lambda=a_0-4b_1-4b_2$
as the parameter controlling the response of axial torsion to the fermionic spin current. At the axial-projectively invariant point, $\lambda=0$, the axial connection equation degenerates and imposes the constraint $J^{5\mu}=0$.
Consequently, an exactly axial-projectively invariant gravitational sector cannot accommodate a generic Dirac configuration with nonvanishing axial current. This also shows that the limit $\lambda\rightarrow0$ is singular and cannot be obtained smoothly from the effective fermionic description. For $\lambda\neq0$, the axial-projective symmetry is explicitly broken and the axial torsion becomes an auxiliary, algebraically determined field, $S^\mu=-\frac{3\kappa}{\lambda}J^{5\mu}$.
Eliminating this non-propagating component of the connection generates the effective interaction
$\mathcal L_{4\psi}
=
-\frac{3\kappa\sqrt{-g}}{16\lambda}
\left(\bar\psi\gamma_\mu\gamma^5\psi\right)
\left(\bar\psi\gamma^\mu\gamma^5\psi\right)$ 
It is well known in MAG community that fermions can be only coupled to axial torsion and this leads to effective four-fermion interaction after integrating out torsion.  The present result means that imposing projective symmetry does not change the main outcome.
Hence, axial torsion mediates a local spin-spin interaction between fermions, whose sign and strength are completely controlled by $\lambda$. The fundamental metric-affine formulation containing the coupling $S_\mu J^{5\mu}$ and the effective formulation containing the four-fermion operator are equivalent descriptions and must not be combined simultaneously.
The distinctive result of this analysis is the identification of a direct relation between axial-projective symmetry breaking and an induced four-fermion interaction. The same coupling combination $\lambda$ simultaneously measures the departure from axial-projective invariance, renders the axial connection equation invertible, determines the response of torsion to the axial current, and fixes the strength of the induced four-fermion interaction. Importantly, this interaction can be derived without solving all the other connection components, because the axial sector decouples.
The construction therefore provides a direct bridge between the projective structure of metric-affine gravity and effective fermionic physics, while clearly separating the singular symmetric phase $\lambda=0$ from the interacting phase $\lambda\neq0$.

\subsection{Vector and U(1) symmetry}\label{sec: vector-u(1)}
Here, we consider the gravitational Lagrangian with the Klein-Gordon scalar field
\begin{equation}
    \begin{aligned}
\mathcal{L}=\mathcal{L}_{grav}+\mathcal{L}_{\phi}=& \frac{\sqrt{-g}}{2\kappa}\Big[a_0 R
+b_1 T_{\rho\mu\nu}T^{\rho\mu\nu}
+b_2 T_{\rho\mu\nu}T^{\mu\nu\rho}
+b_3 T_\rho T^\rho
\notag\\
&+c_1 Q_{\rho\mu\nu}Q^{\rho\mu\nu}
+c_2 Q_{\nu\mu\rho}Q^{\rho\mu\nu}
+c_3 Q^\mu Q_\mu
+c_4 \tilde Q^\mu\tilde Q_\mu
+c_5 Q^\mu\tilde Q_\mu
\notag\\
&+a_1 Q_{\rho\mu\nu}T^{\nu\rho\mu}
+a_2 Q_\mu T^\mu
+a_3 \tilde Q_\mu T^\mu\Big] -\sqrt{-g} [g^{\mu\nu}(\mathcal{D}_\mu \phi)^\dagger (\mathcal{D}_\nu \phi) + m^2 \phi\phi^\dagger]
    \end{aligned}
\end{equation}
As shown in Sec. \ref{sec:scalar field}, we have the Klein-Gordon Lagrangian is invariant under a combined generalized projective transformation of the connection, Eq.\ \eqref{genprojtr}, and a local phase transformation of the scalar field, Eq.\ \eqref{eq: U(1)}, provided that their parameters satisfy
\begin{equation}
\frac{1}{q}\partial_\mu\beta=p_\mu+4q_\mu+r_\mu.
\end{equation}
Under this condition, the change induced by the projective transformation is canceled by the scalar phase variation, so that
\begin{equation}
\delta\mathcal{L}_\phi
=
\delta_{\mathrm{proj}}\mathcal{L}_\phi
+
\delta_\phi\mathcal{L}_\phi
=
0.
\end{equation}
The compensating phase transformation therefore requires the combination $p_\mu+4q_\mu+r_\mu$ to be locally exact. This restriction applies to that combination of the projective parameters, rather than to each covector separately.
The matter variation with respect to the connection allows us to define the scalar hypermomentum as
\begin{equation}
    \Delta_{\lambda}{}^{\mu\nu}=-\frac{2}{\sqrt{-g}}
\frac{\delta \mathcal{L}_{\phi}}
{\delta\Gamma^{\lambda}{}_{\mu\nu}} = 2iq \delta^{\nu}_{\lambda}\left[\phi^{\dagger}\mathcal{D}^{\mu}\phi-
\phi(\mathcal{D}^{\mu}\phi)^{\dagger}\right]= 2q\delta^\nu_\lambda J^\mu,
\end{equation}
where 
\begin{equation}
J^\mu_\phi=i[\phi^{\dagger}\mathcal{D}^{\mu}\phi-
\phi(\mathcal{D}^{\mu}\phi)^{\dagger}],
\end{equation}
thus
\begin{equation}
    \mathcal{P}_{\alpha}{}^{\mu\nu}=2\kappa qJ^{\mu}\delta^{\nu}{}_{\alpha}.
\end{equation}
The gravitational connection equation is sourced only by the trace-type scalar hypermomentum since $\Delta_{\alpha}{}^{\mu\nu}\propto \delta^\nu_\alpha$. 
Now, since the variation of the gravitational action with respect \eqref{genprojtr} is
\begin{equation}
\delta_{proj} S_{\rm grav}
=\frac{1}{2\kappa}
\int d^{4}x\,\sqrt{-g}\,
\mathcal{P}_{\alpha}{}^{\mu\nu}
\delta\Gamma^{\alpha}{}_{\mu\nu}.
\end{equation}
Substituting the projective variation gives
\begin{align}
\delta_{proj}  S_{grav}&=
\frac{1}{2\kappa}
\int d^{4}x\,\sqrt{-g}
\Big[p_{\nu}\mathcal{P}_{\mu}{}^{\mu\nu}+q_{\mu}\mathcal{P}_{\nu}{}^{\mu\nu}+r^{\alpha}g_{\mu\nu}\mathcal{P}_{\alpha}{}^{\mu\nu}+s_{\sigma}\epsilon^{\sigma\alpha}{}_{\mu\nu}\mathcal{P}_{\alpha}{}^{\mu\nu}\Big].
\end{align}
Since $p_\mu$, $q_\mu$, $r_\mu$, and $s_\mu$ are arbitrary
gauge parameters, projective invariance implies the Noether identities, and the traces of the hypermomentum satisfy
\begin{equation}
\Delta_{\mu}{}^{\mu\nu}
=2qJ_{\phi}^{\nu},
\qquad
\Delta_{\nu}{}^{\mu\nu}
=8qJ_{\phi}^{\mu},
\qquad
g_{\mu\nu}
\Delta_{\alpha}{}^{\mu\nu}
=2qJ_{\phi\alpha}.
\end{equation}
On the other hand, the axial projection vanishes
\begin{equation}
\epsilon^{\lambda\alpha}{}_{\mu\nu}
\Delta_{\alpha}{}^{\mu\nu}=2q\epsilon^{\lambda\alpha}{}_{\mu\nu}
\delta^{\nu}_{\alpha}
\mathcal{J}_{\phi}^{\mu}=0.
\end{equation}
Let us consider the full connection field equations
\begin{equation}
\mathcal{P}_{\alpha}{}^{\mu\nu}
=
\kappa\Delta_{\alpha}{}^{\mu\nu}=2\kappa q \delta^\nu_\alpha J^\mu_{\phi};
\end{equation}
the three trace projections of the complete connection equations are consequently
\begin{subequations}\label{eq: sub-hyper-traces}
    \begin{align}
    \mathcal{P}_\mu{}^{\mu\nu}&=\kappa \Delta_{\mu}{}^{\mu\nu} = 2\kappa q J^\nu_{\phi }, \\
    \mathcal{P}_{\nu}{}^{\mu\nu}&= \kappa \Delta_{\nu}{}^{\mu\nu}= 8\kappa q J^{\mu}_\phi,\\
    g_{\mu\nu}\mathcal{P}_\alpha{}^{\mu\nu} &= \kappa g_{\mu\nu} \Delta_{\alpha}{}^{\mu\nu}= 2 \kappa q J_{\phi \alpha}
    \end{align}
\end{subequations}
the scalar sources only the single trace-projective combination $p_\mu+4q_\mu+r_\mu$, while the axial $s_\mu$ direction remains unsourced. As a consequence, we have that since the theory is invariant under the full projective transformation \eqref{genprojtr}; in particular for the $p_\nu$ vector we have
\begin{align}
    (4b_1-2b_2+6b_3-2a_1-8a_2-2a_3) T^\nu+(a_1+3a_2-4c_1-16c_3-2c_5) Q^\nu & \notag \\+(-a_1+3a_3-4c_2-4c_4-8c_5) \tilde Q^\nu &= 2\kappa q J^\nu_{\phi},
\end{align}
and for $q_\mu$
\begin{align}
    (2a_0-4b_1+2b_2-6b_3+a_1-2a_2-5a_3) T^\mu+ (\frac{3}{2}a_0-a_1-3a_2-2c_2-4c_3-5c_5) Q^\mu & \notag \\+ (-3a_0+a_1-3a_3-4c_1-2c_2-10c_4-2c_5)\tilde Q^\mu &= 8\kappa qJ^\mu_{\phi} \,.
\end{align}
Finally, for $r_\alpha$ vector we have
\begin{equation}
    (-2a_0+a_1-2a_2-5a_3)T_\alpha+(-\frac{1}{2}a_0-2c_2-4c_3-5c_5)Q_\alpha+(-a_0-4c_1-2c_2-10c_4-2c_5) \tilde Q_\alpha = 2\kappa q J_{\phi\alpha},
\end{equation}
and consequently, the gravitational action is invariant, up to the Ricci boundary term, precisely when
\begin{align}
a_1&=2a_0+3a_3,&
a_2&=-a_3,
\\b_1&=\frac{3}{4}a_0-b_3,&
b_2&=-\frac{1}{2}a_0+b_3,
\\c_1&=-2c_4+\frac{1}{2}c_5,&
c_2&=-\frac{1}{2}a_0-c_4-2c_5,
\\c_3&=\frac{1}{8}a_0+\frac{1}{2}c_4-\frac{1}{4}c_5 .
\end{align}
Therefore, $a_0,a_3,b_3,c_4,c_5$ may be taken as the five free parameters, and as in the Dirac case, every gravitational coefficient in the three vector-trace equations vanishes identically. Hence, the connection equation imposes
\begin{equation}
J^\mu_\phi=i[\phi^{\dagger}\mathcal{D}^{\mu}\phi-
\phi(\mathcal{D}^{\mu}\phi)^{\dagger}]=0.
\end{equation}
This means that the geometrical vector $\mathcal{G}_{\mu}=T_{\mu}+\frac{1}{2}Q_{\mu}$
belongs to the projective sector of the affine connection. Since this sector is
a gauge direction of the gravitational action, the corresponding connection
equation does not determine a propagating or algebraic gravitational degree of
freedom. Instead, because the scalar field couples to this projective direction, variation with respect to the connection imposes the vanishing of the scalar
current coupled to $\mathcal{G}_{\mu}$.
Moreover since $\frac{1}{q}\partial_\mu \beta=p_\mu+4q_\mu+r_\mu$, the general projective transformation becomes 
\begin{equation}
    \delta\Gamma^\alpha{}_{\mu\nu} = \delta^\alpha_\mu p_\nu + \delta^\alpha_\nu q_\mu +g_{\mu\nu} \left( \frac{1}{q}\partial^\alpha \beta- p^\alpha - 4q^\alpha \right) + \epsilon^{\sigma\alpha}{}_{\mu\nu}s_\sigma.
\end{equation}

\subsubsection{Adopting the metric-trace gauge}

For simplicity, we select a representative of the combined phase–projective transformation choosing $p_\mu=q_\mu=0$, while retain the full gravitational invariance conditions already imposed, so that  the condition given in Eq.\ \eqref{eq: proj as u(1)} reduces to
\begin{equation}
    r_\mu = \frac{1}{q}\partial_{\mu} \beta,
\end{equation}
and therefore the linear affine connection is
\begin{equation}
    \delta\Gamma^\alpha{}_{\mu\nu}=\frac{1}{q}g_{\mu\nu} \partial^\alpha \beta.
\end{equation}
Note that the torsion tensor $T^\rho{}_{\mu\nu}$ is invariant under the transformation of the vector $r^\alpha$, $\delta_r T^{\rho}_{\mu\nu}=0$, having set $p_\mu=q_\mu=0$.
Therefore, the coefficients are such that
\begin{equation}
\begin{cases}
-2a_0+a_1-2a_2-5a_3=0,\\
-\dfrac{1}{2}a_0-2c_2-4c_3-5c_5=0,\\
-a_0-4c_1-2c_2-10c_4-2c_5=0,
\end{cases}
\label{eq:r-projective-coefficient-system}
\end{equation}
whose solutions are
\begin{equation}
\begin{aligned}
a_1&=2a_0+2a_2+5a_3,\\
c_1&=-\frac{1}{8}a_0+c_3-\frac{5}{2}c_4
     +\frac{3}{4}c_5,\\
c_2&=-\frac{1}{4}a_0-2c_3-\frac{5}{2}c_5.
\end{aligned}
\label{eq:r-projective-coefficients}
\end{equation}
As a consequence, we have that the invariant Lagrangian is
\begin{equation}
\begin{aligned}
\mathcal{L}_{\mathrm g}
={}&
a_0 R
+b_1 T_{\rho\mu\nu}T^{\rho\mu\nu}
+b_2 T_{\rho\mu\nu}T^{\mu\nu\rho}
+b_3 T_\mu T^\mu
\\
&+
\left(
-\frac{1}{8}a_0+c_3-\frac{5}{2}c_4+\frac{3}{4}c_5
\right)
Q_{\rho\mu\nu}Q^{\rho\mu\nu}
+
\left(
-\frac{1}{4}a_0-2c_3-\frac{5}{2}c_5
\right)
Q_{\nu\mu\rho}Q^{\rho\mu\nu}
\\
&+
c_3 Q_\mu Q^\mu
+c_4 \tilde{Q}_\mu\tilde{Q}^\mu
+c_5 Q_\mu\tilde{Q}^\mu
+
\left(
2a_0+2a_2+5a_3
\right)
Q_{\rho\mu\nu}T^{\nu\rho\mu}
+a_2 Q_\mu T^\mu
+a_3 \tilde{Q}_\mu T^\mu .
\end{aligned}
\label{eq:r-projective-invariant-lagrangian}
\end{equation}
This action is invariant up to the total-divergence term originating from $a_0R$.
Therefore, the projective symmetry leads to the on-shell constraint $J^\mu_\phi=0$
unless the projective symmetry is broken. So, in order to break the $U(1)$-vector projective symmetry and avoid the constraint of a vanishing scalar current, we require the gravitational parameters not to satisfy the projectively invariant relations. We therefore impose
\begin{align}\label{eq:system r break}
    \lambda_T &= -2a_0+a_1-2a_2-5a_3\\
    \lambda_Q &= -\dfrac{1}{2}a_0-2c_2-4c_3-5c_5,\\
\lambda_{\tilde{Q}}&=-a_0-4c_1-2c_2-10c_4-2c_5.
\end{align}
Now, in order to have one dimensionless constant, in analogy to the Dirac case of Sec.\ \ref{sec: Axial Dirac}, the physically natural choice is to align the breaking with the vector $\mathcal{G}_{\mu}$ which is seen by the scalar. This leads to defining
\begin{equation}\label{eq:symm break r para}
    \lambda_T=\zeta_\phi, \qquad \lambda_Q = \frac{\zeta_\phi}{2}, \qquad \lambda_{\tilde Q}=0,
\end{equation}
therefore the $r_\mu$ projection of the connection equation simply becomes
\begin{equation}
    \zeta_\phi \mathcal{G}_{\mu} =2\kappa qJ_\mu, 
\end{equation}
which means
\begin{equation}\label{eq:broken-projective-current-equation}
    \mathcal{G}_{\mu} =T_{\mu} + \frac{1}{2}Q_{\mu} = \frac{2\kappa q}{\zeta_\phi}J_\mu =\frac{2i\kappa q}{\zeta_\phi}[\phi^{\dagger}\mathcal{D}_{\mu}\phi-
\phi(\mathcal{D}_{\mu}\phi)^{\dagger}].
\end{equation}
The variation of the breaking term under
Eq.\ \eqref{genprojtr} is proportional to
the variation of $\mathcal{G}_\mu$.  Hence every generalized projective
transformation satisfying
\begin{equation}
    p_\mu+4q_\mu+r_\mu=0
    \label{eq:residual-vector-gauge-condition-gp}
\end{equation}
remains an exact vector gauge symmetry of the broken gravitational sector.
The axial parameter $s_\mu$ also remains a gauge parameter because
$\mathcal{G}_\mu$ is independent of $s_\mu$. It is worth noticing that a representative corresponding to a fixed physical value of
$\mathcal{G}_\mu$ must satisfy Eq.\ \eqref{eq: proj Gmu}; this means that since 
$T_\mu$, $Q_\mu$, and $\tilde Q_\mu$ transform as
\begin{equation}
    T_\mu+\frac12Q_\mu
    =
    3(p_\mu-q_\mu)
    +
    \mathcal G_\mu
    -3p_\mu
    +3q_\mu=\mathcal G_\mu,
    \label{eq:G-identity-general-gauge-gp}
\end{equation}
then these vectors vary from one
generalized projective representative to another, but their
matter-sourced combination $\mathcal{G}_\mu$ is fixed.
Under the combined infinitesimal transformation
\begin{equation}
    \delta_{\beta}\phi=i\beta\phi,
    \qquad
    \delta_{\beta}\phi^{\dagger}
    =-i\beta\phi^{\dagger},
    \qquad
    \delta_{\beta}\Gamma^{\rho}{}_{\mu\nu}
    =
    \frac{1}{q}g_{\mu\nu}\partial^{\rho}\beta,
\label{eq:broken-combined-transformation}
\end{equation}
the variation of the gravitational action
is
\begin{equation}
    \delta_{\beta}S_{\mathrm g}
    =
    \frac{\zeta_{\phi}}
    {2\kappa q}
    \int d^4x\,\sqrt{-g}\,
    \mathcal{G}^{\mu}\partial_{\mu}\beta.
\end{equation}
After integration by parts and neglecting the boundary contribution,
this becomes
\begin{equation}
    \delta_{\beta}S_{\mathrm g}
    =
    -\frac{\zeta_{\phi}}
    {2\kappa q}
    \int d^4x\,\sqrt{-g}\,
    \beta\,
    \mathring{\nabla}_{\mu}\mathcal{G}^{\mu}.
\label{eq:variation-broken-gravitational-action}
\end{equation}
Therefore, for $\zeta_{\phi}\neq0$, the local
$U(1)$-vector projective symmetry is explicitly broken. The global
$U(1)$ symmetry, corresponding to $\partial_{\mu}\beta=0$, remains
unbroken.
The surviving global $U(1)$ symmetry implies, upon use of the scalar
field equations,
\begin{equation}
    \mathring{\nabla}_{\mu}J^{\mu}=0.
\label{eq:KG-current-conservation}
\end{equation}
Taking the Levi-Civita divergence of
Eq.\ \eqref{eq:broken-projective-current-equation} gives
\begin{equation}
    \zeta_{\phi}
    \mathring{\nabla}_{\mu}\mathcal{G}^{\mu}
    =
    2\kappa q
    \mathring{\nabla}_{\mu}J^{\mu}
    =0.
\end{equation}
Hence the symmetry-breaking condition allows a non-vanishing
Klein-Gordon current while requiring the sourced vector
$\mathcal{G}_{\mu}$ to be divergence-free on shell.
Thus, with the symmetry breaking conditions of Eqs. \eqref{eq:system r break} and \eqref{eq:symm break r para}, the non-invariant vector part of the gravitational
Lagrangian can be represented, up to $r_{\mu}$-projectively invariant
terms, as
\begin{equation}\label{eq: u(1) lagrangian brok}
    \mathcal{L}_{\mathrm{g,vec}}
    =
    -\frac{\zeta_{\phi}}{2}
    \mathcal{G}_{\mu}\mathcal{G}^{\mu}.
\end{equation}
The expansion of the Klein-Gordon kinetic term gives
\begin{equation}
-g^{\mu\nu}
\left(\mathcal{D}_{\mu}\phi\right)^{\dagger}
\mathcal{D}_{\nu}\phi
=-g^{\mu\nu}
\partial_\mu\phi^{\dagger}
\partial_\nu\phi
+ q\,\mathcal{G}_{\mu}j^{\mu}
-q^{2}\phi^{\dagger}\phi\,
\mathcal{G}_{\mu}\mathcal{G}^{\mu},
\label{eq:KG-expanded}
\end{equation}
where 
\begin{equation}\label{eq:final-current-general-gauge-gp}
    j^\mu=i(\phi^\dagger \partial^\mu \phi - \phi \partial^\mu \phi^\dagger)
\end{equation}
is the connection-independent current.
Consequently, the complete part of the total Lagrangian density that
depends on $\mathcal{G}_{\mu}$ is
\begin{equation}
    \mathcal{L}_{\mathrm{vec}}
    =
    \sqrt{-g}
    \left[
        -\frac{\zeta_{\phi}}{4\kappa}
        \mathcal{G}_{\mu}\mathcal{G}^{\mu}
        + q\mathcal{G}_{\mu}j^{\mu}
        -q^{2}\phi^{\dagger}\phi\,
        \mathcal{G}_{\mu}\mathcal{G}^{\mu}
    \right].
\label{eq:full-vector-lagrangian}
\end{equation}
Eq.\ \eqref{eq:broken-projective-current-equation} is implicit because the
covariant current $J^{\mu}$ itself depends on $\mathcal{G}^{\mu}$. In
terms of the connection-independent current $j^{\mu}$, one instead finds
the explicit solution
\begin{equation}
    \mathcal{G}^{\mu}
    =
    \frac{2\kappa q}
    {\zeta_{\phi}
     +4\kappa q^{2}\phi^{\dagger}\phi}
    j^{\mu}.
\label{eq:G-explicit-solution}
\end{equation}
To complete the square, we define
\begin{equation}
    A(\phi)
    =
    \frac{\zeta_{\phi}}{4\kappa}
    +q^{2}\phi^{\dagger}\phi.
\end{equation}
Equation \eqref{eq:full-vector-lagrangian} can then be written as
\begin{equation}
\mathcal{L}_{\mathrm{vec}}
=
\sqrt{-g}
\Bigg[
-A(\phi)
\left(
    \mathcal{G}_{\mu}
    -
    \frac{ q}{2A(\phi)}j_{\mu}
\right)^2
+\frac{q^{2}}{4A(\phi)}
j_{\mu}j^{\mu}
\Bigg].
\label{eq:vector-completed-square}
\end{equation}
After eliminating the auxiliary vector $\mathcal{G}_{\mu}$, the induced
current-current interaction is
\begin{equation}
    \mathcal{L}_{\mathrm{vec}}^{\mathrm{eff}}
    =
    \sqrt{-g}\,
    \frac{\kappa q^{2}}
    {\zeta_{\phi}
     +4\kappa q^{2}\phi^{\dagger}\phi}
    j_{\mu}j^{\mu}.
\label{eq:effective-current-current}
\end{equation}
Unlike the corresponding Dirac result, the coefficient of the induced
interaction is field dependent, because of the quadratic
term
\begin{equation}
    -q^{2}\phi^{\dagger}\phi\,
    \mathcal{G}_{\mu}\mathcal{G}^{\mu},
\end{equation}
which is necessarily generated by the Klein-Gordon covariant kinetic
term.
In this case, the connection equation fixes only the $r^\alpha$-projective part. Therefore, the  $r^\alpha$-shifted representative of the affine connection for which $\hat{\mathcal{G}}^\alpha=0$, is
\begin{equation}\label{eq: tran G gamma}
\hat{\Gamma}^\alpha{}_{\mu\nu}=\Gamma^{\alpha}{}_{\mu\nu} + g_{\mu\nu}\mathcal{G}^\alpha=\Gamma^{\alpha}{}_{\mu\nu} + \frac{2\kappa q}{\zeta_{\phi}}g_{\mu\nu}J^\alpha,
\end{equation}
being $\zeta_{\phi}\neq0$.
On the nondegenerate branch, the traceless torsion and nonmetricity components vanish through their homogeneous equations. Using the residual generalized projective freedom satisfying Eq.\ \eqref{eq:residual-vector-gauge-condition-gp}, together with the axial $s_\mu$ symmetry, we choose a representative in which the torsion vanishes and the entire sourced vector $\mathcal{G}_\mu$ is carried by the nonmetricity sector, having
\begin{equation}
    T^\rho{}_{\mu\nu}=0,
    \qquad
    T_\mu=0,
    \label{eq:pure-r-torsion-zero-gp}
\end{equation}
while
\begin{equation}
    Q_{\rho\mu\nu}
    =
    g_{\rho\mu}\mathcal G_\nu
    +
    g_{\rho\nu}\mathcal G_\mu,
    \label{eq:pure-r-Q-gp}
\end{equation}
and
\begin{equation}
    Q_\mu=2\mathcal G_\mu,
    \qquad
    \tilde Q_\mu=5\mathcal G_\mu.
    \label{eq:pure-r-Q-traces-gp}
\end{equation}
Therefore the statement $T_\mu=0$ is a property of this representative; it
is not a gauge-independent consequence of the connection field equations.

The scalar sources only the vector combination $\mathcal{G}_\mu$.
The traceless torsion and traceless nonmetricity equations are homogeneous and for a nondegenerate quadratic sector, they vanish.
After eliminating $\mathcal{G}_{\mu}$, using \eqref{eq: tran G gamma} the effective metric-affine
Lagrangian density becomes
\begin{align}
\mathcal{L}_{\mathrm{eff}}
=
\sqrt{-g}\Bigg\{&
\frac{1}{2\kappa}
\Big[
a_0R(g,\hat{\Gamma})
+b_1\hat{T}_{\rho\mu\nu}
       \hat{T}^{\rho\mu\nu}
+b_2\hat{T}_{\rho\mu\nu}
       \hat{T}^{\mu\nu\rho}
+b_3\hat{T}_{\mu}\hat{T}^{\mu}
\nonumber\\
&\quad
+c_1\hat{Q}_{\rho\mu\nu}
       \hat{Q}^{\rho\mu\nu}
+c_2\hat{Q}_{\nu\mu\rho}
       \hat{Q}^{\rho\mu\nu}
+c_3\hat{Q}_{\mu}\hat{Q}^{\mu}
+c_4\hat{\tilde{Q}}_{\mu}
       \hat{\tilde{Q}}^{\mu}
+c_5\hat{Q}_{\mu}
       \hat{\tilde{Q}}^{\mu}
\nonumber\\
&\quad
+a_1\hat{Q}_{\rho\mu\nu}
       \hat{T}^{\nu\rho\mu}
+a_2\hat{Q}_{\mu}\hat{T}^{\mu}
+a_3\hat{\tilde{Q}}_{\mu}
       \hat{T}^{\mu}
\Big]
\nonumber\\
&\quad
-g^{\mu\nu}
 \partial_\mu \phi^\dagger
 \partial_\nu\phi
-m^2\phi^\dagger\phi
+
\frac{\kappa q^2}
{\zeta_\phi+4\kappa q^2\phi^\dagger\phi}
j_{\mu}j^{\mu}
\Bigg\},
\label{eq:effective-MA-lagrangian}
\end{align}
with $\zeta_\phi+4\kappa q^2\phi^\dagger\phi\neq 0$.
After the auxiliary affine vector has been eliminated, the complete effective
Lagrangian is
\begin{equation}
        \mathcal L_{\mathrm{eff}}
        =
        \sqrt{-g}\Bigg[
            \frac{a_0}{2\kappa}\mathring R
            -
            g^{\mu\nu}
            \partial_\mu\phi^\dagger
            \partial_\nu\phi
            -
            m^2\phi^\dagger\phi
           +
            \frac{\kappa q^2}
            {\zeta_\phi+4\kappa q^2\phi^\dagger\phi}
            j_\mu j^\mu
        \Bigg].
    \label{eq:final-effective-Lagrangian-general-gauge-gp}
\end{equation}
Equation \eqref{eq:final-effective-Lagrangian-general-gauge-gp} is independent
of the representative chosen on the residual generalized projective orbit.
A convenient field parametrization is obtained by projectively shifting the invariant gravitational part to a representative with $\hat{\mathcal{G}}_\mu=0$. Keeping
$p_\mu$ and $q_\mu$ nonzero redistributes the same
$\mathcal{G}_\mu$ between the torsion and nonmetricity traces, while a
nonzero $s_\mu$ changes only the unsourced axial gauge sector.  None of
these changes affects the algebraic equation for $\mathcal{G}_\mu$ or the
effective current-current interaction.

The complex scalar sources only the vector sector of the affine connection. The traceless torsion and traceless nonmetricity components vanish on the appropriate nondegenerate branch; the unsourced axial component remains a projective gauge mode and may be gauge-fixed to zero.  The three vector-trace equations then algebraically determine the combination $\mathcal G_\mu$ of $T_\mu,Q_\mu,$ and $\tilde Q_\mu$, with the $r_\mu$ projection directly determining $\mathcal G_\mu=T_\mu+\frac12Q_\mu$; two vector combinations remain gauge-dependent.

In this subsection, we identify the geometrical vector
$\mathcal G_\mu=T_\mu+\tfrac12Q_\mu$ as the component of the affine connection that can play the role of an Abelian connection for a complex scalar field. Under the generalized projective transformation, it changes according to
$\delta\mathcal G_\mu=-p_\mu-4q_\mu-r_\mu$. Consequently, the scalar covariant derivative transforms covariantly under a local $U(1)$ phase transformation provided that
$p_\mu+4q_\mu+r_\mu=q_\phi^{-1}\partial_\mu\beta$. The resulting scalar hypermomentum is purely of vector-trace type: its three traces occur in the fixed ratio \eqref{eq: sub-hyper-traces}, while its axial projection vanishes.
When the gravitational sector is invariant under arbitrary generalized projective transformations, the corresponding projections of its connection equation vanish identically. The scalar coupling then forces the covariant Klein-Gordon current to vanish on shell. This reveals an obstruction to coupling a nonzero scalar current to a completely degenerate projective sector. 
We introduce the one-parameter breaking term
$-\frac12\zeta_\phi\mathcal G_\mu\mathcal G^\mu$. This term explicitly breaks only the projective direction acting on $\mathcal G_\mu$, while the global $U(1)$ symmetry remains intact. If the undeformed gravitational action possesses the full generalized projective symmetry, the transformations satisfying $p_\mu+4q_\mu+r_\mu=0$, together with the axial $s_\mu$ transformation, remain gauge symmetries.
Meanwhile, the unsourced traceless connection components vanish when their algebraic field equations are nondegenerate. Eliminating $\mathcal G_\mu$ consequently generates a field-dependent current–current interaction, while the surviving global $U(1)$ symmetry ensures conservation of the Klein–Gordon current on shell. Unlike the corresponding Dirac construction, the Klein-Gordon kinetic term is quadratic in the geometrical connection and necessarily produces a contribution proportional to
$q^2|\phi|^2\mathcal G_\mu\mathcal G^\mu$. \\
Finally, the scalar sector differs qualitatively from the Dirac case because the
Klein-Gordon kinetic term is already quadratic in the affine vector $G_\mu$.
To make this point explicit, let us locally parametrize the complex scalar field as \cite{Schwabl08}
\begin{equation}
    \phi=\rho e^{i\theta},
    \qquad \rho>0.
\end{equation}
The connection-independent current defined in Eq.\ \eqref{eq:final-current-general-gauge-gp} then becomes
\begin{equation}
j_\mu=i\left(\phi^\dagger\partial_\mu\phi
    -\phi\,\partial_\mu\phi^\dagger\right)
    =-2\rho^2\partial_\mu\theta.
\end{equation}
Using this expression in the effective Lagrangian \eqref{eq:final-effective-Lagrangian-general-gauge-gp}, together with
\begin{equation}
g^{\mu\nu}\partial_\mu\phi^\dagger\partial_\nu\phi
    =(\partial\rho)^2+\rho^2(\partial\theta)^2,
\end{equation}
the scalar sector can be written as
\begin{equation}
\frac{\mathcal L_{\phi,\mathrm{eff}}}{\sqrt{-g}}
=
-(\partial\rho)^2
-
\frac{\zeta_\phi\,\rho^2}
{\zeta_\phi+4\kappa q^2\rho^2}
(\partial\theta)^2
-m^2\rho^2.
\label{eq:scalar-polar-effective-action}
\end{equation}
This form makes the projectively invariant limit particularly transparent.
For $\zeta_\phi=0$ and $\rho>0$, the kinetic term of the phase $\theta$
vanishes. This is consistent with the restoration of the local combined
phase-projective symmetry and with the connection equation
$J^\mu_\phi=0$: in the symmetric phase the scalar phase is not an
independent physical degree of freedom.
It is also important that the algebraic elimination of $G_\mu$ does not
become singular simply at $\zeta_\phi=0$. Rather, the relevant
nondegeneracy condition is
\begin{equation}
    \zeta_\phi+4\kappa q^2\phi^\dagger\phi\neq0.
\end{equation}
Therefore, for $\phi\neq0$, the projectively invariant point
$\zeta_\phi=0$ belongs to a regular branch of the scalar effective theory.
The case $\phi=0$ has to be treated separately, since the polar
decomposition is then ill-defined and the algebraic equation for the
phase sector changes rank.
Finally, the sign of the effective phase kinetic term is controlled by
\begin{equation}
    \frac{\zeta_\phi}
    {\zeta_\phi+4\kappa q^2\rho^2}.
\end{equation}
Consequently, the allowed parameter region should be chosen such that this
ratio has the sign required for a healthy scalar kinetic sector. The $\zeta_\phi$ deformation then makes the projective direction non-gauge and produces the nonpolynomial current-current interaction.
The distinctive result is therefore a symmetry-based identification of the affine vector seen by the scalar, the demonstration of the current constraint produced by complete generalized projective invariance, and the derivation of a minimal symmetry-breaking mechanism that converts this otherwise degenerate connection component into a definite, nonpolynomial current-current interaction.

\section{Conclusions and perspectives}\label{sec: conclusions}

In this work, we investigated the role of generalized projective symmetry in four-dimensional, parity-even metric-affine gravity, restricting the gravitational action to terms linear in the curvature and at most quadratic in torsion and nonmetricity. The central result is that projective symmetry is not merely a redundancy of the affine description. It determines the rank of the blocks of the connection field equation, identifies which components of the connection are gauge variables, constrains the admissible matter hypermomentum, and controls the effective interactions generated when the non-Riemannian connection is eliminated.

Starting from the most general action containing the Ricci scalar together with the eleven independent parity-even quadratic invariants of torsion, nonmetricity, and their mixed contractions, we studied the generalized projective transformation \eqref{genprojtr}.

Requiring the invariance of the gravitational action, up to a Levi-Civita total divergence, produces seven independent relations among the $12$ gravitational coefficients. The completely invariant theory therefore forms a five-parameter family, which may be parametrized by $a_0,a_3,b_3,c_4$, and $c_5$. The mixed torsion-nonmetricity invariants are essential for this result: since generalized projective transformations mix the vector traces of torsion and nonmetricity, the invariant parameter space would not close if these mixed terms were discarded from the outset.

The first methodological contribution of this work is a symmetry-adapted procedure for solving the vacuum connection equations, particularly in the projectively invariant cases in which the connection operator is necessarily singular. The connection equation was treated as a single coupled linear system acting on the irreducible torsion and nonmetricity components. In particular, the nonmetricity, torsion-vector, axial-torsion, and tensor-torsion contributions are different blocks of the same equation; their separate vanishing cannot be assumed because cancellations and mixing between equivalent irreducible representations are possible. We first identified the null directions of this operator through the projective Noether identities, then removed only the corresponding gauge variables, and finally studied the rank of the remaining vector, completely symmetric, and mixed-symmetry blocks.

This procedure was applied to the axial, metric-trace, and fully generalized projective sectors. In the axial-projectively invariant theory, the axial torsion $S_\mu$ remains arbitrary, while all other post-Riemannian components vanish for generic nondegenerate coefficients. In the metric-trace sector, the corresponding nonmetricity vector is a gauge mode, while the remaining torsion and nonmetricity components again vanish on the generic branch. In the completely invariant theory, all three vector traces and the axial torsion belong to the generalized projective gauge orbit. Provided that the remaining tensor blocks are nondegenerate, the vacuum connection takes the form given in Eq.\ \eqref{genprojtr}, where the four covectors parametrize gauge directions. After gauge fixing, the connection reduces to the Levi-Civita connection and the vacuum action reduces, up to a boundary term, to the Einstein-Hilbert action. Consequently, for $a_0\neq0$, the generic vacuum metric equation coincides with the vacuum Einstein equation.

This result must be understood together with the nondegeneracy assumptions. On special hypersurfaces of the parameter space, the vector, symmetric-traceless, or mixed-symmetry blocks may acquire additional zero modes. Such exceptional branches can contain connection components that are not generated by the assumed projective symmetry. The rank-based method developed here therefore does more than reproduce the generic Levi-Civita result: it distinguishes genuine gauge freedom, which follows from a Noether identity, from accidental degeneracy, which occurs only for special combinations of the gravitational coefficients.

The inclusion of matter reveals the physical significance of this distinction. Projective invariance of the gravitational action implies that the corresponding projections of its connection Euler-Lagrange tensor vanish identically. A matter field carrying hypermomentum along one of these null directions cannot generically source a gravitational response algebraic on torsion and nonmetricity tensors. Instead, the associated connection equation imposes a constraint on the matter current. Thus, exact projective symmetry acts as a selection rule for the hypermomentum that can consistently couple to the independent connection.

For a minimally coupled Dirac field, the Hermitian form of the Dirac field is insensitive to the vector traces of torsion and nonmetricity and couples only to the axial torsion, as shown in Eq.\ \eqref{eq: S come j5}. This identifies the axial-projective mode as the affine component seen by the fermionic spin current. At the classical level and for a massless fermion, a local chiral transformation can compensate a locally exact axial-projective transformation according to Eq.\ \eqref{eq: axial as chiral transf}.

The correspondence is therefore between the chiral symmetry and the locally exact axial-projective transformations. It is not a one-to-one identification with an arbitrary axial covector $s_\mu$. The remaining projective transformations generated by $p_\mu$, $q_\mu$, and $r_\mu$ act trivially on the minimally coupled Dirac action.

For the electromagnetic field, the standard Maxwell tensor $\textbf{F}=d\textbf{A}$ is independent of the affine connection. Its action consequently preserves both electromagnetic $U(1)$ invariance and generalized projective invariance without requiring a direct coupling to torsion or nonmetricity. We also considered the genuinely metric-affine alternative $\tilde F_{\mu\nu}=2\nabla_{[\mu}A_{\nu]}$, which contains torsion and therefore is neither electromagnetically gauge invariant nor projectively invariant by itself. A simultaneous completion of both symmetries was constructed by introducing the projectively invariant combinations $\tilde T_\mu=T_\mu-3P_\mu, \tilde S_\mu=S_\mu-12X_\mu$, together with the Stückelberg-invariant vector $B_\mu=A_\mu-\partial_\mu\chi$. This construction makes explicit the distinction between ordinary Maxwell theory, which does not probe the independent connection, and a compensated Maxwell-like extension in which the electromagnetic sector couples directly to the torsional geometry.

For a complex Klein-Gordon field, we introduced the nonminimal affine derivative \eqref{eq: derivative KG field}.
The combination $\mathcal{G}_\mu$ is selected by its transformation law given in Eq.\ \eqref{eq: proj Gmu}.
It can therefore play the role of an Abelian connection for the scalar phase whenever the condition \eqref{eq: proj as u(1)} is respected.
As in the fermionic case, the correspondence involves a locally exact representative of the generalized projective transformations. Moreover, the scalar probes only the single vector combination $p_\mu+4q_\mu+r_\mu$; transformations in its kernel remain invisible to the scalar action. The resulting hypermomentum is purely of vector-trace type, while its axial projection vanishes. When the electromagnetic potential is also included, the scalar can simultaneously carry the projective-induced charge $q$ and the ordinary electromagnetic charge $e$, realizing the product structure $U(1)_{\rm grav}\times U(1)_{\rm EM}$.
The analysis of explicit symmetry breaking provides the physical link between projective geometry and matter interactions. In the Dirac sector, the combination
$\lambda=a_0-4b_1-4b_2$
simultaneously measures the breaking of axial-projective invariance and controls the algebraic response of the axial torsion:
$\lambda S^\mu=-3\kappa J_5^\mu$.
Exactly at the invariant point $\lambda=0$, the axial gravitational equation is a Noether identity and the sourced connection equation imposes $J_5^\mu=0$. Thus, a generic fermionic configuration with nonvanishing axial current cannot be coupled to a completely degenerate axial-projective sector. For $\lambda\neq0$, the symmetry is explicitly broken, the axial equation becomes invertible, and $S_\mu$ is an auxiliary field determined by the fermionic current. Eliminating it generates the contact interaction \eqref{eq: four fermions interaction}.
The same parameter therefore quantifies the departure from projective invariance, lifts the corresponding zero mode, and fixes both the sign and strength of the induced fermionic interaction. The point $\lambda=0$ is a different-rank branch of the theory and should not be obtained by naively taking the limit $\lambda\rightarrow0$ in the effective interaction.

An analogous mechanism operates in the scalar vector sector. Exact generalized projective invariance makes the gravitational equation along $\mathcal{G}_\mu$ degenerate and consequently imposes the vanishing of the covariant scalar current. We lifted this degeneracy through the minimal breaking term given in Eq.\ \eqref{eq: u(1) lagrangian brok}, in the gravitational Lagrangian. This breaks only the projective direction seen by the scalar; when the undeformed action is fully invariant, the transformations satisfying $p_\mu+4q_\mu+r_\mu=0$
and the axial transformations generated by $s_\mu$ remain as residual gauge symmetries. The vector equation becomes $\zeta_\phi\mathcal G^\mu=2\kappa q_\phi J^\mu.$
Writing the covariant current in terms of the connection-independent current \eqref{eq:final-current-general-gauge-gp},
we obtain Eq.\ \eqref{eq:G-explicit-solution} and by 
eliminating the auxiliary vector $\mathcal{G}_\mu$, we generate Eq.\ \eqref{eq:effective-current-current}.
In contrast with the fermionic four-current interaction, its coefficient is field dependent. This dependence is unavoidable because the Klein-Gordon kinetic term is quadratic in the affine vector and produces the term $q_\phi^2|\phi|^2\mathcal G_\mu\mathcal G^\mu.$ Although the local projective-phase symmetry is explicitly broken, the global scalar $U(1)$ symmetry survives and ensures conservation of the corresponding current on-shell.

The common physical mechanism can therefore be summarized as follows. An exact projective symmetry produces a null direction of the gravitational connection operator. If matter carries a current along that direction, the connection equation imposes a constraint on the current rather than determining the connection. A symmetry-breaking quadratic term lifts the null direction and turns the associated connection component into an auxiliary mediator. Integrating out that component then produces a local matter self-interaction. Since the connection has no independent kinetic term in the class of theories considered here, these effects are contact interactions rather than forces mediated by additional propagating fields.

The main novelty of the present construction is consequently the unified chain connecting generalized projective transformations, Noether identities, matter symmetries, hypermomentum sources, and effective interactions. The analysis identifies which affine component is seen by each matter sector: axial torsion by the Dirac axial current, the vector $\mathcal{G}_\mu$ by the Klein-Gordon phase current, and torsional combinations by the compensated Maxwell-like field strength. It also shows that the relevant gravitational coupling combinations have a dual role: they measure the breaking of the corresponding projective symmetry and determine the strength of the matter interaction generated after the connection is eliminated.

The present conclusions apply to a classical, four-dimensional, parity-even theory whose connection remains algebraic. Natural extensions include the study of the exceptional degenerate branches, quantum corrections and the axial anomaly, the dynamics of the compensator fields, and theories containing curvature-squared or derivative terms that can make parts of the affine connection propagate. It would also be interesting to investigate whether the induced fermionic and scalar interactions can leave observable signatures in particle physics, compact objects, or the early Universe. In particular, a further and more restrictive research direction would be to reverse the logic used in this work and ask whether the Equivalence Principle itself can act as a selection criterion for the admissible affine $U(1)$ seen by scalar matter. In the present construction, the combination $G_{\mu}=T_{\mu}+\frac{1}{2}Q_{\mu}$ is selected by its transformation law under the generalized projective symmetry. A broader starting point would be to consider the most general vector built from the affine traces at the same derivative order,
\begin{equation}
\mathcal{G}_{\mu}=\alpha T_{\mu}+\beta Q_{\mu}+\gamma \tilde{Q}_{\mu},
\end{equation}
and to determine which choices of $(\alpha,\beta,\gamma)$ can consistently define a local Abelian phase symmetry once the gravitational action, the scalar hypermomentum, and the electromagnetic sector are treated simultaneously. Projective covariance would provide one restriction, while the requirement that the same local inertial physics be recovered for different scalar species could provide an independent physical restriction. In this way, the particular $U(1)_{\mathrm{grav}}$ found here would no longer be selected only algebraically: one could test whether it is also singled out by a universality requirement on matter.
Overall, our results show that projective symmetry provides a systematic organizing principle for deciding which parts of a metric-affine connection are gauge, which parts can be sourced by matter, and how controlled departures from the symmetry translate the non-Riemannian geometry into effective matter physics.

\section*{Acknowledgments}
M.J.G.\ has been supported by the Estonian Research Council grant PSG910 ``Theoretical frameworks for numerical modified gravity''. L.J.\ was supported by the Estonian Research Council team grant PRG2608 ``Space - Time - Matter'' as well as Estonian Ministry of Education and Research Centre of Excellence TK202U4 ``Foundations of the Universe''. C.F. acknowledges the Istituto Italiano di Fisica Nucleare (INFN) iniziativa specifica QGSKY, the University of Tartu for the hospitality, and the colleagues D. Iosifidis and F.J. Maldonado Torralba for the useful discussions.

\appendix
\section{Appendixes}
\label{Appendix}
The following appendices provide the technical details underlying several derivations presented in the main text. We first derive explicitly the metric field equations associated with the general parity-even quadratic metric-affine action and perform their post-Riemannian expansion in terms of the Levi-Civita connection, torsion, and nonmetricity. We then give a detailed analysis of the connection field equations using the symmetry-adapted approach developed in the main text. In particular, we treat separately the axial-projective, metric-trace, and fully projectively invariant sectors, identifying the corresponding null directions of the connection operator and deriving the conditions under which the remaining irreducible components are uniquely determined.

\section{Derivation of metric field equations}
The direct variation of \eqref{RQTlagrangian} with respect to the metric $\delta g^{\mu\nu}$ leads to the metric field equations, written without performing the post-Riemannian expansion
\begin{equation}
 \begin{aligned}     
\mathcal{M}_{\mu\nu} =& a_0(R_{(\mu\nu)} - \frac{1}{2} g_{\mu\nu}R) - \frac{1}{2}g_{\mu\nu} (L_{Q} + L_{T}+ L_{TQ})  \\ &+b_1(T^{\alpha\beta}{}_{\mu}T_{\alpha\beta\nu} + T^{\alpha\beta}{}_{\nu}T_{\alpha\beta\mu} - T_{\mu\alpha\beta} T_{\nu}{}^{\alpha\beta
    }) + \frac{b_2}{2}(T^{\alpha}{}_{\beta\nu}T^\beta{}_{\mu\alpha} + T^{\alpha}{}_{\beta\mu}T^\beta{}_{\nu\alpha}) + b_3T_{\mu}T_{\nu} \\ & + c_1 (2\nabla_\alpha Q^{\alpha}_{\ \mu\nu}+ 2(T_\alpha +\frac{1}{2} Q_\alpha) Q^{\alpha}{}_{\ \mu\nu}  - Q^{\alpha\beta}{}_{\nu} Q_{\alpha \beta\mu} -Q^{\alpha\beta}{}_{\mu} Q_{\alpha \beta\nu} + Q_{\mu\alpha\beta} Q_{\nu}{}^{\alpha \beta}) \\ &+
c_2 [\nabla_{\alpha}(Q_{\mu}{}^{\alpha}{}_{\nu}+ Q_{\nu}{}^{ \alpha}{}_{\mu})+(T_\alpha + \frac{1}{2}Q_\alpha)(Q_{\mu}{}^{\alpha}{}_{\nu}+ Q_{\nu}{}^{ \alpha}{}_{\mu}) -  \frac{1}{2}(Q^{\alpha\beta}{}_{\mu} Q_{\beta \alpha \nu} + Q^{\alpha\beta}{}_{\nu} Q_{\beta \alpha \mu})] \\ & + c_3(
Q_\mu Q_\nu+2g_{\mu\nu}\nabla_\rho Q^\rho
+2g_{\mu\nu}T_\rho Q^\rho+g_{\mu\nu}Q_\rho Q^\rho)\\ &+ c_4(\nabla_\mu \tilde{Q}_\nu+\nabla_\nu \tilde{Q}_\mu+T_\mu \tilde{Q}_\nu
+T_\nu \tilde{Q}_\mu+\frac{1}{2} Q_\mu \tilde{Q}_\nu
+\frac{1}{2} Q_\nu \tilde{Q}_\mu- \tilde{Q}_\mu\tilde{Q}_\nu) \\ &+ 
\frac{1}{2}c_5 (\nabla_\mu Q_\nu + \nabla_\nu Q_\mu
+ T_\mu Q_\nu +T_\nu Q_\mu
+ Q_\mu Q_\nu
+2g_{\mu\nu}\nabla_\alpha \tilde{Q}^\alpha
+2g_{\mu\nu}T_\alpha \tilde{Q}^\alpha
+g_{\mu\nu}Q_\alpha \tilde{Q}^\alpha) \\ &+ a_1[
\frac{1}{2} Q_{\nu}{}^{\alpha}{}_{\beta}T^{\beta}{}_{\mu\alpha}
+\frac12 Q_{\mu}{}^{\alpha}{}_{\beta}T^{\beta}{}_{\nu\alpha} -\frac{1}{2} Q_{\alpha\mu\beta}T_{\nu}{}^{\alpha\beta}
-\frac{1}{2} Q_{\alpha\nu\beta}T_{\mu}{}^{\alpha\beta}
-\frac{1}{2} T_{\mu\nu}{}^{\alpha}
(T_\alpha+\frac{1}{2} Q_\alpha)
\\ &-\frac{1}{2} T_{\nu\mu}{}^{\alpha}
(T_\alpha+\frac{1}{2} Q_\alpha)-\frac{1}{2} \nabla_\alpha T_{\mu\nu}{}^\alpha
-\frac{1}{2} \nabla_\alpha T_{\nu\mu}{}^\alpha] \\ &+a_2[
\frac{1}{2} Q_\mu T_\nu
+\frac{1}{2} Q_\nu T_\mu
+T^\alpha(T_\alpha+\frac12 Q_\alpha)g_{\mu\nu}
+g_{\mu\nu}\nabla_\alpha T^\alpha] \\ &+a_3[\frac{1}{2} \nabla_\nu T_\mu +\frac{1}{2} \nabla_\mu T_\nu+T_\mu T_\nu+\frac{1}{4} Q_\nu T_\mu+\frac{1}{4} Q_\mu T_\nu].
 \end{aligned}
 \end{equation}
Now, we substitute in the previous equation, the decomposition of the linear affine connection $\Gamma^\alpha{}_{\mu\nu}$ \eqref{eq: general affine connection}
\begin{align}
\mathcal M_{\mu\nu}
={}&a_0\left(R_{(\mu\nu)}-\frac12g_{\mu\nu}R\right)
-\frac12g_{\mu\nu}(L_Q+L_T+L_{TQ})\notag\\
&+b_1\left(
T^{\alpha\beta}{}_\mu T_{\alpha\beta\nu}
+T^{\alpha\beta}{}_\nu T_{\alpha\beta\mu}
-T_{\mu\alpha\beta}T_\nu{}^{\alpha\beta}\right)\notag\\
&+\frac{b_2}{2}\left(
T^\alpha{}_{\beta\nu}T^\beta{}_{\mu\alpha}
+T^\alpha{}_{\beta\mu}T^\beta{}_{\nu\alpha}\right)
+b_3T_\mu T_\nu\notag\\
&+c_1\Bigl[
2\gr{\nabla}_\alpha Q^\alpha{}_{\mu\nu}
+2N^\alpha{}_{\rho\alpha}Q^\rho{}_{\mu\nu}
-2N^\rho{}_{\mu\alpha}Q^\alpha{}_{\rho\nu}
-2N^\rho{}_{\nu\alpha}Q^\alpha{}_{\mu\rho}\notag\\
&\qquad
+2\left(T_\alpha+\frac12Q_\alpha\right)Q^\alpha{}_{\mu\nu}
-Q^{\alpha\beta}{}_\nu Q_{\alpha\beta\mu}
-Q^{\alpha\beta}{}_\mu Q_{\alpha\beta\nu}
+Q_{\mu\alpha\beta}Q_\nu{}^{\alpha\beta}
\Bigr]\notag\\
&+c_2\Bigl[
\gr{\nabla}_\alpha
  \left(Q_\mu{}^\alpha{}_\nu+Q_\nu{}^\alpha{}_\mu\right)
-N^\rho{}_{\mu\alpha}Q_\rho{}^\alpha{}_\nu
+N^\alpha{}_{\rho\alpha}Q_\mu{}^\rho{}_\nu
-N^\rho{}_{\nu\alpha}Q_\mu{}^\alpha{}_\rho\notag\\
&\qquad
-N^\rho{}_{\nu\alpha}Q_\rho{}^\alpha{}_\mu
+N^\alpha{}_{\rho\alpha}Q_\nu{}^\rho{}_\mu
-N^\rho{}_{\mu\alpha}Q_\nu{}^\alpha{}_\rho\notag\\
&\qquad
+\left(T_\alpha+\frac12Q_\alpha\right)
 \left(Q_\mu{}^\alpha{}_\nu+Q_\nu{}^\alpha{}_\mu\right)
-\frac12\left(
Q^{\alpha\beta}{}_\mu Q_{\beta\alpha\nu}
+Q^{\alpha\beta}{}_\nu Q_{\beta\alpha\mu}\right)
\Bigr]\notag\\
&+c_3\Bigl[
Q_\mu Q_\nu
+2g_{\mu\nu}\gr{\nabla}_\rho Q^\rho
+2g_{\mu\nu}N^\rho{}_{\lambda\rho}Q^\lambda
+2g_{\mu\nu}T_\rho Q^\rho
+g_{\mu\nu}Q_\rho Q^\rho
\Bigr]\notag\\
&+c_4\Bigl[
\gr{\nabla}_\mu\tilde Q_\nu
+\gr{\nabla}_\nu\tilde Q_\mu
-N^\rho{}_{\nu\mu}\tilde Q_\rho
-N^\rho{}_{\mu\nu}\tilde Q_\rho\notag\\
&\qquad
+T_\mu\tilde Q_\nu+T_\nu\tilde Q_\mu
+\frac12Q_\mu\tilde Q_\nu
+\frac12Q_\nu\tilde Q_\mu
-\tilde Q_\mu\tilde Q_\nu
\Bigr]\notag\\
&+\frac{c_5}{2}\Bigl[
\gr{\nabla}_\mu Q_\nu+\gr{\nabla}_\nu Q_\mu
-N^\rho{}_{\nu\mu}Q_\rho
-N^\rho{}_{\mu\nu}Q_\rho\notag\\
&\qquad
+T_\mu Q_\nu+T_\nu Q_\mu+Q_\mu Q_\nu
+2g_{\mu\nu}\gr{\nabla}_\alpha\tilde Q^\alpha
+2g_{\mu\nu}N^\alpha{}_{\rho\alpha}\tilde Q^\rho\notag\\
&\qquad
+2g_{\mu\nu}T_\alpha\tilde Q^\alpha
+g_{\mu\nu}Q_\alpha\tilde Q^\alpha
\Bigr]\notag\\
&+a_1\Bigl[
\frac12Q_\nu{}^\alpha{}_\beta T^\beta{}_{\mu\alpha}
+\frac12Q_\mu{}^\alpha{}_\beta T^\beta{}_{\nu\alpha}
-\frac12Q_{\alpha\mu\beta}T_\nu{}^{\alpha\beta}
-\frac12Q_{\alpha\nu\beta}T_\mu{}^{\alpha\beta}\notag\\
&\qquad
-\frac12T_{\mu\nu}{}^\alpha
 \left(T_\alpha+\frac12Q_\alpha\right)
-\frac12T_{\nu\mu}{}^\alpha
 \left(T_\alpha+\frac12Q_\alpha\right)\notag\\
&\qquad
-\frac12\gr{\nabla}_\alpha T_{\mu\nu}{}^\alpha
+\frac12N^\rho{}_{\mu\alpha}T_{\rho\nu}{}^\alpha
+\frac12N^\rho{}_{\nu\alpha}T_{\mu\rho}{}^\alpha
-\frac12N^\alpha{}_{\rho\alpha}T_{\mu\nu}{}^\rho\notag\\
&\qquad
-\frac12\gr{\nabla}_\alpha T_{\nu\mu}{}^\alpha
+\frac12N^\rho{}_{\nu\alpha}T_{\rho\mu}{}^\alpha
+\frac12N^\rho{}_{\mu\alpha}T_{\nu\rho}{}^\alpha
-\frac12N^\alpha{}_{\rho\alpha}T_{\nu\mu}{}^\rho
\Bigr]\notag\\
&+a_2\Bigl[
\frac12Q_\mu T_\nu+\frac12Q_\nu T_\mu
+T^\alpha\left(T_\alpha+\frac12Q_\alpha\right)g_{\mu\nu}
+g_{\mu\nu}\gr{\nabla}_\alpha T^\alpha
+g_{\mu\nu}N^\alpha{}_{\rho\alpha}T^\rho
\Bigr]\notag\\
&+a_3\Bigl[
\frac12\gr{\nabla}_\nu T_\mu
-\frac12N^\rho{}_{\mu\nu}T_\rho
+\frac12\gr{\nabla}_\mu T_\nu
-\frac12N^\rho{}_{\nu\mu}T_\rho\notag\\
&\qquad
+T_\mu T_\nu
+\frac14Q_\nu T_\mu
+\frac14Q_\mu T_\nu
\Bigr].
\end{align}
Using the definition of the curvature scalar \eqref{eq: general R} and of the general Ricci tensor 
\begin{equation}
R_{\mu\nu}= \gr{R}_{\mu\nu}
 +\gr{\nabla}_\alpha N^\alpha{}_{\mu\nu}
 -\gr{\nabla}_\nu N^\alpha{}_{\mu\alpha}
 +N^\alpha{}_{\sigma\alpha}N^\sigma{}_{\mu\nu}
 -N^\alpha{}_{\sigma\nu}N^\sigma{}_{\mu\alpha}\,,
\end{equation}
after having symmetrized it, we get
\begin{align}
\mathcal M_{\mu\nu}
={}&a_0\Bigl[
\gr{R}_{\mu\nu}
+\frac12\gr{\nabla}_\alpha
\bigl(N^\alpha{}_{\mu\nu}+N^\alpha{}_{\nu\mu}\bigr)
-\frac12\gr{\nabla}_\nu N^\alpha{}_{\mu\alpha}
-\frac12\gr{\nabla}_\mu N^\alpha{}_{\nu\alpha}\notag\\
&
+\frac12N^\alpha{}_{\sigma\alpha}
\bigl(N^\sigma{}_{\mu\nu}+N^\sigma{}_{\nu\mu}\bigr)
-\frac12N^\alpha{}_{\sigma\nu}N^\sigma{}_{\mu\alpha}
-\frac12N^\alpha{}_{\sigma\mu}N^\sigma{}_{\nu\alpha}
-\frac12g_{\mu\nu}\gr{R}+
g_{\mu\nu}(-\frac18T_{\rho\alpha\beta}T^{\rho\alpha\beta}
\notag\\
&
-\frac14T_{\rho\alpha\beta}T^{\alpha\rho\beta}
+\frac12T_\alpha T^\alpha
-\gr{\nabla}_\alpha T^\alpha)
-\frac18g_{\mu\nu}Q^{\alpha\rho\sigma}Q_{\alpha\rho\sigma}
+\frac14g_{\mu\nu}Q^{\alpha\rho\sigma}Q_{\rho\sigma\alpha}
+\frac18g_{\mu\nu}Q^\alpha Q_\alpha
-\frac14g_{\mu\nu}\tilde Q^\alpha Q_\alpha\notag\\
&
-\frac12g_{\mu\nu}\gr{\nabla}_\alpha(Q^\alpha-\tilde Q^\alpha)
-\frac12g_{\mu\nu}T_{\rho\alpha\beta}Q^{\alpha\beta\rho}
+\frac12g_{\mu\nu}T_\alpha Q^\alpha
-\frac12g_{\mu\nu}T_\alpha\tilde Q^\alpha
\Bigr]\notag\\
&+a_1\Bigl[
\frac12Q_\nu{}^\alpha{}_\beta T^\beta{}_{\mu\alpha}
+\frac12Q_\mu{}^\alpha{}_\beta T^\beta{}_{\nu\alpha}
-\frac12Q_{\alpha\mu\beta}T_\nu{}^{\alpha\beta}
-\frac12Q_{\alpha\nu\beta}T_\mu{}^{\alpha\beta}\notag\\
&
-\frac12T_{\mu\nu}{}^\alpha
 \left(T_\alpha+\frac12Q_\alpha\right)
-\frac12T_{\nu\mu}{}^\alpha
 \left(T_\alpha+\frac12Q_\alpha\right)\notag\\
&
-\frac12\gr{\nabla}_\alpha T_{\mu\nu}{}^\alpha
+\frac12N^\rho{}_{\mu\alpha}T_{\rho\nu}{}^\alpha
+\frac12N^\rho{}_{\nu\alpha}T_{\mu\rho}{}^\alpha
-\frac12N^\alpha{}_{\rho\alpha}T_{\mu\nu}{}^\rho\notag\\
&
-\frac12\gr{\nabla}_\alpha T_{\nu\mu}{}^\alpha
+\frac12N^\rho{}_{\nu\alpha}T_{\rho\mu}{}^\alpha
+\frac12N^\rho{}_{\mu\alpha}T_{\nu\rho}{}^\alpha
-\frac12N^\alpha{}_{\rho\alpha}T_{\nu\mu}{}^\rho
-\frac12g_{\mu\nu}Q_{\rho\alpha\beta}T^{\beta\rho\alpha}
\Bigr]\notag\\
&+a_2\Bigl[
\frac12Q_\mu T_\nu+\frac12Q_\nu T_\mu
+g_{\mu\nu}T^\alpha T_\alpha
+g_{\mu\nu}\gr{\nabla}_\alpha T^\alpha
+g_{\mu\nu}N^\alpha{}_{\rho\alpha}T^\rho
\Bigr]\notag\\
&+a_3\Bigl[
\frac12\gr{\nabla}_\nu T_\mu
-\frac12N^\rho{}_{\mu\nu}T_\rho
+\frac12\gr{\nabla}_\mu T_\nu
-\frac12N^\rho{}_{\nu\mu}T_\rho\notag\\
&
+T_\mu T_\nu
+\frac14Q_\nu T_\mu
+\frac14Q_\mu T_\nu
-\frac12g_{\mu\nu}\tilde Q_\alpha T^\alpha
\Bigr]\notag\\
&+b_1\Bigl[
T^{\alpha\beta}{}_\mu T_{\alpha\beta\nu}
+T^{\alpha\beta}{}_\nu T_{\alpha\beta\mu}
-T_{\mu\alpha\beta}T_\nu{}^{\alpha\beta}
-\frac12g_{\mu\nu}T_{\rho\alpha\beta}T^{\rho\alpha\beta}
\Bigr]\notag\\
&+\frac{b_2}{2}\Bigl[
T^\alpha{}_{\beta\nu}T^\beta{}_{\mu\alpha}
+T^\alpha{}_{\beta\mu}T^\beta{}_{\nu\alpha}
-g_{\mu\nu}T_{\rho\alpha\beta}T^{\alpha\beta\rho}
\Bigr]\notag\\
&+b_3\Bigl[T_\mu T_\nu-\frac12g_{\mu\nu}T_\alpha T^\alpha\Bigr]\notag\\
&+c_1\Bigl[
2\gr{\nabla}_\alpha Q^\alpha{}_{\mu\nu}
+2N^\alpha{}_{\rho\alpha}Q^\rho{}_{\mu\nu}
-2N^\rho{}_{\mu\alpha}Q^\alpha{}_{\rho\nu}
-2N^\rho{}_{\nu\alpha}Q^\alpha{}_{\mu\rho}\notag\\
&
+2\left(T_\alpha+\frac12Q_\alpha\right)Q^\alpha{}_{\mu\nu}
-Q^{\alpha\beta}{}_\nu Q_{\alpha\beta\mu}
-Q^{\alpha\beta}{}_\mu Q_{\alpha\beta\nu}
+Q_{\mu\alpha\beta}Q_\nu{}^{\alpha\beta}
-\frac12g_{\mu\nu}Q_{\rho\alpha\beta}Q^{\rho\alpha\beta}
\Bigr]\notag\\
&+c_2\Bigl[
\gr{\nabla}_\alpha
  \left(Q_\mu{}^\alpha{}_\nu+Q_\nu{}^\alpha{}_\mu\right)
-N^\rho{}_{\mu\alpha}Q_\rho{}^\alpha{}_\nu
+N^\alpha{}_{\rho\alpha}Q_\mu{}^\rho{}_\nu
-N^\rho{}_{\nu\alpha}Q_\mu{}^\alpha{}_\rho\notag\\
&
-N^\rho{}_{\nu\alpha}Q_\rho{}^\alpha{}_\mu
+N^\alpha{}_{\rho\alpha}Q_\nu{}^\rho{}_\mu
-N^\rho{}_{\mu\alpha}Q_\nu{}^\alpha{}_\rho\notag\\
&
+\left(T_\alpha+\frac12Q_\alpha\right)
 \left(Q_\mu{}^\alpha{}_\nu+Q_\nu{}^\alpha{}_\mu\right)
-\frac12\left(
Q^{\alpha\beta}{}_\mu Q_{\beta\alpha\nu}
+Q^{\alpha\beta}{}_\nu Q_{\beta\alpha\mu}\right)
-\frac12g_{\mu\nu}Q_{\sigma\alpha\beta}Q^{\beta\alpha\sigma}
\Bigr]\notag\\
&+c_3\Bigl[
Q_\mu Q_\nu
+2g_{\mu\nu}\gr{\nabla}_\rho Q^\rho
+2g_{\mu\nu}N^\rho{}_{\lambda\rho}Q^\lambda
+2g_{\mu\nu}T_\rho Q^\rho
+\frac12g_{\mu\nu}Q_\alpha Q^\alpha
\Bigr]\notag\\
&+c_4\Bigl[
\gr{\nabla}_\mu\tilde Q_\nu
+\gr{\nabla}_\nu\tilde Q_\mu
-N^\rho{}_{\nu\mu}\tilde Q_\rho
-N^\rho{}_{\mu\nu}\tilde Q_\rho\notag\\
&
+T_\mu\tilde Q_\nu+T_\nu\tilde Q_\mu
+\frac12Q_\mu\tilde Q_\nu
+\frac12Q_\nu\tilde Q_\mu
-\tilde Q_\mu\tilde Q_\nu
-\frac12g_{\mu\nu}\tilde Q_\alpha\tilde Q^\alpha
\Bigr]\notag\\
&+\frac{c_5}{2}\Bigl[
\gr{\nabla}_\mu Q_\nu+\gr{\nabla}_\nu Q_\mu
-N^\rho{}_{\nu\mu}Q_\rho
-N^\rho{}_{\mu\nu}Q_\rho\notag\\
&\qquad
+T_\mu Q_\nu+T_\nu Q_\mu+Q_\mu Q_\nu
+2g_{\mu\nu}\gr{\nabla}_\alpha\tilde Q^\alpha
+2g_{\mu\nu}N^\alpha{}_{\rho\alpha}\tilde Q^\rho
+2g_{\mu\nu}T_\alpha\tilde Q^\alpha
\Bigr].
\end{align}
Collecting the common terms, we obtain:
\begin{align}
\mathcal M_{\mu\nu}
={}&a_0\Bigl[
\gr{R}_{\mu\nu}
+\frac12\gr{\nabla}_\alpha
\bigl(N^\alpha{}_{\mu\nu}+N^\alpha{}_{\nu\mu}\bigr)
-\frac12\gr{\nabla}_\nu N^\alpha{}_{\mu\alpha}
-\frac12\gr{\nabla}_\mu N^\alpha{}_{\nu\alpha}\notag\\
&\qquad
+\frac12N^\alpha{}_{\sigma\alpha}
\bigl(N^\sigma{}_{\mu\nu}+N^\sigma{}_{\nu\mu}\bigr)
-\frac12N^\alpha{}_{\sigma\nu}N^\sigma{}_{\mu\alpha}
-\frac12N^\alpha{}_{\sigma\mu}N^\sigma{}_{\nu\alpha}
-\frac12g_{\mu\nu}\gr{R}
\Bigr]\notag\\
&+a_1\Bigl[
\frac12Q_\nu{}^\alpha{}_\beta T^\beta{}_{\mu\alpha}
+\frac12Q_\mu{}^\alpha{}_\beta T^\beta{}_{\nu\alpha}
-\frac12Q_{\alpha\mu\beta}T_\nu{}^{\alpha\beta}
-\frac12Q_{\alpha\nu\beta}T_\mu{}^{\alpha\beta}\notag\\
&\qquad
-\frac12T_{\mu\nu}{}^\alpha
 \left(T_\alpha+\frac12Q_\alpha\right)
-\frac12T_{\nu\mu}{}^\alpha
 \left(T_\alpha+\frac12Q_\alpha\right)\notag\\
&\qquad
-\frac12\gr{\nabla}_\alpha T_{\mu\nu}{}^\alpha
+\frac12N^\rho{}_{\mu\alpha}T_{\rho\nu}{}^\alpha
+\frac12N^\rho{}_{\nu\alpha}T_{\mu\rho}{}^\alpha
-\frac12N^\alpha{}_{\rho\alpha}T_{\mu\nu}{}^\rho\notag\\
&\qquad
-\frac12\gr{\nabla}_\alpha T_{\nu\mu}{}^\alpha
+\frac12N^\rho{}_{\nu\alpha}T_{\rho\mu}{}^\alpha
+\frac12N^\rho{}_{\mu\alpha}T_{\nu\rho}{}^\alpha
-\frac12N^\alpha{}_{\rho\alpha}T_{\nu\mu}{}^\rho
\Bigr]\notag\\
&+a_2g_{\mu\nu}N^\alpha{}_{\rho\alpha}T^\rho\notag\\
&+a_3\Bigl[
\frac12\gr{\nabla}_\nu T_\mu
-\frac12N^\rho{}_{\mu\nu}T_\rho
+\frac12\gr{\nabla}_\mu T_\nu
-\frac12N^\rho{}_{\nu\mu}T_\rho
\Bigr]\notag\\
&+b_1\Bigl[
T^{\alpha\beta}{}_\mu T_{\alpha\beta\nu}
+T^{\alpha\beta}{}_\nu T_{\alpha\beta\mu}
-T_{\mu\alpha\beta}T_\nu{}^{\alpha\beta}
\Bigr]\notag\\
&+\frac{b_2}{2}\Bigl[
T^\alpha{}_{\beta\nu}T^\beta{}_{\mu\alpha}
+T^\alpha{}_{\beta\mu}T^\beta{}_{\nu\alpha}
\Bigr]\notag\\
&+c_1\Bigl[
2\gr{\nabla}_\alpha Q^\alpha{}_{\mu\nu}
+2N^\alpha{}_{\rho\alpha}Q^\rho{}_{\mu\nu}
-2N^\rho{}_{\mu\alpha}Q^\alpha{}_{\rho\nu}
-2N^\rho{}_{\nu\alpha}Q^\alpha{}_{\mu\rho}\notag\\
&\qquad
+2\left(T_\alpha+\frac12Q_\alpha\right)Q^\alpha{}_{\mu\nu}
-Q^{\alpha\beta}{}_\nu Q_{\alpha\beta\mu}
-Q^{\alpha\beta}{}_\mu Q_{\alpha\beta\nu}
+Q_{\mu\alpha\beta}Q_\nu{}^{\alpha\beta}
\Bigr]\notag\\
&+c_2\Bigl[
\gr{\nabla}_\alpha
  \left(Q_\mu{}^\alpha{}_\nu+Q_\nu{}^\alpha{}_\mu\right)
-N^\rho{}_{\mu\alpha}Q_\rho{}^\alpha{}_\nu
+N^\alpha{}_{\rho\alpha}Q_\mu{}^\rho{}_\nu
-N^\rho{}_{\nu\alpha}Q_\mu{}^\alpha{}_\rho\notag\\
&\qquad
-N^\rho{}_{\nu\alpha}Q_\rho{}^\alpha{}_\mu
+N^\alpha{}_{\rho\alpha}Q_\nu{}^\rho{}_\mu
-N^\rho{}_{\mu\alpha}Q_\nu{}^\alpha{}_\rho\notag\\
&\qquad
+\left(T_\alpha+\frac12Q_\alpha\right)
 \left(Q_\mu{}^\alpha{}_\nu+Q_\nu{}^\alpha{}_\mu\right)
-\frac12\left(
Q^{\alpha\beta}{}_\mu Q_{\beta\alpha\nu}
+Q^{\alpha\beta}{}_\nu Q_{\beta\alpha\mu}\right)
\Bigr]\notag\\
&+2c_3g_{\mu\nu}N^\rho{}_{\lambda\rho}Q^\lambda\notag\\
&+c_4\Bigl[
\gr{\nabla}_\mu\tilde Q_\nu
+\gr{\nabla}_\nu\tilde Q_\mu
-N^\rho{}_{\nu\mu}\tilde Q_\rho
-N^\rho{}_{\mu\nu}\tilde Q_\rho\notag\\
&\qquad
+T_\mu\tilde Q_\nu+T_\nu\tilde Q_\mu
+\frac12Q_\mu\tilde Q_\nu
+\frac12Q_\nu\tilde Q_\mu
-\tilde Q_\mu\tilde Q_\nu
\Bigr]\notag\\
&+\frac{c_5}{2}\Bigl[
\gr{\nabla}_\mu Q_\nu+\gr{\nabla}_\nu Q_\mu
-N^\rho{}_{\nu\mu}Q_\rho
-N^\rho{}_{\mu\nu}Q_\rho\notag\\
&\qquad
+T_\mu Q_\nu+T_\nu Q_\mu
+2g_{\mu\nu}N^\alpha{}_{\rho\alpha}\tilde Q^\rho
\Bigr]\notag\\
&+(a_3+b_3)T_\mu T_\nu\notag\\
&+\left(\frac{a_2}{2}+\frac{a_3}{4}\right)
\left(Q_\mu T_\nu+Q_\nu T_\mu\right)
+\left(c_3+\frac{c_5}{2}\right)Q_\mu Q_\nu\notag\\
&+g_{\mu\nu}\Biggl[
\left(-\frac{a_0}{8}-\frac{b_1}{2}\right)
T_{\rho\alpha\beta}T^{\rho\alpha\beta}
+\left(\frac{a_0}{4}-\frac{b_2}{2}\right)
T_{\rho\alpha\beta}T^{\alpha\beta\rho}
+\left(a_2+\frac{a_0-b_3}{2}\right)T_\alpha T^\alpha\notag\\
&\qquad
-\left(\frac{a_0}{8}+\frac{c_1}{2}\right)
Q_{\rho\alpha\beta}Q^{\rho\alpha\beta}
+\left(\frac{a_0}{4}-\frac{c_2}{2}\right)
Q_{\sigma\alpha\beta}Q^{\beta\alpha\sigma}
+\left(\frac{a_0}{8}+\frac{c_3}{2}\right)Q_\alpha Q^\alpha
-\frac{c_4}{2}\tilde Q_\alpha\tilde Q^\alpha
-\frac{a_0}{4}Q_\alpha\tilde Q^\alpha\notag\\
&\qquad
-\frac{a_0+a_1}{2}T_{\rho\alpha\beta}Q^{\alpha\beta\rho}
+\left(\frac{a_0}{2}+2c_3\right)T_\alpha Q^\alpha
+\left(c_5-\frac{a_0+a_3}{2}\right)T_\alpha\tilde Q^\alpha\notag\\
&\qquad
+\left(a_2-a_0\right)\gr{\nabla}_\alpha T^\alpha
+\left(2c_3-\frac{a_0}{2}\right)\gr{\nabla}_\alpha Q^\alpha
+\left(c_5+\frac{a_0}{2}\right)
\gr{\nabla}_\alpha\tilde Q^\alpha
\Biggr], 
\end{align}
Expanding the disformation tensor with the definition given in Eq.\ \eqref{eq: disformation}, we infer the full metric field equations \eqref{eq: metric field equations} in Sec.\ \ref{sec:quadratic-MAG},
\begin{align}
\mathcal M_{\mu\nu}
={}&a_0\gr{R}_{\mu\nu}
-\frac12g_{\mu\nu}\Bigl[
a_0\bigl(
\gr{R}+T-Q
+T_{\rho\alpha\beta}Q^{\alpha\beta\rho}
+T_\alpha(\tilde Q^\alpha-Q^\alpha)
\bigr)
+L_T+L_Q+L_{TQ}
\Bigr]\notag\\
&+\gr{\nabla}_\alpha\Bigl[
(a_0+a_1)K^\alpha{}_{(\mu\nu)}
+(a_0-2c_2)L^\alpha{}_{\mu\nu}
+(2c_1+c_2)Q^\alpha{}_{\mu\nu}
\Bigr]\notag\\
&+(a_0+a_3)\gr{\nabla}_{(\mu}T_{\nu)}
+\left(\frac{a_0}{2}+c_5\right)\gr{\nabla}_{(\mu}Q_{\nu)}
+2c_4\gr{\nabla}_{(\mu}\tilde Q_{\nu)}\notag\\
&-\left[
(a_0+a_3)T_\rho
+\left(\frac{a_0}{2}+c_5\right)Q_\rho
+2c_4\tilde Q_\rho
\right]
\bigl(K^\rho{}_{(\mu\nu)}+L^\rho{}_{\mu\nu}\bigr)\notag\\
&-\bigl(K^\rho{}_{\mu\alpha}+L^\rho{}_{\mu\alpha}\bigr)
\left[
\frac{a_0}{2}K^\alpha{}_{\rho\nu}
+a_1K^\alpha{}_{(\rho\nu)}
+(2c_1+c_2)Q^\alpha{}_{\rho\nu}
+\left(\frac{a_0}{2}-2c_2\right)L^\alpha{}_{\rho\nu}
\right]\notag\\
&-\bigl(K^\rho{}_{\nu\alpha}+L^\rho{}_{\nu\alpha}\bigr)
\left[
\frac{a_0}{2}K^\alpha{}_{\rho\mu}
+a_1K^\alpha{}_{(\rho\mu)}
+(2c_1+c_2)Q^\alpha{}_{\rho\mu}
+\left(\frac{a_0}{2}-2c_2\right)L^\alpha{}_{\rho\mu}
\right]\notag\\
&+a_1\left(
Q_{(\mu}{}^\alpha{}_\beta T^\beta{}_{\nu)\alpha}
-Q_{\alpha(\mu|\beta|}T_{\nu)}{}^{\alpha\beta}
\right)\notag\\
&+b_1\left(
2T^{\alpha\beta}{}_{(\mu}T_{\alpha\beta\nu)}
-T_{\mu\alpha\beta}T_\nu{}^{\alpha\beta}
\right)
+b_2T^\alpha{}_{\beta(\mu}T^\beta{}_{\nu)\alpha}\notag\\
&-2c_1Q^{\alpha\beta}{}_{(\mu}Q_{\alpha\beta\nu)}
-c_2Q^{\alpha\beta}{}_{(\mu}Q_{\beta\alpha\nu)}
+c_1Q_{\mu\alpha\beta}Q_\nu{}^{\alpha\beta}\notag\\
&+c_4\left(
2T_{(\mu}\tilde Q_{\nu)}
+Q_{(\mu}\tilde Q_{\nu)}
-\tilde Q_\mu\tilde Q_\nu
\right)
+(a_3+b_3)T_\mu T_\nu\notag\\
&+\left(a_2+\frac{a_3}{2}+c_5\right)Q_{(\mu}T_{\nu)}
+\left(c_3+\frac{c_5}{2}\right)Q_\mu Q_\nu\notag\\
&+g_{\mu\nu}\left[
\left(\frac{a_2-a_0}{2}\right)B_T
+(2c_3+c_5)\gr{\nabla}_\alpha Q^\alpha
-\left(\frac{a_0}{2}+c_5\right)B_Q
\right].
\end{align}

\section{Solving for the affine connection}
Here, we carefully explain the method to solve the connection field equations, using also the invariance under the projective symmetry. We start solving the connection field equations of Section \ref{sub:axial mode}. We proceed explaining the solution of Section \ref{sub: metric-trace}, and we conclude solving the connection field equations of Section \ref{sub:full vacuum solution}.
\subsection{Axial projective sector}\label{appendix: axial proj}
Here, we solve the block connection field equations \eqref{eq:axial-block-operator}.\\
The nonmetricity block of Eq.\ \eqref{eq:axial-block-operator} is
\begin{equation}\label{eq:axial-nonmetricity-block}
    \begin{aligned}
    M_{Q\alpha}{}^{\mu\nu\rho\lambda\kappa} Q_{\rho\lambda\kappa}
=&
\Big\{[4(b_1+b_2)-2c_2]
\delta^\rho_\alpha g^{\mu\lambda}g^{\nu\kappa}
+(a_1-4c_1)
g^{\nu\rho}g^{\mu\lambda}\delta^\kappa_\alpha
\\&-(2c_2+a_1)
g^{\mu\rho}g^{\nu\lambda}\delta^\kappa_\alpha
-
[2(b_1+b_2)+c_5]
g^{\mu\nu}\delta^\rho_\alpha g^{\lambda\kappa}
\\
&+
[2(b_1+b_2)-c_5-a_2]
\delta^\nu_\alpha g^{\mu\rho}g^{\lambda\kappa}
+(a_2-4c_3)
\delta^\mu_\alpha g^{\nu\rho}g^{\lambda\kappa}
\\&-2c_4 g^{\mu\nu}g^{\rho\kappa}\delta^\lambda_\alpha-[4(b_1+b_2)+2c_4+a_3]
\delta^\nu_\alpha g^{\mu\lambda}g^{\rho\kappa}
\\&+(a_3-2c_5)
\delta^\mu_\alpha g^{\nu\lambda}g^{\rho\kappa}\Big\}Q_{\rho\lambda\kappa}.
    \end{aligned}
\end{equation}
Then, the block associated to the torsion vector $T_\rho$ is:
\begin{equation}\label{eq:axial-torsion-vector-block}
    \begin{aligned}
        M_{T\alpha}{}^{\mu\nu\rho}T_\rho =&\Big\{ [
-\frac{8}{3}(b_1+b_2)
+\frac{1}{3}a_1-a_3]
\delta^\rho_\alpha g^{\mu\nu}
\\&+[\frac{4}{3}b_1+\frac{10}{3}b_2-2b_3+\frac13a_1-a_3]g^{\mu\rho}\delta^\nu_\alpha\\&+[\frac43 b_1-\frac{2}{3} b_2+2b_3-\frac23a_1-2a_2]
g^{\nu\rho}\delta^\mu_\alpha \Big\} T_\rho,
    \end{aligned}
\end{equation}
Finally, the tensor-torsion block $t_{\rho\lambda\kappa}$ is:
\begin{equation}
    \begin{aligned}
         M_{t\alpha}{}^{\mu\nu\rho\lambda\kappa}
t_{\rho\lambda\kappa} =&[(-4b_1-6b_2)
g^{\nu\rho}g^{\mu\lambda}\delta^\kappa_\alpha
+(4b_1-4b_2-a_1)
\delta^\rho_\alpha g^{\nu\lambda}g^{\mu\kappa}
\\&+
(2b_2-a_1)
g^{\mu\rho}g^{\nu\lambda}\delta^\kappa_\alpha]t_{\rho\lambda\kappa}, 
    \end{aligned}
\end{equation}
 We have to solve the complete system simultaneously.
In four dimensions, $Q_{\rho\lambda\kappa}$ has forty independent
components, $T_\rho$ has four components, and
$t_{\rho\lambda\kappa}$ has $16$ components. Hence the combined
non-axial variable has $60$ components, but four of them are identically dependent by
virtue of \eqref{eq:axial-Noether-identity}.
It is worth noticing that, although the nonmetricity traces no longer appear explicitly in this
representation \eqref{eq:axial-nonmetricity-block}, they have not been discarded. They are contained in
the contractions of $Q_{\rho\lambda\kappa}$ with the metric tensors and Kronecker deltas entering $M_Q$. 
Similarly, the totally symmetric and mixed-symmetry parts are
contained in the different permutations of
$Q_{\rho\lambda\kappa}$ appearing in $M_Q Q$. In particular,
\begin{equation}
Q_{(\alpha\mu\nu)}
=
\frac{1}{3}
\left(
Q_{\alpha\mu\nu}
+
Q_{\mu\nu\alpha}
+
Q_{\nu\alpha\mu}
\right).
\end{equation}
Consequently, $Q_\alpha$, $\tilde Q_\alpha$, and
$Q_{(\alpha\mu\nu)}$ need not be introduced as additional unknown
fields. They are linear projections of the single nonmetricity tensor $Q_{\rho\lambda\kappa}$.
The block decomposition \eqref{eq:axial-block-operator}
can therefore be treated directly as one homogeneous linear system.
After taking account of the symmetry
$Q_{\rho\lambda\kappa}=Q_{\rho\kappa\lambda}$, the nonmetricity block
acts on forty independent components. Together with the four
components of $T_\rho$ and the $16$  components of
$t_{\rho\lambda\kappa}$, the combined operator acts on sixty
non-axial connection components.
The axial Noether identity
\begin{equation}
\epsilon^{\sigma\alpha}{}_{\mu\nu}
\mathcal P_\alpha{}^{\mu\nu}
\equiv0
\end{equation}
provides four dependent equations. Thus, after removing these four
identities, the connection equation defines a linear map between two
sixty-dimensional spaces. If the combined reduced operator has rank
sixty, then its kernel is trivial and $Q_{\rho\lambda\kappa}=0$, $T_\rho=0$, and $t_{\rho\lambda\kappa}$=0.
If the rank is smaller than sixty, then the kernel contains
additional non-axial connection configurations. In that case,
determining which components remain undetermined requires decomposing the kernel into its trace, totally symmetric, and mixed-symmetry parts.
In order to  determine its kernel \cite{Guzman:2020kgh,Ferrara:2026hbq, Iosifidis:2021tvx},
we first introduce
\begin{equation}
\mathcal P_{\alpha\mu\nu}
=
g_{\mu\rho}g_{\nu\sigma}
\mathcal P_\alpha{}^{\rho\sigma}
\end{equation}
and take the three independent contractions
\begin{equation}
g^{\mu\nu}\mathcal P_{\alpha\mu\nu}=0,
\qquad
g^{\alpha\mu}\mathcal P_{\alpha\mu\nu}=0,
\qquad
g^{\alpha\nu}\mathcal P_{\alpha\mu\nu}=0.
\end{equation}
These contractions give the coupled vector system
\begingroup
\scriptsize
\begin{equation}
\begin{aligned}
&\left(
\begin{array}{cc}
-2(b_1+b_2)-2c_2-4c_3-5c_5
&
-4(b_1+b_2)-4c_1-2c_2-10c_4-2c_5
\\
a_1+3a_2-4c_1-16c_3-2c_5
&
-a_1+3a_3-4c_2-4c_4-8c_5
\\
6(b_1+b_2)-a_1-3a_2-2c_2-4c_3-5c_5
&
-12(b_1+b_2)+a_1-3a_3-4c_1-2c_2-10c_4-2c_5
\end{array}
\right.
\\[1ex]
&\hspace{7cm}\left.
\begin{array}{c}
-8(b_1+b_2)+a_1-2a_2-5a_3\\
4b_1-2b_2+6b_3-2a_1-8a_2-2a_3\\
4b_1+10b_2-6b_3+a_1-2a_2-5a_3
\end{array}
\right)
\left(
\begin{array}{c}
Q_\alpha\\
\tilde Q_\alpha\\
T_\alpha
\end{array}
\right)
=0.
\end{aligned}
\label{eq:axial-vector-block}
\end{equation}
\endgroup
If the determinant of the matrix in
\eqref{eq:axial-vector-block} is nonzero, we have
\begin{equation}
Q_\alpha=0,
\qquad
\tilde Q_\alpha=0,
\qquad
T_\alpha=0.
\label{eq:axial-vector-solution}
\end{equation}
Once the vector traces vanish, the totally symmetric projection of
the connection equation gives
\begin{equation}
4\left(b_1+b_2-c_1-c_2\right)
Q_{(\alpha\mu\nu)}=0.
\label{eq:axial-symmetric-nonmetricity-projection}
\end{equation}
Therefore,
\begin{equation}
Q_{(\alpha\mu\nu)}=0
\end{equation}
provided that
\begin{equation}
b_1+b_2-c_1-c_2\neq0.
\label{eq:axial-symmetric-nonmetricity-condition}
\end{equation}
After removing its traces and its totally symmetric part, the
remaining nonmetricity has mixed symmetry and satisfies
\begin{equation}
Q_{\alpha\mu\nu}
+
Q_{\mu\nu\alpha}
+
Q_{\nu\alpha\mu}=0.
\label{eq:hook-nonmetricity-cyclic-identity}
\end{equation}
The irreducible tensor torsion satisfies
\begin{equation}
t_{\alpha\mu\nu}
+
t_{\mu\nu\alpha}
+
t_{\nu\alpha\mu}=0.
\label{eq:tensor-torsion-cyclic-identity}
\end{equation}
The symmetric and antisymmetric projections of the remaining
connection equation can be written as
\begin{equation}
\left(
\begin{array}{cc}
4(b_1+b_2)+2c_1-c_2
&
-4(b_1+b_2)-a_1
\\[1mm]
\frac{3}{2}(a_1-2c_1+c_2)
&
-6b_1+\frac{3}{2}a_1
\end{array}
\right)
\left(
\begin{array}{c}
Q_{\alpha\mu\nu}
\\[1mm]
\displaystyle
\frac{1}{2}
\left(
t_{\alpha\mu\nu}+2t_{\mu\nu\alpha}
\right)
\end{array}
\right)
=0.
\label{eq:axial-mixed-symmetry-block}
\end{equation}
In \eqref{eq:axial-mixed-symmetry-block},
$Q_{\alpha\mu\nu}$ denotes the traceless mixed-symmetry part remaining
after the previous projections. The second row is obtained from the
antisymmetric projection after rewriting it with the same mixed
symmetry as the first row. This cyclic rearrangement is invertible on
the sixteen-dimensional tensor-torsion sector.

The mixed-symmetry block has a trivial kernel when
\begin{equation}
\left[4(b_1+b_2)+2c_1-c_2\right]
(-6b_1+\frac{3}{2}a_1)
+\frac{3}{2}\left[4(b_1+b_2)+a_1\right]
\left(a_1-2c_1+c_2\right)
\neq0.
\label{eq:axial-mixed-symmetry-condition}
\end{equation}
Under this condition,
\begin{equation}
Q_{\alpha\mu\nu}=0,
\qquad
t_{\alpha\mu\nu}=0
\end{equation}
in the mixed-symmetry sector.
Consequently, the reduced non-axial connection operator has maximal rank sixty if the determinant of the matrix in
\eqref{eq:axial-vector-block} is nonzero and the conditions
\eqref{eq:axial-symmetric-nonmetricity-condition} and
\eqref{eq:axial-mixed-symmetry-condition} hold. 
\subsection{Metric-trace projective sector}\label{appendix: metric-trace}
In this Section, we solve the connection field equations \eqref{eq:r-combined-block-equation} of Section \ref{sub: metric-trace}
The nonmetricity column is
\begin{equation}
\begin{aligned}
M_{Q\alpha}{}^{\mu\nu\rho\lambda\kappa}
Q_{\rho\lambda\kappa}
=
\Big[&
\left(
a_0-2c_2
\right)
\delta^\rho_\alpha
g^{\mu\lambda}g^{\nu\kappa}
\\
&+
\left(
3a_0+2a_2+5a_3
+2c_2+10c_4+2c_5
\right)
g^{\nu\rho}g^{\mu\lambda}
\delta^\kappa_\alpha
\\
&-
\left(
2a_0+2a_2+5a_3+2c_2
\right)
g^{\mu\rho}g^{\nu\lambda}
\delta^\kappa_\alpha
\\
&-
\left(
\tfrac{1}{2}a_0+c_5
\right)
g^{\mu\nu}
\delta^\rho_\alpha g^{\lambda\kappa}
\\
&+
\left(
\tfrac{1}{2}a_0-c_5-a_2
\right)
\delta^\nu_\alpha
g^{\mu\rho}g^{\lambda\kappa}
\\
&+
\left(
\tfrac{1}{2}a_0+2c_2+5c_5+a_2
\right)
\delta^\mu_\alpha
g^{\nu\rho}g^{\lambda\kappa}
\\
&-
2c_4g^{\mu\nu}
g^{\rho\kappa}\delta^\lambda_\alpha
\\
&-
\left(
a_0+2c_4+a_3
\right)
\delta^\nu_\alpha
g^{\mu\lambda}g^{\rho\kappa}
\\
&+
\left(
a_3-2c_5
\right)
\delta^\mu_\alpha
g^{\nu\lambda}g^{\rho\kappa}
\Big]
Q_{\rho\lambda\kappa}.
\end{aligned}
\label{eq:r-nonmetricity-block}
\end{equation}
The nonmetricity traces are already contained in this expression
through contractions of the complete tensor. For example,
\begin{equation}
\delta^\rho_\alpha g^{\lambda\kappa}
Q_{\rho\lambda\kappa}
=
Q_\alpha,
\qquad
g^{\rho\kappa}\delta^\lambda_\alpha
Q_{\rho\lambda\kappa}
=
\tilde Q_\alpha.
\end{equation}
Likewise, the totally symmetric and mixed-symmetry parts of
nonmetricity are contained in the different permutations of
$Q_{\rho\lambda\kappa}$. 
The torsion-vector column is
\begin{equation}
\begin{aligned}
M_{T\alpha}{}^{\mu\nu\rho}T_\rho
=
\Bigg[&
\frac{2}{3}
\left(
a_2+a_3
\right)
\delta^\rho_\alpha g^{\mu\nu}
\\
&+
\left(
\frac{4}{3}a_0
-\frac{4}{3}b_1
+\frac{2}{3}b_2
-2b_3
+\frac{2}{3}a_2
+\frac{2}{3}a_3
\right)
g^{\mu\rho}\delta^\nu_\alpha
\\
&+
\left(
-\frac{4}{3}a_0
+\frac{4}{3}b_1
-\frac{2}{3}b_2
+2b_3
-\frac{10}{3}a_2
-\frac{10}{3}a_3
\right)
g^{\nu\rho}\delta^\mu_\alpha
\Bigg]
T_\rho.
\end{aligned}
\label{eq:r-torsion-vector-block}
\end{equation}

The axial-torsion column is
\begin{equation}
M_{S\alpha}{}^{\mu\nu\rho}S_\rho
=
\frac{1}{6}
\left(
a_0-4b_1-4b_2
\right)
g^{\beta\mu}
\epsilon^\nu{}_{\sigma\alpha\beta}
S^\sigma.
\label{eq:r-axial-torsion-block}
\end{equation}
Finally, the irreducible tensor-torsion column is
\begin{equation}
\begin{aligned}
M_{t\alpha}{}^{\mu\nu\rho\lambda\kappa}
t_{\rho\lambda\kappa}
=
\Bigg[&
\left(
-a_0+2b_2
\right)
g^{\nu\rho}g^{\mu\lambda}
\delta^\kappa_\alpha
\\
&+
\left(
4b_1-2a_0-2a_2-5a_3
\right)
\delta^\rho_\alpha
g^{\nu\lambda}g^{\mu\kappa}
\\
&-
\left(
2b_2+2a_0+2a_2+5a_3
\right)
g^{\mu\rho}g^{\nu\lambda}
\delta^\kappa_\alpha
\Bigg]
t_{\rho\lambda\kappa}.
\end{aligned}
\label{eq:r-tensor-torsion-block}
\end{equation}
Since $r^\alpha$ is arbitrary, the connection Euler-Lagrange tensor
satisfies the off-shell Noether identity
\begin{equation}
g_{\mu\nu}\mathcal P_\alpha{}^{\mu\nu}
\equiv0.
\label{eq:r-Noether-identity}
\end{equation}
Thus, four components of the connection equation vanish identically.
The corresponding right null direction of the combined connection
operator lies entirely inside the nonmetricity column and is given by Eq.\ \eqref{eq: r-nonmetr}.
Substituting the traces of the transformed nonmetricity \eqref{eq: transfo traces Q} and $\delta_r Q_{\alpha\mu\nu}$ into the complete nonmetricity block yields
\begin{equation}
M_{Q\alpha}{}^{\mu\nu\rho\lambda\kappa}
Q^{(r)}_{\rho\lambda\kappa}
=0.
\label{eq:r-right-kernel}
\end{equation}
This identity results from a cancellation among all the terms in
\eqref{eq:r-nonmetricity-block}; it does not require their individual coefficients to vanish.
Now, we can project the connection field equations, using the Noether identities \cite{Iosifidis:2021tvx}.
The axial projection of the connection equation gives
\begin{equation}
\epsilon^{\sigma\alpha}{}_{\mu\nu}
\mathcal P_\alpha{}^{\mu\nu}=0
\quad\Longrightarrow\quad
\left(
a_0-4b_1-4b_2
\right)S^\sigma=0.
\label{eq:r-axial-equation}
\end{equation}
Therefore, for generic parameters satisfying
\begin{equation}
a_0-4b_1-4b_2\neq0,
\label{eq:r-axial-nondegeneracy}
\end{equation}
the axial torsion vanishes\footnote{If $a_0-4b_1-4b_2=0$, the theory also possesses the axial projective symmetry considered in the preceding subsection, and
$S^\sigma$ becomes an additional gauge mode.}:
\begin{equation}
S^\sigma=0.
\label{eq:r-axial-solution}
\end{equation}
The metric contraction of the connection equation does not give an
independent vector equation, because it is precisely the Noether
identity \eqref{eq:r-Noether-identity}. Two independent vector
equations can instead be obtained from
\begin{equation}
\mathcal P_\alpha{}^{\alpha\nu}=0,
\qquad
\mathcal P_\alpha{}^{\nu\alpha}=0.
\end{equation}
They take the form
\begin{equation}
\begin{aligned}
\left(
4b_1-2b_2+6b_3
-4a_0-12a_2-12a_3
\right)T^\nu
+
\left(
-2a_0-2a_2-2a_3
-4c_2-4c_4-8c_5
\right)
\left(
\tilde Q^\nu-\frac{5}{2}Q^\nu
\right)&=0,
\\
\left(
4a_0-4b_1+2b_2-6b_3
\right)T^\nu+
2
\left(
a_2+a_3
\right)
\left(
\tilde Q^\nu-\frac{5}{2}Q^\nu
\right)&=0.
\end{aligned}
\label{eq:r-vector-system}
\end{equation}
The combination
\begin{equation}
\tilde Q^\nu-\frac{5}{2}Q^\nu
\end{equation}
is invariant under the transformation, as shown in Eq.\ \eqref{eq:metric-trace-invariant-nonmetricity-vector}.
The orthogonal vector combination corresponds to the projective gauge
direction and cannot be determined by the connection equation.

The system \eqref{eq:r-vector-system} is nondegenerate when
\begin{equation}
\begin{aligned}
&2
\left(
a_2+a_3
\right)
\left(
4b_1-2b_2+6b_3
-4a_0-12a_2-12a_3
\right)
\\
&\hspace{3cm}-
\left(
-2a_0-2a_2-2a_3
-4c_2-4c_4-8c_5
\right)
\left(
4a_0-4b_1+2b_2-6b_3
\right)\neq 0.
\end{aligned}
\label{eq:r-vector-nondegeneracy}
\end{equation}
For parameters satisfying this condition, the unique solution in the
non-gauge vector subspace is
\begin{equation}
T^\nu=0,
\qquad
\tilde Q^\nu-\frac{5}{2}Q^\nu=0.
\label{eq:r-vector-solution}
\end{equation}
Hence
\begin{equation}
\tilde Q^\nu=\frac{5}{2}Q^\nu.
\label{eq:r-trace-relation}
\end{equation}
The remaining vector $Q^\nu$ is not determined because it
parametrizes the projective kernel.
Now, we use the irreducible decomposition of the nonmetricity tensor \eqref{eq:irreducible Q} given in Sec.\ \ref{sec: Metric_affine}, which is now used only as a projection tool for solving
the single system \eqref{eq:r-combined-block-equation}.
After imposing \eqref{eq:r-vector-solution}, the nonmetricity tensor
can be written as
\begin{equation}
Q_{\alpha\mu\nu}
=
\frac{1}{2}
\left(
Q_\mu g_{\alpha\nu}
+
Q_\nu g_{\alpha\mu}
\right)
+
\Omega_{\alpha\mu\nu}.
\label{eq:r-nonmetricity-after-vector-solution}
\end{equation}
The first term is the projective direction
\eqref{eq: r-nonmetr}, while $\Omega_{\alpha\mu\nu}$ contains the remaining nonmetricity tensor
components. Since the first term is annihilated by the complete
nonmetricity block, the tensorial part of the connection equation
reduces to
\begin{equation}
\begin{aligned}
&\left(
a_0-2c_2
\right)
\Omega_\alpha{}^{\mu\nu}
\\
&+
\left(
3a_0+2a_2+5a_3
+2c_2+10c_4+2c_5
\right)
\Omega^{\nu\mu}{}_\alpha
\\
&-
\left(
2a_0+2a_2+5a_3+2c_2
\right)
\Omega^{\mu\nu}{}_\alpha
\\
&+
\left(
-a_0+2b_2
\right)
t^{\nu\mu}{}_\alpha
\\
&+
\left(
4b_1-2a_0-2a_2-5a_3
\right)
t_\alpha{}^{\nu\mu}
\\
&-
\left(
2b_2+2a_0+2a_2+5a_3
\right)
t^{\mu\nu}{}_\alpha=0.
\end{aligned}
\label{eq:r-combined-tensor-system}
\end{equation}
This remains one coupled equation for the traceless nonmetricity and
tensor-torsion sectors. In particular, the coefficients multiplying
the different index permutations in
\eqref{eq:r-combined-tensor-system} must not be required to vanish
separately.
The totally symmetric projection gives
\begin{equation}
2
\left(
a_0-c_2+5c_4+c_5
\right)
\Omega_{(\alpha\mu\nu)}
=0.
\label{eq:r-symmetric-nonmetricity-equation}
\end{equation}
Thus, if
\begin{equation}
a_0-c_2+5c_4+c_5\neq0,
\label{eq:r-symmetric-nonmetricity-nondegeneracy}
\end{equation}
then
\begin{equation}
\Omega_{(\alpha\mu\nu)}=0.
\end{equation}

The remaining mixed-symmetry part of
$\Omega_{\alpha\mu\nu}$ is coupled to
$t_{\alpha\mu\nu}$ by
\eqref{eq:r-combined-tensor-system}. For generic parameter values,
the restriction of the combined operator
\begin{equation}
\begin{pmatrix}
M_Q & M_t
\end{pmatrix}
\end{equation}
to the doubly traceless nonmetricity and irreducible tensor-torsion
subspace is nonsingular. Under this final nondegeneracy assumption,
\begin{equation}
\Omega_{\alpha\mu\nu}=0,
\qquad
t_{\alpha\mu\nu}=0.
\label{eq:r-tensor-solution}
\end{equation}
If the restricted tensor operator is singular, additional
mixed-symmetry nonmetricity or tensor-torsion modes can remain
undetermined. These exceptional parameter choices constitute
separate degenerate branches of the theory.
\subsection{Fully projectively invariant Lagrangian}\label{appendix: full proj}
Here, we solve the reduced connection field equations \eqref{eq:full-reduced-combined-equation} of Section \ref{sub:full vacuum solution}.\\
The torsion-vector column of Eq.\ \eqref{eq:vanishing-vector-axial-columns} is
\begin{align}
M_{T\alpha}{}^{\mu\nu\rho}T_\rho
={}&
\left(
-\frac23a_0+\frac13a_1-a_3
\right)
T_\alpha g^{\mu\nu}
\nonumber\\
&+
\left(
\frac23a_0+\frac13a_1-a_3
-\frac43b_1+\frac23b_2-2b_3
\right)
T^\mu\delta^\nu_\alpha
\nonumber\\
&+
\left(
-\frac23a_1-2a_2
+\frac43b_1-\frac23b_2+2b_3
\right)
T^\nu\delta^\mu_\alpha.
\label{eq:general-torsion-vector-column}
\end{align}
We rewrite this general expression for the torsion column, the relations among the coefficients $a_1$, $a_2$, $b_1$, and $b_2$, so that the expression of $T_\alpha g^{\mu\nu}$ becomes
\begin{equation}
-\frac23a_0
+\frac13(2a_0+3a_3)
-a_3=0.
\end{equation}
The coefficient of $T^\mu\delta^\nu_\alpha$ becomes
\begin{align}
\frac23a_0
+\frac13(2a_0+3a_3)-a_3
-\frac43\left(\frac34a_0-b_3\right)
+\frac23\left(-\frac12a_0+b_3\right)-2b_3
=0,
\end{align}
and the coefficient of $T^\nu\delta^\mu_\alpha$ becomes
\begin{align}
-\frac23(2a_0+3a_3)-2(-a_3)
+\frac43\left(\frac34a_0-b_3\right)
-\frac23\left(-\frac12a_0+b_3\right)+2b_3
=0.
\end{align}
Thus
\begin{equation}
M_{T\alpha}{}^{\mu\nu\rho}T_\rho\equiv0.
\label{eq:torsion-vector-column-zero}
\end{equation}
This cancellation is required because $T_\mu$ is shifted arbitrarily by the
full projective symmetry.
The axial contribution is proportional to
\begin{equation}
M_{S\alpha}{}^{\mu\nu\rho}S_\rho
=
\frac16
\left(a_0-4b_1-4b_2\right)
g^{\beta\mu}\epsilon^\nu{}_{\sigma\alpha\beta}S^\sigma.
\label{eq:axial-column-before-constraints}
\end{equation}
The full symmetry relations give
\begin{equation}
b_1+b_2=\frac14a_0,
\end{equation}
and hence
\begin{equation}
M_{S\alpha}{}^{\mu\nu\rho}S_\rho\equiv0.
\label{eq:axial-column-zero}
\end{equation}
The nonmetricity
column is
\begin{align}
M_{Q\alpha}{}^{\mu\nu\rho\lambda\kappa}
Q_{\rho\lambda\kappa}
=&
\left(2a_0+2c_4+4c_5\right)Q_\alpha{}^{\mu\nu}+
\left(2a_0+3a_3+8c_4-2c_5\right)
Q^{\nu\mu}{}_\alpha
+\left(-a_0-3a_3+2c_4+4c_5\right)
Q^{\mu\nu}{}_\alpha
\nonumber\\
&-
\left(\frac12a_0+c_5\right)
Q_\alpha g^{\mu\nu}+
\left(\frac12a_0+a_3-c_5\right)
Q^\mu\delta^\nu_\alpha+
\left(-\frac12a_0-a_3-2c_4+c_5\right)
Q^\nu\delta^\mu_\alpha
\nonumber\\
&-2c_4\tilde Q_\alpha g^{\mu\nu}-
\left(a_0+2c_4+a_3\right)
\tilde Q^\mu\delta^\nu_\alpha+
\left(a_3-2c_5\right)
\tilde Q^\nu\delta^\mu_\alpha.
\label{eq:full-nonmetricity-column}
\end{align}
Although the traces occur explicitly in
\eqref{eq:full-nonmetricity-column}, the complete column annihilates the
vector-generated part of nonmetricity.  
In this particular case, in order to analyse the contributions of the vectorial and the tensorial part of the nonmetricity, we use again the nonmetricity decomposition \eqref{eq:irreducible Q}, that here we report for simplicity:
\begin{align}
Q_{\alpha\mu\nu}
&=
\frac{1}{18}
\left(
5Q_\alpha g_{\mu\nu}
-Q_\mu g_{\alpha\nu}
-Q_\nu g_{\alpha\mu}
\right)
+
\frac19
\left(
-\tilde Q_\alpha g_{\mu\nu}
+2\tilde Q_\mu g_{\alpha\nu}
+2\tilde Q_\nu g_{\alpha\mu}
\right)
+\Omega_{\alpha\mu\nu}.
\label{eq:full-irreducible-nonmetricity}
\end{align}
Substituting only the vectorial part of Eq.\ \eqref{eq:full-irreducible-nonmetricity}
\begin{align}
Q^{\mathrm{vector}}_{\alpha\mu\nu}
={}&
\frac1{18}
\left(
5Q_\alpha g_{\mu\nu}
-Q_\mu g_{\alpha\nu}
-Q_\nu g_{\alpha\mu}
\right)
+
\frac19
\left(
-\tilde Q_\alpha g_{\mu\nu}
+2\tilde Q_\mu g_{\alpha\nu}
+2\tilde Q_\nu g_{\alpha\mu}
\right)
\label{eq:vector-generated-nonmetricity}
\end{align}
into \eqref{eq:full-nonmetricity-column} gives
\begin{equation}
M_{Q\alpha}{}^{\mu\nu\rho\lambda\kappa}
Q^{\mathrm{vector}}_{\rho\lambda\kappa}
=0.
\label{eq:nonmetricity-vector-kernel}
\end{equation}
Thus, the nonmetricity block acts only on
$\Omega_{\alpha\mu\nu}$.
Finally, after imposing the full-projective relations, the tensor-torsion column can
be kept in the explicit form
\begin{align}
M_{t\alpha}{}^{\mu\nu\rho\lambda\kappa}
t_{\rho\lambda\kappa}
=&a_0g^{\beta\mu}t^\nu{}_{\alpha\beta}
+4\left(\frac34a_0-b_3\right)t_\alpha{}^{\nu\mu}
+2\left(-\frac12a_0+b_3\right)
\left(t^\mu{}_{\alpha}{}^\nu-t^\nu{}_{\alpha}{}^\mu\right)
\nonumber\\
&-\left(2a_0+3a_3\right)
\left(t_\alpha{}^{\nu\mu}+t^{\mu\nu}{}_\alpha\right).
\label{eq:full-tensor-torsion-column}
\end{align}
Equations \eqref{eq:full-nonmetricity-column} and
\eqref{eq:full-tensor-torsion-column} must be solved as a coupled system.
Their separate vanishing is not required.
In order to project the connection field equations, we define
\begin{equation}
\mathcal P_{\alpha\mu\nu}
=
g_{\mu\rho}g_{\nu\sigma}
\mathcal P_\alpha{}^{\rho\sigma}.
\label{eq:lowered-connection-equation}
\end{equation}
The three vector contractions
\begin{equation}
g^{\mu\nu}\mathcal P_{\alpha\mu\nu},
\qquad
g^{\alpha\mu}\mathcal P_{\alpha\mu\nu},
\qquad
g^{\alpha\nu}\mathcal P_{\alpha\mu\nu}
\label{eq:three-vector-traces}
\end{equation}
do not determine $T_\mu,Q_\mu,\tilde Q_\mu$.  In the fully invariant
theory, they are precisely the vector Noether identities.  Similarly, the
axial contraction is the fourth identity and does not determine $S_\mu$.
We may therefore choose the projective gauge
\begin{equation}
T_\mu=0,
\qquad
Q_\mu=0,
\qquad
\tilde Q_\mu=0,
\qquad
S_\mu=0.
\label{eq:full-projective-gauge-choice}
\end{equation}
This is a gauge choice, not a consequence of the connection field equations.
After \eqref{eq:full-projective-gauge-choice}, nonmetricity reduces to
\begin{equation}
Q_{\alpha\mu\nu}
=
\Omega_{\alpha\mu\nu}.
\label{eq:Omega-after-gauge}
\end{equation}
The completely symmetric part of the connection equation is
\begin{equation}
\mathcal P_{(\alpha\mu\nu)}=0.
\label{eq:symmetric-connection-projection}
\end{equation}
Consequently, equation \eqref{eq:symmetric-connection-projection} reduces to an equation dependent only on $\Omega_{\alpha\mu\nu}$, since the torsion tensor does not have any totally symmetric part: 
\begin{equation}
-\frac32
\left(a_0+4c_4+2c_5\right)
\Omega_{(\alpha\mu\nu)}=0.
\label{eq:Omega-symmetric-field-equation}
\end{equation}
Therefore,
\begin{equation}
\Omega_{(\alpha\mu\nu)}=0
\label{eq:Omega-symmetric-solution}, 
\end{equation}
being $a_0+4c_4+2c_5\neq0$.
After \eqref{eq:Omega-symmetric-solution}, the tensor
$\Omega_{\alpha\mu\nu}$ has also the following properties:
\begin{equation}
\Omega_{(\alpha\mu\nu)}=0,
\qquad
\Omega_{\alpha\mu}{}^\mu=0,
\qquad
\Omega^\alpha{}_{\alpha\mu}=0.
\label{eq:Omega-mixed-properties}
\end{equation}
Therefore, there are other $16$ remaining components, which couple to
$t_{\alpha\mu\nu}$. In particular, the part
of the fully invariant Lagrangian of the action \eqref{eq:fully-invariant-expanded-action} containing the mixed tensor terms of torsion and nonmetricity can be written as
\begin{align}
&2(a_0-b_3)
\left[
\frac12\left(t_{\alpha\mu\nu}+2t_{\mu\nu\alpha}\right)
\right]
\left[
\frac12\left(t^{\alpha\mu\nu}+2t^{\mu\nu\alpha}\right)
\right]
\nonumber\\
&\hspace{3cm}
+\frac34(a_0-2c_4+2c_5)
\Omega_{\alpha\mu\nu}\Omega^{\alpha\mu\nu}
-3(a_0+a_3)\Omega_{\alpha\mu\nu}
\frac12\left(t^{\alpha\mu\nu}+2t^{\mu\nu\alpha}\right).
\label{eq:mixed-tensor-quadratic-form}
\end{align}
Varying \eqref{eq:mixed-tensor-quadratic-form} with respect to its two
independent mixed-symmetry tensors gives a factor $4(a_0-b_3)$ in the
first diagonal entry, a factor
$\frac32(a_0-2c_4+2c_5)$ in the second diagonal entry, and the same
off-diagonal factor $-3(a_0+a_3)$ in both equations.
The two independent projections of the complete connection equation onto
the equivalent sixteen-component mixed-symmetry representations give
\begin{equation}
\begin{pmatrix}
4(a_0-b_3)&-3(a_0+a_3)\\[1mm]
-3(a_0+a_3)&\dfrac32(a_0-2c_4+2c_5)
\end{pmatrix}
\begin{pmatrix}
\dfrac12\left(t_{\alpha\mu\nu}+2t_{\mu\nu\alpha}\right)\\[3mm]
\Omega_{\alpha\mu\nu}
\end{pmatrix}
=0.
\label{eq:full-mixed-tensor-block}
\end{equation}
The determinant of the matrix in
\eqref{eq:full-mixed-tensor-block} is
\begin{align}
|M_{\mathcal{QT}}^{tens}|=&4(a_0-b_3)\frac32(a_0-2c_4+2c_5)
-9(a_0+a_3)^2
=3\left[
2(a_0-b_3)(a_0-2c_4+2c_5)
-3(a_0+a_3)^2
\right],
\label{eq:mixed-tensor-determinant}
\end{align}
thus, this block has only the trivial solution when
\begin{equation}
2(a_0-b_3)(a_0-2c_4+2c_5)
-3(a_0+a_3)^2\neq0.
\label{eq:mixed-tensor-nondegeneracy}
\end{equation}
Under the non-degeneracy condition of
\eqref{eq:mixed-tensor-nondegeneracy}, and after the symmetric projection has
given the condition \eqref{eq:Omega-symmetric-solution}, we obtain that the tensorial part of the nonmetricity tensor is trivial,
\begin{equation}
\Omega_{\alpha\mu\nu}=0,
\qquad
\frac12\left(t_{\alpha\mu\nu}+2t_{\mu\nu\alpha}\right)=0.
\label{eq:mixed-tensor-trivial-solution}
\end{equation}
The second equation in \eqref{eq:mixed-tensor-trivial-solution} implies
$t_{\alpha\mu\nu}=0$.  To see this explicitly, it gives
\begin{equation}
t_{\alpha\mu\nu}=-2t_{\mu\nu\alpha}.
\end{equation}
Cyclic application yields
\begin{equation}
t_{\alpha\mu\nu}
=-2t_{\mu\nu\alpha}
=4t_{\nu\alpha\mu}
=-8t_{\alpha\mu\nu},
\end{equation}
and hence
\begin{equation}
9t_{\alpha\mu\nu}=0.
\end{equation}
Thus
\begin{equation}
t_{\alpha\mu\nu}=0.
\label{eq:tensor-torsion-solution}
\end{equation}
Finally, under the parameters satisfying
\begin{equation}
a_0+4c_4+2c_5\neq0
\label{eq:first-generic-condition}
\end{equation}
and
\begin{equation}
2(a_0-b_3)(a_0-2c_4+2c_5)
-3(a_0+a_3)^2\neq0,
\label{eq:second-generic-condition}
\end{equation}
all non-gauge tensor components vanish:
\begin{equation}
\Omega_{\alpha\mu\nu}=0,
\qquad
t_{\alpha\mu\nu}=0.
\label{eq:all-nongauge-components-zero}
\end{equation}

\bibliographystyle{utphys}
\bibliography{sample.bib}

@book{Will1993,
    author = "Will, Clifford M.",
    title = "{Theory and Experiment in Gravitational Physics}",
    year = "1993",
    doi = "10.1017/9781316338612",
    publisher = "Cambridge University Press"
}

@article{Ni_2016,
    author = "Ni, Wei-Tou",
    title = "{Solar-system tests of the relativistic gravity}",
    doi = "10.1142/S0218271816300032",
    journal = "Int. J. Mod. Phys. D",
    volume = "25",
    number = "14",
    pages = "1630003",
    year = "2016"
}

@article{De_Marchi_2020,
    author = "De Marchi, Fabrizio and Cascioli, Gael",
    title = "{Testing general relativity in the solar system: present and future perspectives}",
    doi = "10.1088/1361-6382/ab6ae0",
    journal = "Class. Quant. Grav.",
    volume = "37",
    number = "9",
    pages = "095007",
    year = "2020"
}

@article{Abbott2017-NS,
    author = "Abbott, B. P. and others",
    collaboration = "LIGO Scientific and Virgo",
    title = "{GW170817: Observation of Gravitational Waves from a Binary Neutron Star Inspiral}",
    doi = "10.1103/PhysRevLett.119.161101",
    journal = "Phys. Rev. Lett.",
    volume = "119",
    number = "16",
    pages = "161101",
    year = "2017"
}

@article{LIGOScientific2021psn,
    author = "Abbott, R. and others",
    collaboration = "LIGO Scientific, Virgo and KAGRA",
    title = "{Population of Merging Compact Binaries Inferred Using Gravitational Waves through GWTC-3}",
    eprint = "2111.03634",
    archivePrefix = "arXiv",
    primaryClass = "astro-ph.HE",
    reportNumber = "LIGO-P2100239",
    doi = "10.1103/PhysRevX.13.011048",
    journal = "Phys. Rev. X",
    volume = "13",
    number = "1",
    pages = "011048",
    year = "2023"
}

@article{Abbott2022,
    author = "Abbott, R. and others",
    collaboration = "LIGO Scientific, Virgo and KAGRA",
    title = "{GWTC-3: Compact Binary Coalescences Observed by LIGO and Virgo during the Second Part of the Third Observing Run}",
    eprint = "2111.03606",
    archivePrefix = "arXiv",
    primaryClass = "gr-qc",
    reportNumber = "LIGO-P2000318",
    doi = "10.1103/PhysRevX.13.041039",
    journal = "Phys. Rev. X",
    volume = "13",
    number = "4",
    pages = "041039",
    year = "2023"
}

@article{Stairs_2003,
    author = "Stairs, Ingrid H.",
    title = "{Testing General Relativity with Pulsar Timing}",
    doi = "10.12942/lrr-2003-5",
    journal = "Living Rev. Rel.",
    volume = "6",
    number = "1",
    pages = "5",
    year = "2003"
}

@article{Kramer_2021,
    author = "Kramer, M. and Stairs, I. H. and Manchester, R. N. and Wex, N. and Deller, A. T. and Coles, W. A. and Ali, M. and Burgay, M. and Camilo, F. and Cognard, I. and Damour, T. and Desvignes, G. and Ferdman, R. D. and Freire, P. C. C. and Grondin, S. and Guillemot, L. and Hobbs, G. B. and Janssen, G. and Karuppusamy, R. and Lorimer, D. R. and Lyne, A. G. and McKee, J. W. and McLaughlin, M. and M{\"u}nch, L. E. and Perera, B. B. P. and Pol, N. and Possenti, A. and Sarkissian, J. and Stappers, B. W. and Theureau, G.",
    title = "{Strong-Field Gravity Tests with the Double Pulsar}",
    doi = "10.1103/PhysRevX.11.041050",
    journal = "Phys. Rev. X",
    volume = "11",
    number = "4",
    pages = "041050",
    year = "2021"
}

@article{Hooft1974,
    author = "{'t Hooft}, G. and Veltman, M.",
    title = "{One-loop divergencies in the theory of gravitation}",
    journal = "Ann. Inst. H. Poincare Phys. Theor.",
    volume = "20",
    number = "1",
    pages = "69--94",
    year = "1974"
}

@article{Obukhov:2017pxa,
    author = "Obukhov, Yuri N.",
    title = "{Gravitational waves in Poincar{\'e} gauge gravity theory}",
    eprint = "1702.05185",
    archivePrefix = "arXiv",
    primaryClass = "gr-qc",
    doi = "10.1103/PhysRevD.95.084028",
    journal = "Phys. Rev. D",
    volume = "95",
    number = "8",
    pages = "084028",
    year = "2017"
}

@proceedings{Giulini:2003tc,
    editor = "Giulini, Domenico J. W. and Kiefer, Claus and Lammerzahl, Claus",
    title = "{Proceedings,  271st WE-Heraeus Seminar on Aspects of Quantum Gravity: From Theory to Experiment Search}: {Bad Honnef, Germany, February 25-March 1, 2002}",
    doi = "10.1007/b13561",
    volume = "631",
    pages = "pp.1--242",
    year = "2003"
}

@book{Rovelli:2004tv,
    author = "Rovelli, Carlo",
    title = "{Quantum gravity}",
    doi = "10.1017/CBO9780511755804",
    publisher = "Univ. Pr.",
    address = "Cambridge, UK",
    series = "Cambridge Monographs on Mathematical Physics",
    year = "2004"
}

@article{Perivolaropoulos:2021jda,
    author = "Perivolaropoulos, Leandros and Skara, Foteini",
    title = "{Challenges for {\ensuremath{\Lambda}}CDM: An update}",
    eprint = "2105.05208",
    archivePrefix = "arXiv",
    primaryClass = "astro-ph.CO",
    doi = "10.1016/j.newar.2022.101659",
    journal = "New Astron. Rev.",
    volume = "95",
    pages = "101659",
    year = "2022"
}

@article{DESI:2024mwx,
    author = "Adame, A. G. and others",
    collaboration = "DESI",
    title = "{DESI 2024 VI: cosmological constraints from the measurements of baryon acoustic oscillations}",
    eprint = "2404.03002",
    archivePrefix = "arXiv",
    primaryClass = "astro-ph.CO",
    reportNumber = "FERMILAB-PUB-24-0154-PPD",
    doi = "10.1088/1475-7516/2025/02/021",
    journal = "JCAP",
    volume = "02",
    pages = "021",
    year = "2025"
}

@article{Aloni:2021eaq,
    author = "Aloni, Daniel and Berlin, Asher and Joseph, Melissa and Schmaltz, Martin and Weiner, Neal",
    title = "{A Step in understanding the Hubble tension}",
    eprint = "2111.00014",
    archivePrefix = "arXiv",
    primaryClass = "astro-ph.CO",
    doi = "10.1103/PhysRevD.105.123516",
    journal = "Phys. Rev. D",
    volume = "105",
    number = "12",
    pages = "123516",
    year = "2022"
}

@article{DiValentino:2021izs,
    author = "Di Valentino, Eleonora and Mena, Olga and Pan, Supriya and Visinelli, Luca and Yang, Weiqiang and Melchiorri, Alessandro and Mota, David F. and Riess, Adam G. and Silk, Joseph",
    title = "{In the realm of the Hubble tension{\textemdash}a review of solutions}",
    eprint = "2103.01183",
    archivePrefix = "arXiv",
    primaryClass = "astro-ph.CO",
    reportNumber = "IPPP/20/108",
    doi = "10.1088/1361-6382/ac086d",
    journal = "Class. Quant. Grav.",
    volume = "38",
    number = "15",
    pages = "153001",
    year = "2021"
}

@article{Pantos:2026koc,
    author = "Pantos, Ioannis and Perivolaropoulos, Leandros",
    title = "{Status of the $S_8$ Tension: A 2026 Review of Probe Discrepancies}",
    eprint = "2602.12238",
    archivePrefix = "arXiv",
    primaryClass = "astro-ph.CO",
    doi = "10.1016/j.dark.2026.102286",
    journal = "Phys. Dark Univ.",
    volume = "52",
    pages = "102286",
    year = "2026"
}

@article{Ferraris:1982wci,
    author = "Ferraris, M. and Francaviglia, M. and Reina, C.",
    title = "{Variational formulation of general relativity from 1915 to 1925 {\textquotedblleft}Palatini's method{\textquotedblright} discovered by Einstein in 1925}",
    doi = "10.1007/BF00756060",
    journal = "Gen. Rel. Grav.",
    volume = "14",
    number = "3",
    pages = "243--254",
    year = "1982"
}

@article{Hehl:1994ue,
    author = "Hehl, Friedrich W. and McCrea, J. Dermott and Mielke, Eckehard W. and Ne'eman, Yuval",
    title = "{Metric affine gauge theory of gravity: Field equations, Noether identities, world spinors, and breaking of dilation invariance}",
    eprint = "gr-qc/9402012",
    archivePrefix = "arXiv",
    reportNumber = "TAUP-N192-94, TAUP-192-94",
    doi = "10.1016/0370-1573(94)00111-F",
    journal = "Phys. Rept.",
    volume = "258",
    pages = "1--171",
    year = "1995"
}

@article{Mayet:2016zxu,
    author = "Mayet, F. and others",
    title = "{A review of the discovery reach of directional Dark Matter detection}",
    eprint = "1602.03781",
    archivePrefix = "arXiv",
    primaryClass = "astro-ph.CO",
    doi = "10.1016/j.physrep.2016.02.007",
    journal = "Phys. Rept.",
    volume = "627",
    pages = "1--49",
    year = "2016"
}

@article{Capozziello:2017rvz,
    author = "Capozziello, S. and Jovanovi{\'c}, P. and Jovanovi{\'c}, V. Borka and Borka, D.",
    title = "{Addressing the missing matter problem in galaxies through a new fundamental gravitational radius}",
    eprint = "1702.03430",
    archivePrefix = "arXiv",
    primaryClass = "gr-qc",
    doi = "10.1088/1475-7516/2017/06/044",
    journal = "JCAP",
    volume = "06",
    pages = "044",
    year = "2017"
}

@article{Arun:2017uaw,
    author = "Arun, Kenath and Gudennavar, S. B. and Sivaram, C.",
    title = "{Dark matter, dark energy, and alternate models: A review}",
    eprint = "1704.06155",
    archivePrefix = "arXiv",
    primaryClass = "physics.gen-ph",
    doi = "10.1016/j.asr.2017.03.043",
    journal = "Adv. Space Res.",
    volume = "60",
    pages = "166--186",
    year = "2017"
}

@article{Carroll:2000fy,
    author = "Carroll, Sean M.",
    title = "{The Cosmological constant}",
    eprint = "astro-ph/0004075",
    archivePrefix = "arXiv",
    reportNumber = "EFI-2000-13",
    doi = "10.12942/lrr-2001-1",
    journal = "Living Rev. Rel.",
    volume = "4",
    pages = "1",
    year = "2001"
}

@article{Planck:2018vyg,
    author = "Aghanim, N. and others",
    collaboration = "Planck",
    title = "{Planck 2018 results. VI. Cosmological parameters}",
    eprint = "1807.06209",
    archivePrefix = "arXiv",
    primaryClass = "astro-ph.CO",
    doi = "10.1051/0004-6361/201833910",
    journal = "Astron. Astrophys.",
    volume = "641",
    pages = "A6",
    year = "2020",
    note = "[Erratum: Astron.Astrophys. 652, C4 (2021)]"
}

@article{Guth:2005zr,
    author = "Guth, Alan H. and Kaiser, David I.",
    title = "{Inflationary cosmology: Exploring the Universe from the smallest to the largest scales}",
    eprint = "astro-ph/0502328",
    archivePrefix = "arXiv",
    reportNumber = "MIT-CTP-3594",
    doi = "10.1126/science.1107483",
    journal = "Science",
    volume = "307",
    pages = "884--890",
    year = "2005"
}

@article{Einstein1916,
    author = "Einstein, A.",
    title = "{Die Grundlage der allgemeinen Relativit{\"a}tstheorie}",
    journal = "Annalen Phys.",
    year = "1916",
    volume = "354",
    number = "7",
    pages = "769--822",
    doi = "10.1002/andp.19163540702"
}

@article{BeltranJimenez:2019esp,
    author = "Beltr{\'a}n Jim{\'e}nez, Jose and Heisenberg, Lavinia and Koivisto, Tomi S.",
    title = "{The Geometrical Trinity of Gravity}",
    eprint = "1903.06830",
    archivePrefix = "arXiv",
    primaryClass = "hep-th",
    doi = "10.3390/universe5070173",
    journal = "Universe",
    volume = "5",
    number = "7",
    pages = "173",
    year = "2019"
}

@article{BeltranJimenez:2019odq,
    author = "Beltr{\'a}n Jim{\'e}nez, Jose and Heisenberg, Lavinia and Iosifidis, Damianos and Jim{\'e}nez-Cano, Alejandro and Koivisto, Tomi S.",
    title = "{General teleparallel quadratic gravity}",
    eprint = "1909.09045",
    archivePrefix = "arXiv",
    primaryClass = "gr-qc",
    doi = "10.1016/j.physletb.2020.135422",
    journal = "Phys. Lett. B",
    volume = "805",
    pages = "135422",
    year = "2020"
}

@article{Capozziello:2022zzh,
    author = "Capozziello, Salvatore and De Falco, Vittorio and Ferrara, Carmen",
    title = "{Comparing equivalent gravities: common features and differences}",
    eprint = "2208.03011",
    archivePrefix = "arXiv",
    primaryClass = "gr-qc",
    doi = "10.1140/epjc/s10052-022-10823-x",
    journal = "Eur. Phys. J. C",
    volume = "82",
    number = "10",
    pages = "865",
    year = "2022"
}

@article{Blixt:2023kyr,
    author = "Blixt, Daniel and Golovnev, Alexey and Guzman, Maria-Jose and Maksyutov, Ramazan",
    title = "{Geometry and covariance of symmetric teleparallel theories of gravity}",
    eprint = "2306.09289",
    archivePrefix = "arXiv",
    primaryClass = "gr-qc",
    doi = "10.1103/PhysRevD.109.044061",
    journal = "Phys. Rev. D",
    volume = "109",
    number = "4",
    pages = "044061",
    year = "2024"
}

@article{Guzman:2020kgh,
    author = "Guzman, Maria-Jose and Khaled Ibraheem, Shymaa",
    title = "{Classification of primary constraints for new general relativity in the premetric approach}",
    eprint = "2009.13430",
    archivePrefix = "arXiv",
    primaryClass = "gr-qc",
    doi = "10.1142/S021988782140003X",
    journal = "Int. J. Geom. Meth. Mod. Phys.",
    volume = "18",
    number = "supp01",
    pages = "2140003",
    year = "2021"
}

@article{Iosifidis:2023mot,
    author        = {Iosifidis, Damianos and Hehl, Friedrich W.},
    title         = {Motion of Test Particles in Spacetimes with Torsion and Nonmetricity},
    journal       = {Phys. Lett. B},
    volume        = {850},
    pages         = {138498},
    year          = {2024},
    doi           = {10.1016/j.physletb.2024.138498},
    eprint        = {2310.15595},
    archivePrefix = {arXiv},
    primaryClass  = {gr-qc}
}

@article{Sauro:2022proj,
  author        = {Sauro, Dario and Martini, Riccardo and Zanusso, Omar},
  title         = {Projective transformations in metric-affine and Weylian geometries},
  journal       = {Int. J. Geom. Meth. Mod. Phys.},
  volume        = {20},
  number        = {13},
  pages         = {2350237},
  year          = {2023},
  doi           = {10.1142/S0219887823502377},
  eprint        = {2208.10872},
  archivePrefix = {arXiv},
  primaryClass  = {hep-th}
}

@article{Percacci:2020ddy,
  author        = {Percacci, Roberto and Sezgin, Ergin},
  title         = {New class of ghost- and tachyon-free metric affine gravities},
  journal       = {Phys. Rev. D},
  volume        = {101},
  number        = {8},
  pages         = {084040},
  year          = {2020},
  doi           = {10.1103/PhysRevD.101.084040},
  eprint        = {1912.01023},
  archivePrefix = {arXiv},
  primaryClass  = {hep-th},
  note          = {Erratum: Phys. Rev. D 111 (2025) 109902}
}

@article{Barker:2024dhb,
    author = "Barker, Will and Zell, Sebastian",
    title = "{Consistent particle physics in metric-affine gravity from extended projective symmetry}",
    eprint = "2402.14917",
    archivePrefix = "arXiv",
    primaryClass = "hep-th",
    month = "2",
    year = "2024"
}

@article{Afonso:2017bxr,
    author = "Afonso, Victor I. and Bejarano, Cecilia and Beltran Jimenez, Jose and Olmo, Gonzalo J. and Orazi, Emanuele",
    title = "{The trivial role of torsion in projective invariant theories of gravity with non-minimally coupled matter fields}",
    eprint = "1705.03806",
    archivePrefix = "arXiv",
    primaryClass = "gr-qc",
    reportNumber = "IFIC-17-25",
    doi = "10.1088/1361-6382/aa9151",
    journal = "Class. Quant. Grav.",
    volume = "34",
    number = "23",
    pages = "235003",
    year = "2017"
}

@article{Bahamonde:2024efl,
    author = "Bahamonde, Sebastian and Gigante Valcarcel, Jorge",
    title = "{Stability in cubic metric-affine gravity}",
    eprint = "2411.12954",
    archivePrefix = "arXiv",
    primaryClass = "gr-qc",
    doi = "10.1103/PhysRevD.111.084058",
    journal = "Phys. Rev. D",
    volume = "111",
    number = "8",
    pages = "084058",
    year = "2025"
}

@article{Ferrara:2026hbq,
    author = "Ferrara, Carmen and Golovnev, Alexey and Guzm{\'a}n, Mar{\'\i}a Jos{\'e}",
    title = "{Primary Constraints of Newer General Relativity}",
    eprint = "2605.30221",
    archivePrefix = "arXiv",
    primaryClass = "gr-qc",
    doi = "10.1142/S021988782650297X",
    journal = "Int. J. Geom. Meth. Mod. Phys.",
    volume = "23",
    pages = "2650297",
    year = "2026"
}

@article{BeltranJimenez:2019acz,
    author = "Beltr{\'a}n Jim{\'e}nez, Jose and Delhom, Adria",
    title = "{Ghosts in metric-affine higher order curvature gravity}",
    eprint = "1901.08988",
    archivePrefix = "arXiv",
    primaryClass = "gr-qc",
    doi = "10.1140/epjc/s10052-019-7149-x",
    journal = "Eur. Phys. J. C",
    volume = "79",
    number = "8",
    pages = "656",
    year = "2019"
}

@article{Aoki:2023sum,
    author = "Aoki, Katsuki and Bahamonde, Sebastian and Gigante Valcarcel, Jorge and Gorji, Mohammad Ali",
    title = "{Cosmological perturbation theory in metric-affine gravity}",
    eprint = "2310.16007",
    archivePrefix = "arXiv",
    primaryClass = "gr-qc",
    reportNumber = "YITP-23-135",
    doi = "10.1103/PhysRevD.110.024017",
    journal = "Phys. Rev. D",
    volume = "110",
    number = "2",
    pages = "024017",
    year = "2024"
}

@article{Bahamonde:2022kwg,
    author = "Bahamonde, Sebastian and Chevrier, Johann and Gigante Valcarcel, Jorge",
    title = "{New black hole solutions with a dynamical traceless nonmetricity tensor in Metric-Affine Gravity}",
    eprint = "2210.05998",
    archivePrefix = "arXiv",
    primaryClass = "gr-qc",
    doi = "10.1088/1475-7516/2023/02/018",
    journal = "JCAP",
    volume = "02",
    pages = "018",
    year = "2023"
}

@article{Iosifidis:2019fsh,
    author = "Iosifidis, Damianos",
    title = "{Linear Transformations on Affine-Connections}",
    eprint = "1911.04535",
    archivePrefix = "arXiv",
    primaryClass = "gr-qc",
    doi = "10.1088/1361-6382/ab778d",
    journal = "Class. Quant. Grav.",
    volume = "37",
    number = "8",
    pages = "085010",
    year = "2020"
}

@article{BeltranJimenez:2019hrm,
    author = "Beltr{\'a}n Jim{\'e}nez, Jose and Maldonado Torralba, Francisco Jos{\'e}",
    title = "{Revisiting the stability of quadratic Poincar{\'e} gauge gravity}",
    eprint = "1910.07506",
    archivePrefix = "arXiv",
    primaryClass = "gr-qc",
    doi = "10.1140/epjc/s10052-020-8163-8",
    journal = "Eur. Phys. J. C",
    volume = "80",
    number = "7",
    pages = "611",
    year = "2020"
}

@article{delaCruz-Dombriz:2019tge,
    author = "de la Cruz-Dombriz, {\'A}lvaro and Maldonado Torralba, Francisco Jos{\'e} and Mazumdar, Anupam",
    title = "{Ghost-free higher-order theories of gravity with torsion}",
    eprint = "1911.08846",
    archivePrefix = "arXiv",
    primaryClass = "gr-qc",
    doi = "10.1140/epjc/s10052-021-09019-6",
    journal = "Eur. Phys. J. C",
    volume = "81",
    number = "3",
    pages = "240",
    year = "2021"
}

@article{Andrei:2024vvy,
    author = {Andrei, Ilaria and Iosifidis, Damianos and J{\"a}rv, Laur and Saal, Margus},
    title = "{Friedmann cosmology with hyperfluids}",
    eprint = "2411.19127",
    archivePrefix = "arXiv",
    primaryClass = "gr-qc",
    doi = "10.1103/PhysRevD.111.064063",
    journal = "Phys. Rev. D",
    volume = "111",
    number = "6",
    pages = "064063",
    year = "2025"
}

@article{Iosifidis:2021tvx,
    author = "Iosifidis, Damianos",
    title = "{Quadratic metric-affine gravity: solving for the affine-connection}",
    eprint = "2109.13293",
    archivePrefix = "arXiv",
    primaryClass = "gr-qc",
    doi = "10.1140/epjc/s10052-022-10499-3",
    journal = "Eur. Phys. J. C",
    volume = "82",
    number = "7",
    pages = "577",
    year = "2022"
}

@article{Matveev:2020wif,
    author = "Matveev, Vladimir S. and Scholz, Erhard",
    title = "{Light cone and Weyl compatibility of conformal and projective structures}",
    eprint = "2001.01494",
    archivePrefix = "arXiv",
    primaryClass = "math.DG",
    doi = "10.1007/s10714-020-02716-9",
    journal = "Gen. Rel. Grav.",
    volume = "52",
    number = "7",
    pages = "66",
    year = "2020"
}

@article{Obukhov:2021uor,
    author = "Obukhov, Yuri N. and Puetzfeld, Dirk",
    title = "{Demystifying autoparallels in alternative gravity}",
    eprint = "2105.08428",
    archivePrefix = "arXiv",
    primaryClass = "gr-qc",
    doi = "10.1103/PhysRevD.104.044031",
    journal = "Phys. Rev. D",
    volume = "104",
    number = "4",
    pages = "044031",
    year = "2021"
}

@article{Kleinert:1997hr,
    author = "Kleinert, Hagen and Shabanov, Sergei V.",
    title = "{Spaces with torsion from embedding and the special role of autoparallel trajectories}",
    eprint = "gr-qc/9709067",
    archivePrefix = "arXiv",
    doi = "10.1016/S0370-2693(98)00421-3",
    journal = "Phys. Lett. B",
    volume = "428",
    pages = "315--321",
    year = "1998"
}

@article{Csillag:2026kdc,
    author = "Csillag, Lehel and Voicu, Nicoleta and Elgendi, Salah and Pfeifer, Christian",
    title = "{On the Finsler variational nature of autoparallels in metric-affine geometry}",
    eprint = "2603.18416",
    archivePrefix = "arXiv",
    primaryClass = "math-ph",
    month = "3",
    year = "2026"
}

@article{Iosifidis:2018zwo,
    author = "Iosifidis, Damianos and Koivisto, Tomi",
    title = "{Scale transformations in metric-affine geometry}",
    eprint = "1810.12276",
    archivePrefix = "arXiv",
    primaryClass = "gr-qc",
    reportNumber = "NORDITA-2018-109",
    doi = "10.3390/universe5030082",
    journal = "Universe",
    volume = "5",
    pages = "82",
    year = "2019"
}

@article{Kleinert:1996yi,
    author = "Kleinert, H. and Pelster, A.",
    title = "{Lagrange mechanics in spaces with curvature and torsion}",
    eprint = "gr-qc/9605028",
    archivePrefix = "arXiv",
    doi = "10.1023/A:1026701613987",
    journal = "Gen. Rel. Grav.",
    volume = "31",
    pages = "1439",
    year = "1999"
}

@article{Fiziev:1995te,
    author = "Fiziev, P. and Kleinert, H.",
    title = "{New action principle for classical particle trajectories in spaces with torsion}",
    eprint = "hep-th/9503074",
    archivePrefix = "arXiv",
    doi = "10.1209/epl/i1996-00555-0",
    journal = "EPL",
    volume = "35",
    pages = "241--246",
    year = "1996"
}

@article{Rigouzzo:2023sbb,
    author = "Rigouzzo, Claire and Zell, Sebastian",
    title = "{Coupling metric-affine gravity to the standard model and dark matter fermions}",
    eprint = "2306.13134",
    archivePrefix = "arXiv",
    primaryClass = "gr-qc",
    doi = "10.1103/PhysRevD.108.124067",
    journal = "Phys. Rev. D",
    volume = "108",
    number = "12",
    pages = "124067",
    year = "2023"
}

@book{Peskin:1995ev,
    author = "Peskin, Michael E. and Schroeder, Daniel V.",
    title = "{An Introduction to quantum field theory}",
    doi = "10.1201/9780429503559",
    isbn = "978-0-201-50397-5, 978-0-429-50355-9, 978-0-429-49417-8",
    publisher = "Addison-Wesley",
    address = "Reading, USA",
    year = "1995"
}

@article{Delhom:2020hkb,
    author = "Delhom, Adri{\`a}",
    title = "{Minimal coupling in presence of non-metricity and torsion}",
    eprint = "2002.02404",
    archivePrefix = "arXiv",
    primaryClass = "gr-qc",
    doi = "10.1140/epjc/s10052-020-8330-y",
    journal = "Eur. Phys. J. C",
    volume = "80",
    number = "8",
    pages = "728",
    year = "2020"
}

@book{Schwabl08,
  title={Advanced Quantum Mechanics},
  author={Schwabl, Franz},
  edition={4th},
  year={2008},
  publisher={Springer},
  isbn={978-3-540-85061-8}
}

@article{Ruegg:2003ps,
    author = "Ruegg, Henri and Ruiz-Altaba, Marti",
    title = "{The Stueckelberg field}",
    eprint = "hep-th/0304245",
    archivePrefix = "arXiv",
    reportNumber = "UGVA-DPT-2003-04-1106, UGVA-DPT-04-1106",
    doi = "10.1142/S0217751X04019755",
    journal = "Int. J. Mod. Phys. A",
    volume = "19",
    pages = "3265--3348",
    year = "2004"
}

@article{Karananas:2021zkl,
    author = "Karananas, Georgios K. and Shaposhnikov, Mikhail and Shkerin, Andrey and Zell, Sebastian",
    title = "{Matter matters in Einstein-Cartan gravity}",
    eprint = "2106.13811",
    archivePrefix = "arXiv",
    primaryClass = "hep-th",
    reportNumber = "LMU-ASC 18/21, FTPI-MINN-21-10, UMN-TH-4017/21",
    doi = "10.1103/PhysRevD.104.064036",
    journal = "Phys. Rev. D",
    volume = "104",
    number = "6",
    pages = "064036",
    year = "2021"
}

@article{Marzo:2021esg,
    author = "Marzo, Carlo",
    title = "{Ghost and tachyon free propagation up to spin 3 in Lorentz invariant field theories}",
    eprint = "2108.11982",
    archivePrefix = "arXiv",
    primaryClass = "hep-ph",
    reportNumber = "MI-TH-1941",
    doi = "10.1103/PhysRevD.105.065017",
    journal = "Phys. Rev. D",
    volume = "105",
    number = "6",
    pages = "065017",
    year = "2022"
}

@article{Barnich:1993vg,
    author = "Barnich, Glenn and Henneaux, Marc",
    title = "{Consistent couplings between fields with a gauge freedom and deformations of the master equation}",
    eprint = "hep-th/9304057",
    archivePrefix = "arXiv",
    reportNumber = "ULB-PMIF-93-01",
    doi = "10.1016/0370-2693(93)90544-R",
    journal = "Phys. Lett. B",
    volume = "311",
    pages = "123--129",
    year = "1993"
}

@article{Deser:1963zzc,
    author = "Deser, Stanley and Arnowitt, R.",
    title = "{Interaction Among Gauge Vector Fields}",
    doi = "10.1016/0029-5582(63)90081-6",
    journal = "Nucl. Phys.",
    volume = "49",
    pages = "133--143",
    year = "1963"
}

@article{Jarv:2018bgs,
    author = {J{\"a}rv, Laur and R{\"u}nkla, Mihkel and Saal, Margus and Vilson, Ott},
    title = "{Nonmetricity formulation of general relativity and its scalar-tensor extension}",
    eprint = "1802.00492",
    archivePrefix = "arXiv",
    primaryClass = "gr-qc",
    doi = "10.1103/PhysRevD.97.124025",
    journal = "Phys. Rev. D",
    volume = "97",
    number = "12",
    pages = "124025",
    year = "2018"
}

@article{Utiyama:1956sy,
    author = "Utiyama, Ryoyu",
    editor = "Hsu, Jong-Ping and Fine, D.",
    title = "{Invariant theoretical interpretation of interaction}",
    doi = "10.1103/PhysRev.101.1597",
    journal = "Phys. Rev.",
    volume = "101",
    pages = "1597--1607",
    year = "1956"
}

@inproceedings{Pich:2012sx,
    author = "Pich, Antonio",
    title = "{The Standard Model of Electroweak Interactions}",
    booktitle = "{2010 European School of High Energy Physics}",
    eprint = "1201.0537",
    archivePrefix = "arXiv",
    primaryClass = "hep-ph",
    reportNumber = "IFIC-11-73, FTUV-12-0102",
    pages = "1--50",
    month = "1",
    year = "2012"
}

@article{Hayashi:1979wj,
    author = "Hayashi, Kenji and Shirafuji, Takeshi",
    title = "{Gravity from Poincare Gauge Theory of the Fundamental Particles. 1. Linear and Quadratic Lagrangians}",
    reportNumber = "UT-KOMABA-79-13",
    doi = "10.1143/PTP.64.866",
    journal = "Prog. Theor. Phys.",
    volume = "64",
    pages = "866",
    year = "1980",
    note = "[Erratum: Prog.Theor.Phys. 65, 2079 (1981)]"
}

@article{Brensinger:2024udu,
    author = "Brensinger, Samuel J. and Vecera, Patrick",
    title = "{Yang-Mills fields, spin, and torsion in dynamical projective gravitation}",
    eprint = "2404.02243",
    archivePrefix = "arXiv",
    primaryClass = "gr-qc",
    doi = "10.1103/PhysRevD.110.124017",
    journal = "Phys. Rev. D",
    volume = "110",
    number = "12",
    pages = "124017",
    year = "2024"
}

@article{Vignolo:2021frk,
    author = "Vignolo, Stefano and Carloni, Sante and Cianci, Roberto and Esposito, Fabrizio and Fabbri, Luca",
    title = "{Spinor fields in f(Q)-gravity}",
    eprint = "2112.03628",
    archivePrefix = "arXiv",
    primaryClass = "gr-qc",
    doi = "10.1088/1361-6382/ac36e3",
    journal = "Class. Quant. Grav.",
    volume = "39",
    number = "1",
    pages = "015009",
    year = "2022"
}

@article{Macias:1995hf,
    author = "Macias, A. and Mielke, E. W. and Morales-Tecotl, H. A. and Tresguerres, R.",
    title = "{Torsion and Weyl convector in metric affine models of gravity}",
    doi = "10.1063/1.531225",
    journal = "J. Math. Phys.",
    volume = "36",
    pages = "5868--5876",
    year = "1995"
}

@article{Higgs:1964ia,
    author = "Higgs, Peter W.",
    title = "{Broken symmetries, massless particles and gauge fields}",
    doi = "10.1016/0031-9163(64)91136-9",
    journal = "Phys. Lett.",
    volume = "12",
    pages = "132--133",
    year = "1964"
}

@article{Grover:2024whe,
    author = "Grover, Tyler and Stiffler, Kory and Vecera, Patrick",
    title = "{Covariant and manifestly projective invariant formulation of Thomas-Whitehead gravity}",
    eprint = "2404.04813",
    archivePrefix = "arXiv",
    primaryClass = "hep-th",
    doi = "10.1103/PhysRevD.110.084058",
    journal = "Phys. Rev. D",
    volume = "110",
    number = "8",
    pages = "084058",
    year = "2024"
}

@article{Blixt:2020ekl,
    author = "Blixt, Daniel and Guzm{\'a}n, Mar{\'\i}a-Jos{\'e} and Hohmann, Manuel and Pfeifer, Christian",
    title = "{Review of the Hamiltonian analysis in teleparallel gravity}",
    eprint = "2012.09180",
    archivePrefix = "arXiv",
    primaryClass = "gr-qc",
    doi = "10.1142/S0219887821300051",
    journal = "Int. J. Geom. Meth. Mod. Phys.",
    volume = "18",
    number = "supp01",
    pages = "2130005",
    year = "2021"
}

@article{ParticleDataGroup:2026mpi,
    author = "Takahashi, F. and others",
    collaboration = "Particle Data Group",
    title = "{Review of Particle Physics}",
    doi = "10.1142/s0217751x26300115",
    journal = "Int. J. Mod. Phys. A",
    volume = "41",
    number = "22",
    pages = "2630011",
    year = "2026"
}

@article{Einstein1905,
  author = {Einstein, Albert},
  title = {{\"U}ber einen die Erzeugung und Verwandlung des Lichtes betreffenden heuristischen Gesichtspunkt},
  journal = {Annalen der Physik},
  volume = {17},
  number = {6},
  pages = {132--148},
  year = {1905},
  doi = {10.1002/andp.19053220607}
}

@article{Dirac:1927dy,
    author = "Dirac, Paul A. M.",
    title = "{Quantum theory of emission and absorption of radiation}",
    doi = "10.1098/rspa.1927.0039",
    journal = "Proc. Roy. Soc. Lond. A",
    volume = "114",
    pages = "243",
    year = "1927"
}

@article{Weyl29,
author = {Hermann Weyl },
title = {GRAVITATION AND THE ELECTRON},
journal = {Proceedings of the National Academy of Sciences},
volume = {15},
number = {4},
pages = {323-334},
year = {1929},
doi = {10.1073/pnas.15.4.323},
URL = {https://www.pnas.org/doi/abs/10.1073/pnas.15.4.323},
eprint = {https://www.pnas.org/doi/pdf/10.1073/pnas.15.4.323}}

@article{Schwinger:1948iu,
    author = "Schwinger, Julian S.",
    title = "{On Quantum electrodynamics and the magnetic moment of the electron}",
    doi = "10.1103/PhysRev.73.416",
    journal = "Phys. Rev.",
    volume = "73",
    pages = "416--417",
    year = "1948"
}

@article{Feynman:1949zx,
    author = "Feynman, R. P.",
    editor = "Brown, L. M.",
    title = "{Space - time approach to quantum electrodynamics}",
    doi = "10.1103/PhysRev.76.769",
    journal = "Phys. Rev.",
    volume = "76",
    pages = "769--789",
    year = "1949"
}

@article{Dyson:1949bp,
    author = "Dyson, F. J.",
    title = "{The Radiation theories of Tomonaga, Schwinger, and Feynman}",
    doi = "10.1103/PhysRev.75.486",
    journal = "Phys. Rev.",
    volume = "75",
    pages = "486--502",
    year = "1949"
}

@article{Yang:1954ek,
    author = "Yang, Chen-Ning and Mills, Robert L.",
    editor = "Hsu, Jong-Ping and Fine, D.",
    title = "{Conservation of Isotopic Spin and Isotopic Gauge Invariance}",
    doi = "10.1103/PhysRev.96.191",
    journal = "Phys. Rev.",
    volume = "96",
    pages = "191--195",
    year = "1954"
}

@article{Feynman:1958ty,
    author = "Feynman, R. P. and Gell-Mann, Murray",
    editor = "Brown, L. M.",
    title = "{Theory of Fermi interaction}",
    doi = "10.1103/PhysRev.109.193",
    journal = "Phys. Rev.",
    volume = "109",
    pages = "193--198",
    year = "1958"
}

@article{Glashow:1961tr,
    author = "Glashow, S. L.",
    title = "{Partial Symmetries of Weak Interactions}",
    doi = "10.1016/0029-5582(61)90469-2",
    journal = "Nucl. Phys.",
    volume = "22",
    pages = "579--588",
    year = "1961"
}

@article{Goldstone:1962es,
    author = "Goldstone, Jeffrey and Salam, Abdus and Weinberg, Steven",
    title = "{Broken Symmetries}",
    doi = "10.1103/PhysRev.127.965",
    journal = "Phys. Rev.",
    volume = "127",
    pages = "965--970",
    year = "1962"
}

@article{Englert:1964et,
    author = "Englert, F. and Brout, R.",
    editor = "Taylor, J. C.",
    title = "{Broken Symmetry and the Mass of Gauge Vector Mesons}",
    doi = "10.1103/PhysRevLett.13.321",
    journal = "Phys. Rev. Lett.",
    volume = "13",
    pages = "321--323",
    year = "1964"
}

@article{Gell-Mann:1964ewy,
    author = "Gell-Mann, Murray",
    title = "{A Schematic Model of Baryons and Mesons}",
    doi = "10.1016/S0031-9163(64)92001-3",
    journal = "Phys. Lett.",
    volume = "8",
    pages = "214--215",
    year = "1964"
}

@article{Greenberg:1964pe,
    author = "Greenberg, O. W.",
    title = "{Spin and Unitary Spin Independence in a Paraquark Model of Baryons and Mesons}",
    doi = "10.1103/PhysRevLett.13.598",
    journal = "Phys. Rev. Lett.",
    volume = "13",
    pages = "598--602",
    year = "1964"
}

@article{Kibble:1967sv,
    author = "Kibble, T. W. B.",
    editor = "Taylor, J. C.",
    title = "{Symmetry breaking in nonAbelian gauge theories}",
    doi = "10.1103/PhysRev.155.1554",
    journal = "Phys. Rev.",
    volume = "155",
    pages = "1554--1561",
    year = "1967"
}

@article{Weinberg:1967tq,
    author = "Weinberg, Steven",
    title = "{A Model of Leptons}",
    doi = "10.1103/PhysRevLett.19.1264",
    journal = "Phys. Rev. Lett.",
    volume = "19",
    pages = "1264--1266",
    year = "1967"
}

@article{Kobayashi:1973fv,
    author = "Kobayashi, Makoto and Maskawa, Toshihide",
    title = "{CP Violation in the Renormalizable Theory of Weak Interaction}",
    reportNumber = "KUNS-242",
    doi = "10.1143/PTP.49.652",
    journal = "Prog. Theor. Phys.",
    volume = "49",
    pages = "652--657",
    year = "1973"
}

@article{Gross1973UltravioletBO,
  title={Ultraviolet Behavior of Non-Abelian Gauge Theories},
  author={David J. Gross and Frank Wilczek},
  journal={Physical Review Letters},
  year={1973},
  volume={30},
  pages={1343-1346},
  url={https://api.semanticscholar.org/CorpusID:9130789}
}

@article{tHooft:1971akt,
    author = "'t Hooft, Gerard",
    title = "{Renormalization of Massless Yang-Mills Fields}",
    doi = "10.1016/0550-3213(71)90395-6",
    journal = "Nucl. Phys. B",
    volume = "33",
    pages = "173--199",
    year = "1971"
}

@article{tHooft:1971qjg,
    author = "'t Hooft, Gerard",
    editor = "Taylor, J. C.",
    title = "{Renormalizable Lagrangians for Massive Yang-Mills Fields}",
    doi = "10.1016/0550-3213(71)90139-8",
    journal = "Nucl. Phys. B",
    volume = "35",
    pages = "167--188",
    year = "1971"
}

@article{GargamelleNeutrino:1974khg,
    author = "Hasert, F. J. and others",
    collaboration = "Gargamelle Neutrino",
    title = "{Observation of Neutrino Like Interactions without Muon or Electron in the Gargamelle Neutrino Experiment}",
    reportNumber = "CERN-TC-L-INT-74-1",
    doi = "10.1016/0550-3213(74)90038-8",
    journal = "Nucl. Phys. B",
    volume = "73",
    pages = "1--22",
    year = "1974"
}

@article{ALEPH:2005ab,
    author = "Schael, S. and others",
    collaboration = "ALEPH, DELPHI, L3, OPAL, SLD, LEP Electroweak Working Group, SLD Electroweak Group, SLD Heavy Flavour Group",
    title = "{Precision electroweak measurements on the $Z$ resonance}",
    eprint = "hep-ex/0509008",
    archivePrefix = "arXiv",
    reportNumber = "SLAC-R-774",
    doi = "10.1016/j.physrep.2005.12.006",
    journal = "Phys. Rept.",
    volume = "427",
    pages = "257--454",
    year = "2006"
}

@article{Weinberg:1978kz,
    author = "Weinberg, Steven",
    editor = "Deser, S.",
    title = "{Phenomenological Lagrangians}",
    reportNumber = "HUTP-78-A051A",
    doi = "10.1016/0378-4371(79)90223-1",
    journal = "Physica A",
    volume = "96",
    number = "1-2",
    pages = "327--340",
    year = "1979"
}

@article{Fritzsch:1973pi,
    author = "Fritzsch, H. and Gell-Mann, Murray and Leutwyler, H.",
    title = "{Advantages of the Color Octet Gluon Picture}",
    reportNumber = "CALT-68-409",
    doi = "10.1016/0370-2693(73)90625-4",
    journal = "Phys. Lett. B",
    volume = "47",
    pages = "365--368",
    year = "1973"
}

@article{Pauli:1934xm,
    author = "Pauli, W. and Weisskopf, V. F.",
    title = "{On Quantization of the Scalar Relativistic Wave Equation. (In German)}",
    journal = "Helv. Phys. Acta",
    volume = "7",
    pages = "709--731",
    year = "1934"
}

@article{Ivanenko:1983fts,
    author = "Ivanenko, D. and Sardanashvily, G.",
    title = "{The Gauge Treatment of Gravity}",
    doi = "10.1016/0370-1573(83)90046-7",
    journal = "Phys. Rept.",
    volume = "94",
    pages = "1--45",
    year = "1983"
}

\end{document}